\documentclass[preprint,11pt,authoryear]{article}

\usepackage{amssymb}
\usepackage{amsmath}

\usepackage{geometry}
\usepackage{amsfonts}
\usepackage{bm}
\usepackage{algorithm}
\usepackage{algorithmicx}
\usepackage[noend]{algpseudocode}
\usepackage{varwidth}
\usepackage{xspace}
\makeatletter
\def\BState{\State\hskip-\ALG@thistlm}
\makeatother

\usepackage{lineno}

\usepackage[english]{babel}
\makeatletter
\def\thmhead@plain#1#2#3{%
  \thmname{#1}\thmnumber{\@ifnotempty{#1}{ }\@upn{#2}}%
  \thmnote{ {\the\thm@notefont#3}}}
\let\thmhead\thmhead@plain
\makeatother

\usepackage{amsthm}
\usepackage{amssymb}
\usepackage{amsmath}
\usepackage{bm}
\usepackage{amsbsy}
\usepackage{bigints} %
\usepackage{relsize}
\usepackage{mathrsfs}
\usepackage{mathtools}
\usepackage{accents}
\usepackage{enumitem}
\usepackage{amsfonts}
\allowdisplaybreaks

\usepackage{hyperref}

 \usepackage{hhline}
\theoremstyle{plain}
\newtheorem{theorem}{Theorem}[section]

\newtheorem{proposition}[theorem]{Proposition}

\theoremstyle{definition}

\theoremstyle{remark}
\newtheorem*{remark}{Remark}

\usepackage{array}
\newcolumntype{?}{!{\vrule width 0.8pt}}

\makeatletter
\newcommand*\mysize{%
  \@setfontsize\mysize{5.5}{9}%
}
\makeatother

\makeatletter
\newcommand*\mysizeb{%
  \@setfontsize\mysizeb{6.5}{11}%
}
\makeatother

\usepackage{changepage}

\def\muhat{{\widehat\mu}}
\def\psihat{{\widehat\psi}}
\def\zetahat{{\widehat\zeta}}
\def\bzetatilde{{\widetilde\bzeta}}
\def\ba{\boldsymbol{a}}

\def\bh{\boldsymbol{h}}

\def\bs{\boldsymbol{s}}
\def\bt{\boldsymbol{t}}

\def\bx{\boldsymbol{x}}
\def\by{\boldsymbol{y}}

\def\bA{\boldsymbol{A}}

\def\bC{\boldsymbol{C}}
\def\bD{\boldsymbol{D}}

\def\bH{\boldsymbol{H}}
\def\bI{\boldsymbol{I}}

\def\bM{\boldsymbol{M}}

\def\bO{\boldsymbol{O}}

\def\bQ{\boldsymbol{Q}}

\def\bT{\boldsymbol{T}}
\def\bU{\boldsymbol{U}}
\def\bV{\boldsymbol{V}}

\def\Cov{\mathbb{C}\text{ov}}

\DeclareMathOperator*{\blockdiag}{blockdiag}

\def\Cov{\mathbb{C}\text{ov}}
\def\indsim{\stackrel{\text{ind.}}{\sim}}
\DeclareMathOperator*{\diag}{diag}

\DeclareMathOperator{\E}{\mathbb{E}}

\DeclareMathOperator{\Var}{\mathbb{V}ar}
\DeclareMathOperator{\ind}{\mathbb{I}}
\DeclareMathOperator*{\tr}{tr}
\DeclareMathOperator{\vect}{vec}
\DeclareMathOperator{\vech}{vech}
\DeclareMathOperator*{\stack}{stack}

\newcommand{\T}[1]{#1^{\intercal}}

\newcommand{\msg}[2]{m_{#1 \rightarrow #2}}

\newcommand{\np}[2]{\boldsymbol\eta_{#1 \rightarrow #2}}

\newcommand{\npq}[1]{\boldsymbol\eta_{q (#1)}}

\def\bzero{\boldsymbol{0}}
\def\bone{\boldsymbol{1}}

\def\bepsilon{\boldsymbol{\varepsilon}}

\def\bzeta{\boldsymbol{\zeta}}
\def\bdeta{\boldsymbol{\eta}}
\def\btheta{\boldsymbol{\theta}}

\def\bmu{\boldsymbol{\mu}}
\def\bnu{\boldsymbol{\nu}}

\def\bsigma{\boldsymbol{\sigma}}

\def\bpsi{\boldsymbol{\psi}}

\def\bLambda{\boldsymbol{\Lambda}}

\def\bXi{\boldsymbol{\Xi}}

\def\bSigma{\boldsymbol{\Sigma}}

\def\bPsi{\boldsymbol{\Psi}}

\def\Hsc{{\mathcal H}}

\def\zetatilde{{\widetilde\zeta}}
\def\normal{\text{N}}
\def\hc{\text{Half-Cauchy}}

\def\invchisq{\text{Inverse–}\chi^2}

\def\const{\text{const.}}

\def\const{\text{const.}}

\newcommand{\sigsqeps}[1]{\sigma^{(#1) 2}_{\epsilon}}
\newcommand\aeps[1]{a_{\epsilon}^{(#1)}}

\newcommand\Sigmamu[1]{\bSigma_\mu^{(#1)}}
\newcommand\Sigmapsi[2]{\bSigma_{\psi_{#1}}^{(#2)}}
\newcommand{\betamu}[2]{\beta_{\mu, #1}^{(#2)}}
\newcommand{\umu}[2]{u_{\mu, #1}^{(#2)}}
\newcommand\betapsi[3]{\beta_{\psi_{#1}, #2}^{(#3)}}
\newcommand\upsi[3]{u_{\psi_{#1}, #2}^{(#3)}}
\newcommand\numu[1]{\bnu_\mu^{(#1)}}
\newcommand\numuT[1]{\bnu_\mu^{(#1) \intercal}}
\newcommand\nupsi[2]{\bnu_{\psi_{#1}}^{(#2)}}
\newcommand\nupsiT[2]{\bnu_{\psi_{#1}}^{(#2) \intercal}}
\newcommand\C[2]{\bC_{#1}^{(#2)}}
\newcommand\CT[2]{\bC_{#1}^{(#2) \intercal}}

\newcommand\Vpsi[1]{\bV_\psi^{(#1)}}
\newcommand\VpsiT[1]{\bV_\psi^{(#1) \intercal}}
\newcommand\sigsqmu[1]{\sigma_{\mu}^{(#1) 2}}
\newcommand\amu[1]{a_{\mu}^{(#1)}}

\newcommand\sigsqpsi[2]{\sigma_{\psi_{#1}}^{(#2) 2}}
\newcommand\apsi[2]{a_{\psi_{#1}}^{(#2)}}

\newcommand\hmu[2]{h_{\mu, #1}^{(#2)}}

\newcommand\hmupsi[2]{\bh_{\mu \psi, #1}^{(#2)}}
\newcommand\hmupsiT[2]{\bh_{\mu \psi, #1}^{(#2) \intercal}}
\newcommand\Hpsi[2]{\bH_{\psi, #1}^{(#2)}}

\newcommand\tni[1]{\text{terms not involving $#1$}}
\def\xhat{{\widehat x}}

\algnewcommand\algorithmicinput{\textbf{Data Inputs:}}
\algnewcommand\DataInputs{\item[\algorithmicinput]}
\algnewcommand\HyperparameterInputs{\item[{\textbf{Hyperparameter Inputs:}}]}
\algnewcommand\ParameterInputs{\item[{\textbf{Parameter Inputs:}}]}
\algnewcommand\ParameterOutputs{\item[{\textbf{Parameter Outputs:}}]}
\algnewcommand\Inputs{\item[{\textbf{Inputs:}}]}
\algnewcommand\Outputs{\item[{\textbf{Outputs:}}]}
\algnewcommand\Updates{\item[{\textbf{Updates:}}]}
\algnewcommand\Initialise{\item[{\textbf{Initialise:}}]}
\algnewcommand\Initialize{\item[{\textbf{Initialize:}}]}

\usepackage{authblk}
\usepackage{xcolor}

\usepackage[normalem]{ulem}

\usepackage[round]{natbib}

\title{Efficient Bayesian functional principal component analysis of irregularly-observed multivariate curves} %
\author[1,2]{Tui H. Nolan}
\author[1]{Sylvia Richardson}
\author[1]{H\'{e}l\`{e}ne Ruffieux\footnote{\noindent Corresponding author: \href{mailto:helene.ruffieux@mrc-bsu.cam.ac.uk}{helene.ruffieux@mrc-bsu.cam.ac.uk}.}}
\affil[1]{MRC Biostatistics Unit, University of Cambridge, United Kingdom}
\affil[2]{University of Technology Sydney, Australia}

\begin{document}

\maketitle

\section*{\centering Abstract}
{\footnotesize
The analysis of multivariate functional curves %
has the potential to 
yield  %
important scientific discoveries in %
domains such as healthcare, medicine, economics and social sciences. 
However, it is common for real-world settings to present longitudinal data that are both irregularly and sparsely observed, which introduces important challenges for the current functional data methodology. %
A Bayesian hierarchical framework for multivariate functional principal component analysis is proposed, %
which accommodates the intricacies of such irregular observation settings by flexibly %
pooling information across subjects and correlated curves. %
The model represents %
common latent dynamics %
via shared functional principal component 
scores, 
thereby effectively 
borrowing strength across curves while circumventing the computationally challenging task of estimating covariance matrices. These scores also provide a parsimonious representation of the major modes of joint variation of the curves and constitute interpretable scalar summaries that can be employed in follow-up analyses. Estimation is carried out using %
variational %
inference, which combines efficiency, modularity and approximate posterior density estimation, enabling the joint analysis of %
large datasets with parameter uncertainty quantification. %
Detailed simulations assess the %
effectiveness of the approach in sharing information from sparse and irregularly sampled multivariate curves. 
The methodology is also exploited %
to estimate the %
molecular disease courses of individual patients with SARS-CoV-2 infection %
and 
characterise patient heterogeneity in  recovery outcomes; %
this study reveals key %
coordinated dynamics across the immune, inflammatory and metabolic systems, which are associated with survival and long-COVID symptoms up to one year post disease onset. The approach is implemented in the R package \texttt{bayesFPCA}.
}

\section{Introduction}
\label{sec:intro}

The availability of longitudinal datasets is rising sharply, benefitting from the progress of technologies and monitoring tools  in domains where cross-sectional studies were previously the norm. This paradigm shift calls for principled statistical approaches that can flexibly model curves and %
uncover shared dynamics across them %
to enhance statistical power and interpretation. 

Functional data analysis (FDA) has enjoyed increasing applicability in various areas, including neuroimaging
\citep{wangwwo19} and wearable technology \citep{goldsmith15}. Functional principal components analysis
(FPCA) is a key technique in this field that is used for dimensionality reduction on the inherently infinite-dimensional
functional data. The resulting principal components and scores can then be used for further analysis, such
as an orthogonal basis and uncorrelated covariates in functional regression models. Such techniques, however,
are classically restricted to univariate datasets, but more realistic problems in biomedical research involve numerous functional observations
for each subject. In fact, the current article was inspired by longitudinal measurements on
several molecular markers to characterise systemic recovery from SARS-CoV-2 infection \citep{bergamaschi2021longitudinal, ruffieux2023patient}. 

The multivariate setting for FPCA, which we call multivariate FPCA (mFPCA), was investigated by
\citet[Chapter~8.5]{ramsay05} for dense functional data, where they proposed concatenating the multivariate
functional data and applying multivariate PCA to the resulting sample of long vectors. However, real-world multivariate
functional data generally involves complicated observation settings, with sparse and irregular
data on correlated variables, possibly also irregularly sampled across different subjects. %
For irregular multivariate functional data, extensions of covariance-based FPCA methods
\citep{yao05} have become popular, permitting adaptations to multivariate longitudinal datasets. \citet{happ18}
estimate covariances and cross-covariances via the scores estimated from univariate FPCA, however
accurate covariance estimation may be difficult for sparse functional data because scores from univariate
FPCA are shrunk towards zero. \citet{li19} estimate the covariance and cross-covariance function
via B-spline smoothing through a tensor product formulation, which provides fast and accurate estimation
in the presence of sparse functional data. Once an estimate of the covariance function has been attained,
eigenfunctions and scores can be gathered via an eigendecomposition.

Although covariance-based mFPCA methods have provided a means of analysing sparse functional
data, the bivariate smoothing operation can be a computationally challenging task for large datasets.
In the case of
univariate FPCA, Bayesian implementations allow for direct inference on the eigenfunctions and
scores without requiring an estimate of the covariance function
as all quantities are considered unknown and estimated jointly. Such approaches have built on or are
similar to the probabilistic PCA framework that was introduced by \citet{tipping99}
and \citet{bishop99}. \citet{james00} used an expectation maximisation algorithm for estimation and inference in
the context of sparsely observed curves. Variational inference for FPCA was introduced by \citet{vanderlinde08} via a
generative model with a factorised approximation of the full posterior density function. \citet{goldsmith15} introduced
a fully Bayes framework for multilevel function-on-scalar regression models with FPCA applied to two levels of residuals.
\citet{nolan23} introduced a variational message passing framework for Bayesian inference on the model
parameters. The parameters in the latter approach are updated by messages passed between computational units
known as fragments \citep{wand17}, facilitating efficient extensions of univariate Bayesian FPCA to more elaborate models,
which \citet{nolan23} achieved for the multilevel setting. Finally, selecting the numbers of functional principal components and basis functions for representing the latent components %
is also an important problem. Although such choices are typically not the object of extensive sensitivity analysis  in the FPCA literature, various approaches have been described, based on %
cross-validation \citep[e.g.,][]{huang2008functional}, information criteria \citep[e.g., AIC or BIC;][]{yao05},  %
estimation of the variance explained by the components \citep[e.g.,][]{greven11}, or Bayesian model selection by placing prior distributions on these numbers \citep[e.g.,][]{suarez17}. %
In some of these approaches, inference is performed with an upper bound on the number of components, relying on regularisation to adaptively discard the superfluous components. %

While Bayesian mFPCA remains underdeveloped, Bayesian methodologies for various univariate and multivariate FDA
models have been established. In particular, in the context of Bayesian functional latent factor models, a substantial focus has been placed on achieving ordered shrinkage and rank selection. Notable approaches include the use of \emph{multiplicative gamma process shrinkage priors}, as proposed by \citet{bhattacharya2011sparse, montagna2012bayesian}, and \emph{cumulative shrinkage process priors}, as proposed by \citet{legramanti2020bayesian, kowal2023semiparametric}. These approaches employ prior distributions on factors or loading coefficients to encourage increasing shrinkage on higher-index factors, %
effectively removing irrelevant components and encoding ordering constraints.

In this article, we present a variational mFPCA framework based on a Bayesian hierarchical model that allows borrowing information across related variables, from sparsely or irregularly sampled functional curves. We employ a generalisation of the Karhunen–Loève theorem for %
multivariate Hilbert spaces \citep{happ18} to implement a direct representation of the curves as multivariate Karhunen–Loève expansions, with joint estimation of %
all model parameters. %
Specifically, we formulate a hierarchical functional factor model for multivariate curves, which enforces the necessary regularisation on variable-specific factor loading curves. Rather than encoding identifiability constraints into the model itself, such as via the above-mentioned dedicated prior formulations proposed in the context of FDA, we estimate the components up to a rotation of the parameter space and subsequently restore orthonormalisation and ordered contributions to the data variability, required by the mFPCA decomposition. %
We develop variational message passing (VMP) and mean-field variational Bayes (MFVB) algorithms for our proposed model. Variational inference constitutes a scalable alternative to MCMC inference whose computational burden can be prohibitive for the 
large multivariate functional settings we are interested in. Crucially, it also permits approximating the posterior distribution of all parameters, unlike other approximate Bayesian inference approaches, such as the expectation-maximisation algorithm, which only provide point estimates. We evaluate the statistical and computational performance of our approach in simulations emulating real-data settings, benchmarking it against MCMC inference on the same model, as well as against the frequentist mFPCA approach of \citet{happ18}, the Bayesian semiparametric latent factor model approach of \citet{kowal2023semiparametric} and separate applications of Bayesian univariate FPCA \citep{nolan23}. Doing so, we illustrate the effectiveness of our joint framework in pulling information across sparsely and irregularly sampled curves to improve estimation of subject-level trajectories, principal component scores and latent functions, and we assess the impact of model misspecification. We then exploit our approach to clarify the latent dynamics driving patient-to-patient variability in  recovery from COVID-19 using detailed longitudinal data covering one year post infection.

This article is organised as follows. Section~\ref{sec:motivation} presents the COVID-19 study, and motivates the need for new methodology to tackle biomedical research questions based on complex longitudinal measurement settings. %
Section~\ref{sec:mfpca} recalls the multivariate Karhunen–Loève decomposition and presents our joint hierarchical model for mFPCA. Section~\ref{sec:vbi} details our variational inference framework and %
describes a post-variational procedure to obtain orthonormal eigenfunctions and uncorrelated scores. Section~\ref{sec:simul} presents the results of a series of simulation studies. Section~\ref{sec:application} applies our approach to the COVID-19 study and discusses the possible biomedical implications of our findings, focusing on %
how %
the immune, metabolic and inflammatory systems jointly coordinate organismal recovery. %
Section~\ref{sec:conclusion} summarises our work and suggests further methodological developments. We provide a software implementation for our approach as an R package called \texttt{bayesFPCA}.

\section{Data and motivating example}\label{sec:motivation}

COVID-19 is a systemic disease, causing widespread dysregulation across the immune, metabolic and inflammatory systems  \citep{lucas2020longitudinal, masuda2021integrative}. While biological alterations resolve for most individuals soon after the acute phase, they can also be associated with short- and long-term complications, such as ICU admission, prolonged symptoms (``long COVID'') or death. %
Evolution of clinical and molecular parameters over time has also been shown to be very heterogeneous between patients  \citep{holmes2021incomplete, peluso2021markers}, but the mechanisms underlying the different disease dynamics, and the coordination of these dynamics across biological systems, remain largely unclear. Such an understanding %
could guide the development therapeutic strategies to anticipate and prevent %
serious disease trajectories, %
as well as help formulate individualised recommendations for patients with %
 long COVID. It could also provide a basis for studying other infectious diseases, such as caused by the Epstein-Barr virus (EBV) for example.

The numerous studies conducted over the past years have achieved varying degrees of success. %
These research efforts have however underscored the necessity of coupling access to detailed data, obtained from %
patient samples assayed repeatedly over months post infection, and use of principled statistical approaches tailored to longitudinal multi-parameter settings. %
Such approaches %
should be equipped to %
(i)~jointly model several blood markers across different systems, in order to quantify coordinated alterations in organismal functions; (ii)~disentangle the inter- and intra-patient temporal covariation of these alterations over the course of illness; (iii)~reconstruct the marker trajectories at the patient level; (iv)~uncover latent dynamics driving incomplete recovery in order to relate them to biological pathways influencing the risk of death and of long COVID. 

Here we propose to develop new methodology based on these desiderata, and apply it  to the analysis of %
data comprising longitudinal measurements of polar metabolites, serum cytokines and C-reactive protein levels from a cohort of $215$ infected subjects with different clinical severities \citep{bergamaschi2021longitudinal, ruffieux2023patient}. Specifically, we will model the disease trajectories of symptomatic SARS-CoV-2 PCR-positive patients admitted to Cambridge Hospitals (CITIID-NIHR BioResource COVID-
19 Collaboration), and compare them with measurements from uninfected individuals as controls. 

This problem requires analysis tools adapted to sparse multivariate functional settings, %
since multiple markers are quantified for each patient over time, and some are scarcely observed (the number of observations per patient and marker may be as low as two). %
Moreover, the observation grid is irregular, as it differs both across markers for a same patient, and across patients for a same marker. Finally, the dataset is relatively large: it comprises longitudinal measurements across multiple cellular and molecular markers, for tens of subjects. %
As explained before, these characteristics (large-data setting with sparse observations on an irregular grid) are poorly handled by covariance-based multivariate frequentist FPCA approaches, as they induce unwanted shrinkage and computational intractabilities. This, and the need to flexibly borrow strength between markers from different biological systems, motivates the development of a Bayesian joint FPCA approach that can effectively pull information to estimate shared latent dynamics from related, yet sparsely observed curves. 

We will %
detail our methodology in the next sections, and employ it on the COVID-19 data in Section~\ref{sec:application} to estimate %
individual disease trajectories as well as patient-level scores summarising the disease dynamics of each patient. This will hopefully allow us to identify %
latent kinetics that are common to multiple markers and that drive patient heterogeneity, thereby clarifying how the different biological systems coordinate the organismal response to infection. We will also exploit survival data and patient questionnaires on long-COVID symptoms to illustrate the added value of our joint approach compared to separate univariate FPCA analyses of individual markers.

\section{Multivariate functional principal components analysis}
\label{sec:mfpca}

\subsection{Karhunen–Loève representation of multivariate functional data}\label{sec:mfpca:kl}

Univariate FPCA is concerned with dimensionality reduction of independent realisations of a random function $x \in L^2 ([0, 1])$
into a finite eigenbasis that is constructed from the leading eigenfunctions of the covariance operator of $x$. In the multivariate
setting, random functions are replaced by row vectors of random functions $x (\bt) = [ \begin{array}{ccc} x^{(1)} (t_1) & \cdots &
x^{(p)} (t_p) \end{array} ]$ for $\bt = \T{(t_1, \dots, t_p)} \in [0, 1]^p$. The vector random functions are elements of the Hilbert
space $\Hsc \equiv L^2 ([0, 1])^p$, %
which is equipped with the norm defined by the inner product:
\begin{equation}
	\langle f, g \rangle_{\Hsc} \equiv \sum_{j = 1}^p \int_0^1 f^{(j)} (t) g^{(j)} (t) dt, \quad f, g \in \Hsc.
\label{H_inn_prod}
\end{equation}
Henceforth, the elements of $\Hsc$ will be referred to simply as \emph{random functions}.

For $x \in \Hsc$, the mean function is defined as $\mu (\bt) \equiv [ \begin{array}{ccc} \E \{ x^{(1)} (t_1) \} & \cdots &
\E \{ x^{(p)} (t_p) \} \end{array} ]$. Next, define the matrix of covariances $C (\bs, \bt)$ for $\bs, \bt \in [0, 1]^p$,  
with $(j, j')$-entry
\begin{equation}
	C_{jj'} (s_j, t_{j'})= \Cov \{ x^{(j)}(s_j), x^{(j')}(t_{j'}) \}, \quad s_j, t_{j'} \in [0, 1]. %
\label{mult_cov_mat}
\end{equation}
Finally,
we define the covariance operator $\Sigma: \Hsc \rightarrow \Hsc$ with $j'$th element of $\Sigma f$, $f \in \Hsc$
given by

\[
	(\Sigma f)^{(j')} (t_{j'}) \equiv \sum_{j = 1}^p \int_0^1 C_{jj'} (s_{j}, t_{j'}) f^{(j)}(s_{j}) ds_{j}, \quad t_{j'} \in [0, 1].
\]

\noindent According to the technical details of Proposition 2 of \citet{happ18}, there exists a complete orthonormal
basis of eigenfunctions $\psi_l \in \Hsc$, for $l \in \mathbb{N}$, of $\Sigma$ such that $\Sigma \psi_l = \lambda_l \psi_l$
with $\lambda_l \rightarrow 0$ as $l \rightarrow \infty$.

These are the main ingredients for establishing the multivariate version of the FPCA decomposition. First, we have
the multivariate version of Mercer's Theorem, which states that 
$C_{jj}(s_{j}, t_{j}) = \sum_{l=1}^\infty \lambda_l \psi_l^{(j)} (s_{j})
\psi_l^{(j)} (t_{j})$, for $j = 1, \ldots, p$ and $s_j, t_{j} \in [0, 1]$. 
Next, for a set of independent realisations $x_i(\bt)$, $i = 1, \ldots, n,$ the multivariate Karhunen–Lo\`{e}ve decomposition
is the basis for the mFPCA expansion \citep{yao05}:
\begin{equation}
	x_i (\bt) = \mu (\bt) + \sum_{l=1}^\infty \zeta_{il} \psi_l (\bt), \quad i = 1, \dots, n, \quad \bt \in [0, 1]^p,
\label{kl_expansion}
\end{equation}
 where $\zeta_{il} = \langle \{ x_i - \mu \}, \psi_l \rangle_\Hsc$ are the principal component
scores. The $\zeta_{il}$ are independent across $i$ and uncorrelated across $l$, with $\E (\zeta_{il}) = 0$
and $\Var (\zeta_{il}) = \lambda_l$. The decay of the eigenvalues with increasing $l$ ensures that we can
truncate the sum in \eqref{kl_expansion} for large enough $L$, such that
\begin{equation}
	\xhat_i (\bt) = \mu (\bt) + \sum_{l=1}^L \zeta_{il} \psi_l (\bt), \quad i = 1, \dots, n, \quad \bt \in [0, 1]^p.
\label{mfpca_decomp}
\end{equation}
Each observation $x_i$ can then be represented by its vector of scores $\bzeta_i =
\T{(\zeta_{i1}, \dots, \zeta_{iL})}$ and used for further analysis, such as regression \citep{muller05}. The key
difference between mFPCA and applying univariate FPCA to each of the $p$ variables is that there is only
one score $\zeta_{il}$ for the $l$th multivariate eigenfunction in the former approach, whereas there would
be a separate score for each element %
of the $l$th eigenfunction in the latter approach. In this way, mFPCA
inherently accounts for correlations between the variables, thereby borrowing information across them and producing a common score for each component $l = 1, \ldots, L$. %

As in univariate FPCA \citep[Section~2]{nolan23}, expansions similar to \eqref{mfpca_decomp} are also possible,
where
\begin{equation}
	\xhat_i (\bt) \equiv \mu (\bt) + \sum_{l=1}^L z_{il} \ h_l (\bt), \quad i = 1, \dots, n, \quad \bt \in [0, 1]^p,
\label{xhat_not_orthogonal}
\end{equation}
where $z_{il}$ are correlated across $l$, but remain independent across $i$, and the $h_l$ are not
orthonormal. Theorem \ref{thm:orth_basis} is a generalisation of Theorem 2.1 of \citet{nolan23}, and it
shows that an orthogonal decomposition of the resulting basis functions
and weights is sufficient for establishing the appropriate estimates \eqref{mfpca_decomp} from
\eqref{xhat_not_orthogonal}.

\begin{theorem}
	
	Given the decomposition in \eqref{xhat_not_orthogonal}, there exists a unique set of orthonormal
	eigenfunctions $\psi_1, \dots, \psi_L$ and an uncorrelated set of scores $\zeta_{i1}, \dots, \zeta_{iL}$,
	$i = 1, \dots, n$, such that $\xhat_i (\bt) = \mu (\bt) + \sum_{l=1}^L \zeta_{il} \ \psi_l (\bt)$, $\bt \in [0, 1]^p$.
	
\label{thm:orth_basis}
\end{theorem}

\begin{remark}
	
	The form of the proof of Theorem \ref{thm:orth_basis} is identical to that of Theorem 2.1 of \citet{nolan23},
	with inner products and norms in $L^2$ replaced by those in $\Hsc$. Therefore, we refer the reader to
	the proof of Theorem 2.1 of \citet{nolan23} in Section A of their online supplementary material.
	
\label{rmk:orth_basis}
\end{remark}

Theorem \ref{thm:orth_basis} permits direct estimation of the scores and eigenfunctions in the multivariate
Karhunen–Lo\`{e}ve decomposition \eqref{kl_expansion}, without initially estimating a covariance function
as in \citet{happ18} and \citet{li19}. According to Theorem \ref{thm:orth_basis}, we can simply orthogonalise
the basis functions and decorrelate the weights to gather estimate of the orthonormal eigenfunctions and
uncorrelated scores. There are several advantages in this method in that it does not require
estimation or smoothing of a large covariance and can more directly handle sparse or irregular functional data.

\subsection{A Bayesian hierarchical model for mFPCA}
\label{sec:bayes_mod}

In practice, the functional data are collected as a set of noisy observations over discrete points in time. Let the
set of design points for the $i$th subject's measurements on the $j$th variable be summarised by the vector
$\bt_i^{(j)} \equiv \T{(t_{i1}^{(j)}, \dots, t_{in_i^{(j)}}^{(j)})}$. Then, the corresponding vector of observations
is given by $\bx_i^{(j)} \equiv x_i^{(j)} (\bt_i^{(j)}) + \bepsilon_i^{(j)}$, where $\bepsilon_i^{(j)} \indsim \normal (\bzero_{n_i^{(j)}},
\sigsqeps{j} \bI_{n_i^{(j)}})$. In addition, we set $\bmu_i^{(j)} \equiv \mu^{(j)} (\bt_i^{(j)})$ and
$\bpsi_{il}^{(j)} \equiv \psi_l^{(j)} (\bt_i^{(j)})$. Then, the multivariate Karhunen–Lo\`{e}ve decomposition
in \eqref{kl_expansion} takes the form:
\begin{equation}
	\bx_i^{(j)} = \bmu_i^{(j)} + \sum_{l = 1}^L \zeta_{il} \bpsi_{il}^{(j)} + \bepsilon_i^{(j)},
	\quad i = 1, \dots, n, \ j = 1, \dots, p.
\label{resp_mod}
\end{equation}

As already alluded to in Section~\ref{sec:mfpca:kl}, a key feature of this decomposition is the shared score parametrisation, $\bzeta_i = \left(\zeta_{i1}, \ldots, \zeta_{iL}\right)$, which enables borrowing strength across the $p$ variables and %
provides a parsimonious subject-level representation of the variation in the data. %
This, combined with the flexible accommodation of variable- and subject-specific time grids, tailors it to irregular and sparse functional data settings, where estimation from scarcely observed curves particularly benefits from pulling information across all curves at different time points. 
Moreover, while the scores are common to all $p$ variables, the latent functions $\psi_l^{(j)}(t)$ are variable-specific which enables flexible modelling tailored each variable's dynamics, whose complexity may differ across %
 dimensions %
of a same component~$l$. %

Another interesting modelling treatment of multivariate curves, in the broader functional data analysis field, %
is to allow for variable-specific scores, $\zeta_{il}^{(j)}$, but enforce common latent functions across the $p$ variables, $\psi_l(t)$, on the FPC scores, as for instance considered by \citet{kowal2017bayesian}. Under this %
perspective, %
the score for the $j$th %
variable %
determines how much %
the latent function contributes to its variability. Hence, such a setting can be useful for problems where one expects a common %
process for the $p$ variables and can provide nuanced insights into how each variable uniquely relates to this %
process. An example in biology would be the modelling of expression levels of genes from a given molecular pathway (for instance related to inflammation), measured over time. It then may be reasonable to expect that the genes all ``feed into'' a single latent dynamics reflecting the same biological process or its disruption. Such a framework could retain interpretability through the inspection of gene-specific scores reflecting their relative contributions to this dynamics, possibly offering insights into the biological mechanisms of disease. 

Finally, assuming variable-specific scores \emph{and} latent functions is most flexible, and can be advised in cases where the underlying latent dynamics are not expected to share much across the $p$ variables (as we will illustrate in our simulations of Section~\ref{sec:sim:miss}). However, if the dynamics are related, such a specification will fail %
to exploit shared structures. %
\citet{kowal2017bayesian} also assume variable-specific functions and scores, but borrow information at a higher level in the model hierarchy, which may be seen as an intermediate approach, between a multivariate and a fully univariate treatment. %

Hence, these %
model specifications %
each involve different assumptions about the nature of the latent processes underlying the multivariate curves; %
here, because the focus is on proposing a novel Bayesian treatment of multivariate FPCA, we consider the assumptions of shared scores and variable-specific latent functions in line with the multivariate Karhunen--Loève expansion (Equation~\ref{resp_mod}). 

We represent continuous curves from discrete observations via nonparametric regression \citep{ruppert03, ruppert09},
with the mixed model-based penalised spline basis function representation, as in \citet{durban05}. The
representations for the $j$th elements of the mean function and latent functions are: %
\begin{equation}\label{eq_mean_eigen}\mu^{(j)} (t) \approx \betamu{0}{j} + \betamu{1}{j} t + \sum_{k=1}^K \umu{k}{j} z_{k}(t) \quad\mbox{and}\quad
\psi_l^{(j)} (t) \approx \betapsi{l}{0}{j} + \betapsi{l}{1}{j} t + \sum_{k = 1}^K \upsi{l}{k}{j} z_{k}(t),
\end{equation} for 
$j = 1, \dots, p$ and $l = 1, \dots, L$, where $\{ z_k (\cdot) \}_{1 \le k \le K}$ is a suitable set of
basis functions. Splines and wavelet families are the most common choices for the $z_k$; in our simulations, we
use O'Sullivan penalised splines, which are described in Section 4 of \citet{wand08}.
Next, set $\numu{j} \equiv \T{(\betamu{0}{j}, \betamu{1}{j}, \umu{1}{j}, \dots,
\umu{K}{j})}$, $\nupsi{l}{j} \equiv
\T{(\betapsi{l}{0}{j}, \betapsi{l}{1}{j}, \upsi{l}{1}{j}, \dots, \upsi{l}{K}{j})}$ and
$\C{i}{j} \equiv [ \begin{array}{ccccc} \bone_{n_i^{(j)}} & \bt_i^{(j)} & z_1(\bt_i^{(j)}) & \cdots & z_K(\bt_i^{(j)}) \end{array} ]$.
For notational
convenience, the dependence that any matrix or vector has on the vector of observations times $\bt_i^{(j)}$ will be
understood, rather than shown explicitly. For example, we use $\C{i}{j}$, as opposed to $\bC_i^{(j)} (\bt_i^{(j)})$.

With these notational definitions at hand, we have $\bmu_i^{(j)} \approx \bC_i^{(j)} \numu{j}$ and $\bpsi_{i,l}^{(j)} \approx
\bC_i^{(j)} \nupsi{l}{j}$. Then simple derivations that stem from \eqref{resp_mod} show that the vector of observations on
each response curve satisfies the representation $\bx_i^{(j)} = \C{i}{j} (\numu{j} + \sum_{l=1}^L \zeta_{il} \nupsi{l}{j}) +
\bepsilon_{i}^{(j)}$. %
In the following, we set $\bnu^{(j)} \equiv \T{(\numuT{j}, \nupsiT{1}{j}, \dots, \nupsiT{L}{j})}$ for $j = 1, \dots, p$. We are now in a position to introduce the Bayesian model: 
\begin{equation}
\begin{gathered}
	\bx_i^{(j)} \mid \bnu^{(j)}, \bzeta_i, \sigsqeps{j} \indsim \normal \left\{
		\bC_i^{(j)} \left( \numu{j} + \sum_{l=1}^L \zeta_{il} \nupsi{l}{j} \right), \sigsqeps{j} \bI_{n_i^{(j)}}
	\right\}, 
\quad i = 1, \dots, n,\quad j = 1, \dots, p,\\
	\begin{bmatrix}
		\numu{j} \\
		\nupsi{l}{j}
	\end{bmatrix} \Bigg\rvert \sigsqmu{j}, \sigsqpsi{l}{j} \indsim \normal \left(
		\begin{bmatrix}
			\bzero_{K + 2} \\
			\bzero_{K + 2}
		\end{bmatrix}, \begin{bmatrix}
			\Sigmamu{j} & \bO \\
			\bO & \Sigmapsi{l}{j}
		\end{bmatrix}
	\right), \quad \bzeta_i \indsim \normal (\bzero_L, \bI_L),\hspace{1.15cm} l = 1, \dots, L, \\
	\sigsqmu{j} \mid \amu{j} \indsim \invchisq (1, 1/\amu{j}), \quad
	\amu{j} \indsim \invchisq (1, 1/A^2), \\
	\sigsqpsi{l}{j} \mid \apsi{l}{j} \indsim \invchisq (1, 1/\apsi{l}{j}), \quad
	\apsi{l}{j} \indsim \invchisq (1, 1/A^2), \\
	\sigsqeps{j} \mid \aeps{j} \indsim \invchisq (1, 1/\aeps{j}), \quad
	\aeps{j} \indsim \invchisq (1, 1/A^2),
\end{gathered}
\label{bayes_mfpca}
\end{equation}
 with $\bzeta_i \equiv \T{(\zeta_{i,1}, \dots, \zeta_{i,L})}$, 
$\Sigmamu{j} \equiv \blockdiag (\sigma_\beta^2 \bI_2, \sigsqmu{j} \bI_K)$ and
$\Sigmapsi{l}{j} \equiv \blockdiag (\sigma_\beta^2 \bI_2, \sigsqpsi{l}{j} \bI_K)$, where $\blockdiag_{i = 1, \dots, d} (\bM_i)$ is the block diagonal matrix with matrices $\bM_i$, $i = 1, \ldots, d$, arranged on the diagonal. %
Furthermore, $\sigma_\beta^2$ and $A$ are user-specified hyperparameters; following \citet{wand08},  we recommend setting them to large values resulting in %
diffuse prior specifications.  %
For instance, choosing $\sigma_{\beta} = A = 10^5$ yields reliable estimates in our numerical experiments (Section~\ref{sec:simul}).   %
Note that the iterated inverse-$\chi^2$ prior specification on each $\sigsqmu{j}$,
which involves an inverse-$\chi^2$ prior on the auxiliary variable $\amu{j}$, is equivalent to $\sigsqmu{j} \sim
\hc (A)$. This hierarchical construction based on auxiliary variables facilitates arbitrarily non-informative
priors on standard deviation parameters \citep{gelman06}. The same comments apply to the iterated inverse-$\chi^2$
specifications for each $\sigsqpsi{1}{j}, \dots, \sigsqpsi{L}{j}$, and $\sigsqeps{j}$.

In model (\ref{bayes_mfpca}), the prior variance of the scores is set to unity, which helps prevent undesired compensation effects between the scores and the latent functions' %
spline coefficients, given that the variances of the latter are inferred. While %
estimating the score variances instead may provide a more direct account of the relevance of each component %
in explaining the variation in the functional data, this information can easily be regained in a \emph{post-hoc} manner as we will explain in Section~\ref{sec:post_vmp_orth}. Moreover, estimating the variable- and component-specific spline coefficient variances with our proposed specification enables flexible representation of the latent functions, %
adapting to variations in complexity across different components and variables. Such a specification also ensures that sufficient %
regularisation is induced, shrinking the %
coefficients of latent functions irrelevant to specific components and variables %
to zero. %

The latent functions in model \eqref{bayes_mfpca} are only identifiable up to rotation. %
Moreover, a unique and interpretable ordering these functions is also not enforced. 
Recall that, to obtain a multivariate FPCA representation, we 
require orthonormal multivariate eigenfunctions with respect to the $\Hsc$ inner product and uncorrelated scores with non-increasing variances. %
These orthonormality constraints imply that the latent functions can be seen as an orthonormal basis for the functional observations $\bx_i^{(j)}$, thereby %
preventing information redundancy among components. %
 In Section~\ref{sec:post_vmp_orth}, we will detail a post-processing treatment of the unconstrained posterior estimates to restore %
 these orthogonality constraints 
 along with the ordering of the components in capturing the variability in the functional data. %

\section{Variational Bayes inference}
\label{sec:vbi}

\subsection{Background}

For notational convenience, we set
$\bsigma_\epsilon^2 \equiv \{ \sigsqeps{j} \}_{j = 1, \dots, p}$ and similarly for $\bsigma_\mu^2$, $\bsigma_{\psi_1}^2, \dots,
\bsigma_{\psi_L}^2$,
$\ba_\epsilon$, $\ba_\mu$ and $\ba_{\psi_1}, \dots, \ba_{\psi_L}$. We also write $\bnu^{(j)} \equiv \T{(\bnu_\mu^{(j) \intercal}, \bnu_{\psi_1}^{(j) \intercal},
..., \bnu_{\psi_L}^{(j) \intercal})}$, $\bnu \equiv \T{(\bnu^{(1)}, \dots, \bnu^{(p)})}$,
and $\bx_i \equiv \T{(\bx_i^{(1) \intercal}, \dots, \bx_i^{(p) \intercal})}$, %
$\bx \equiv \T{(\T{\bx_1}, \dots, \T{\bx_n})}$. The full Bayesian inference on the model parameters requires estimating %
the posterior density %
$p (\bnu, \bzeta_1, \dots, \bzeta_n, \bsigma_\epsilon^2, \ba_\epsilon, \bsigma_\mu^2,
\ba_\mu, \bsigma_{\psi_1}^2, \dots, \bsigma_{\psi_L}^2, \ba_{\psi_1}, \dots, \ba_{\psi_L} \mid \bx)$. %
As will be shown in Section~\ref{sec:simul}, classical inference via Markov chain Monte Carlo (MCMC) 
methods %
for model \eqref{bayes_mfpca} can be very slow, even for moderate values of $\bnu$. %
Variational inference is a fast alternative to MCMC methods \citep{ormerod10, blei2017variational}. In this article,
the intractability of the full posterior density function is handled by using the following product density approximation, also called \emph{mean-field approximation}:
\begin{align}
\begin{split}
	&p (
		\bnu, \bzeta_1, \dots, \bzeta_n, \bsigma_\epsilon^2, \ba_\epsilon, \bsigma_\mu^2,
		\ba_\mu, \bsigma_{\psi_1}^2, \dots, \bsigma_{\psi_L}^2, \ba_{\psi_1}, \dots, \ba_{\psi_L} \mid \bx
	) \approx \\
		& \qquad q (\bnu) \prod_{i = 1}^n q (\bzeta_i) \prod_{j = 1}^p \left[q (\sigsqeps{j}) q (\aeps{j})
		q (\sigsqmu{j}) q (\amu{j}) \prod_{l = 1}^L \left\{q (\sigsqpsi{l}{j}) q (\apsi{l}{j})\right\} \right],
\end{split}
\label{mf_restrn}
\end{align}
where each $q$ represents an approximate density function that is specified by its argument. In variational inference,
the $q$-density functions are chosen to minimise the Kullback–Leibler divergence of the left-hand side of \eqref{mf_restrn} from its right-hand side (``reverse'' Kullback–Leibler divergence). Classical variational algorithms rely on the observation that minimising the reverse Kullback–Leibler divergence amounts to maximising a lower bound on the marginal log-likelihood, called the ELBO, for \emph{\underline{e}vidence \underline{l}ower \underline{bo}und} \citep{blei2017variational}. Because the expression of the ELBO does not involve the marginal likelihood, it can conveniently be used as objective function. 
In \eqref{mf_restrn}, we have assumed posterior independence of the mean and latent functions, the global parameters,
from the scores, the subject-specific parameters. The posterior independence of the variance parameters and their associated
hyperparameters is a consequence of incorporating asymptotic independence between regression coefficients and
variance parameters \citep{menictas13} and induced factorisations based on graph theoretic
results \citep[Section~10.2.5]{bishop06}.
Proposition \ref{propn:nu} permits further factorisations for the complete vector of spline coefficients $\bnu$. Its proof is
provided in Appendix~\ref{app:proof_propn_nu}. %

\begin{proposition}
	
	The approximate $q$-density function for the full vector of spline coefficients $\bnu$ in \eqref{mf_restrn} factorises
	according to $q(\bnu) = \prod_{j = 1}^p q(\bnu^{(j)})$.
	
\label{propn:nu}
\end{proposition}

The parameters for each of the $q$-density functions in \eqref{mf_restrn} are interrelated, but can be determined via a coordinate
ascent algorithm \citep[Algorithm~1]{ormerod10}. This corresponds to the classical \emph{\underline{m}ean-\underline{f}ield \underline{v}ariational \underline{B}ayes} (MFVB) approach. Classical MFVB requires the derivation of all approximate posterior
density functions, and %
does not take advantage of the fragment-based variational message passing (VMP) set-up of the Bayesian model for univariate
FPCA in \citet{nolan23}, where the authors presented convenient model extensions based on a factor graph
approach. %

\subsection{Variational message passing}
\label{sec:vmp}

We next provide a brief overview of VMP tailored towards mFPCA, before detailing our new multivariate functional principal component Gaussian likelihood fragment. %
For a deeper exposition to VMP, we %
refer the reader to \citet{minka05} and \citet{wand17}.

\begin{figure}
		\centering
		\includegraphics[scale=1.15]{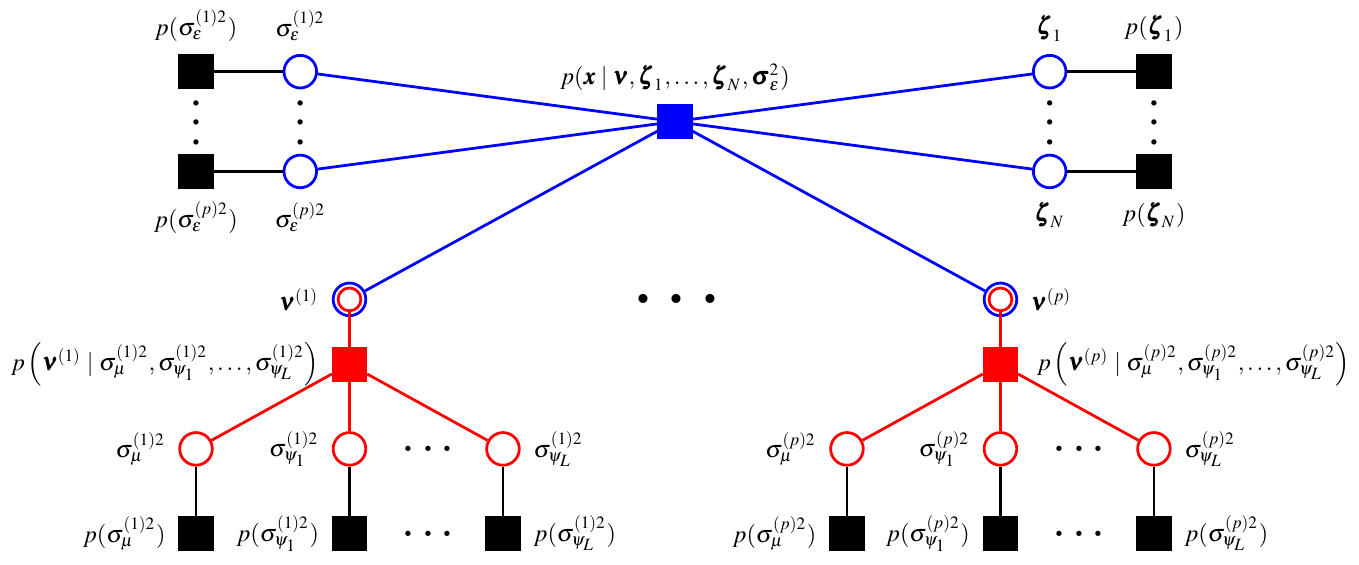}	
\caption{
The factor graph representation of the Bayesian model for mFPCA~\eqref{bayes_mfpca}. The \emph{multivariate functional principal component Gaussian likelihood fragment} is represented in blue. It consists of the likelihood specification as the blue square node and all parameters contributing to that specification as blue circles connected to the square node.
}
\label{fig:fg_mfpca}
	\end{figure}

VMP for arbitrary Bayesian models relies on identifying fragments within the model. Each fragment
consists of one probabilistic specification and all model parameters within that specification. For instance,
the fragment for the likelihood specification in model \eqref{bayes_mfpca} is presented in blue in Figure~\ref{fig:fg_mfpca}.
The central blue square node, the \emph{factor}, represents the likelihood specification, while the circular nodes represent the parameters $\{ \bnu^{(j)} \}_{j = 1, \dots, p}$, $\{ \bzeta_i \}_{i = 1, \dots, n}$
and $\{ \sigsqeps{j} \}_{j = 1, \dots, p}$ that are arguments for the likelihood. Notice that the parameters are separated
according to the product density restriction in \eqref{mf_restrn} and that Proposition \ref{propn:nu} permits further
parameter decomposition of $\bnu$.

Throughout the VMP iterations, approximate posterior densities are updated according to
messages passed between factors and stochastic nodes. The messages have the general form $\msg{f}{\btheta} (\btheta)$,
where $f$ represents an arbitrary factor and $\btheta$ represents an arbitrary parameter vector. The arrow in the
subscript represents the direction of the message, while the message itself is a function of the stochastic node
that participates in the update. In order to infer the multivariate latent functions and scores, we have to determine
the $q$-density functions for $\bnu^{(1)}, \dots, \bnu^{(j)}$ and $\bzeta_1 ,\dots, \bzeta_n$. These can be expressed as
\begin{align}
\begin{split}
	q (\bnu^{(j)})
		&\propto
			\msg{p ( \bx \mid \bnu, \bzeta_1, \dots, \bzeta_n, \sigsqeps{1}, \dots, \sigsqeps{p} )}{\bnu^{(j)}} (\bnu^{(j)}) \
			\msg{p (\bnu^{(j)} \mid \sigsqmu{j}, \sigsqpsi{1}{j}, \dots, \sigsqpsi{L}{j})}{\bnu^{(j)}} (\bnu^{(j)}), \hspace{0.25cm}
		j = 1, \dots, p, \\
	q (\bzeta_i)
		&\propto
			\msg{p ( \bx \mid \bnu, \bzeta_1, \dots, \bzeta_n, \sigsqeps{1}, \dots, \sigsqeps{p} )}{\bzeta_i} (\bzeta_i) \
			\msg{p (\bzeta_i)}{\bzeta_i} (\bzeta_i), \quad
		i = 1, \dots, n.\label{nu_zeta_q_dens_funcs}
\end{split}
\end{align}

A key step in developing the message passing framework is to express density functions in exponential family
form: $p (\bx) \propto \exp \{ \T{\bT(x)} \bdeta \}$, where $\bT (x)$ is a vector of sufficient
statistics that identify the distributional family, and $\bdeta$ is the natural parameter vector;
the messages in \eqref{nu_zeta_q_dens_funcs} are typically in the exponential family of density functions.
Under these circumstances, the associated natural parameter vector updates for the approximate posterior
densities in \eqref{nu_zeta_q_dens_funcs} take the form:
\begin{align}
\begin{split}
	\npq{\bnu^{(j)}} &=
		\np{p ( \bx \mid \bnu, \bzeta_1, \dots, \bzeta_n, \sigsqeps{1}, \dots, \sigsqeps{p} )}{\bnu^{(j)}} +
		\np{p (\bnu^{(j)} \mid \sigsqmu{j}, \sigsqpsi{1}{j}, \dots, \sigsqpsi{L}{j})}{\bnu^{(j)}}, \quad j = 1, \dots, p, \\
	\npq{\bzeta_i} &=
		\np{p ( \bx \mid \bnu, \bzeta_1, \dots, \bzeta_n, \sigsqeps{1}, \dots, \sigsqeps{p} )}{\bzeta_i} +
		\np{p (\bzeta_i)}{\bzeta_i} , \quad i = 1, \dots, n.
\end{split}
\label{etaq}
\end{align}
We outline the exponential family forms of the normal and inverse chi-squared density functions in
Appendix~\ref{app:exp_fam_form}. %

The fragment-based representation of the Bayesian model is the key ingredient in taking advantage of previous
algebraic derivations and computer coding. Although the likelihood fragment (blue in Figure \ref{fig:fg_mfpca}) is specific
for Bayesian mFPCA and requires derivation,
all other fragments in model~\eqref{bayes_mfpca} have been identified
and derived in previous publications; this is a major advantage in obtaining VMP updates for the mFPCA
model, compared to MFVB updates \citep[Algorithm~1]{ormerod10}. %
The Gaussian prior specifications for the each $p (\bzeta_i)$, $i = 1, \dots, n$,
are examples of \emph{Gaussian prior fragments} \citep[Section~4.1.1]{wand17}; the inverse chi-squared prior specifications
on all the subscripted $a^{(j)}$ parameters, $j = 1, \dots, p$, are univariate \emph{inverse G-Wishart prior fragments}
\citep[Algorithm~1]{maestrini20}; similarly, the subscripted variance parameter specifications of the form $\sigma^{(j) 2} \mid
a^{(j)} \sim \invchisq (1, 1/a^{(j)})$, $j = 1, \dots, p$, are instances of the univariate \emph{iterated inverse G-Wishart fragment}
\citep[Algorithm~2]{maestrini20}.
Finally, each of the $p$ fragments representing the penalised specifications on a given $\bnu^{(j)}$ is an example of
the \emph{multiple Gaussian penalisation fragment} derived in Algorithm 2 of \citet{nolan23} %
(red in Figure \ref{fig:fg_mfpca}) and which %
have proven
to be central computational units in VMP for Bayesian FPCA. %

We name the new fragment for the likelihood specification in \eqref{bayes_mfpca} the \emph{multivariate
functional principal component Gaussian likelihood fragment} and outline its updates in the next section.

\subsection{Multivariate functional principal component Gaussian likelihood fragment}
\label{sec:mfpca_lik_frag}

For each $j = 1, \dots, p$,
the message from $p (\bx \mid \bnu, \bzeta_1, \dots, \bzeta_n, \sigsqeps{1}, \dots, \sigsqeps{p})$ to each $\bnu^{(j)}$
can be shown to be proportional to a multivariate normal density function, with natural parameter vector
\begin{equation}
	\np{p ( \bx \mid \bnu, \bzeta_1, \dots, \bzeta_n, \sigsqeps{1}, \dots, \sigsqeps{p} )}{\bnu^{(j)}}
		\longleftarrow
			\begin{bmatrix}
				\E_q (1/\sigsqeps{j}) \displaystyle\sum_{i=1}^n \T{\left\{
					\T{\E_q (\bzetatilde_i)} \otimes \C{i}{j}
				\right\}} \bx_i^{(j)} \\
				-\frac12 \E_q (1/\sigsqeps{j}) \displaystyle\sum_{i=1}^n \vect \left\{
					\E_q (\bzetatilde_i \T{\bzetatilde_i}) \otimes (\CT{i}{j} \C{i}{j})
				\right\}
			\end{bmatrix},
\label{np_lik_nu}
\end{equation}
where  $\otimes$ is the Kronecker product, $\bzetatilde_i \equiv \T{(1, \T{\bzeta_i})}, \ i = 1, \dots, n$, and, for a $d_1 \times d_2$ matrix $A$,  $\vect(A)$ concatenates the columns of $A$ from left to right.  %

For each $i = 1, \dots, n$, the message from $p (\bx \mid \bnu, \bzeta_1, \dots, \bzeta_n, \sigsqeps{1}, \dots, \sigsqeps{p})$
to $\bzeta_i$ is proportional to a multivariate normal density function, with natural parameter vector
\begin{equation}
	\np{p (\bx \mid \bnu, \bzeta_1, \dots, \bzeta_n, \sigsqeps{1}, \dots, \sigsqeps{p})}{\bzeta_i}
		\longleftarrow
			\begin{bmatrix}
				\sum_{j = 1}^p \E_q (1/\sigsqeps{j}) \left\{
					\T{\E_q (\Vpsi{j})} \CT{i}{j} \bx_i^{(j)} - \E_q (\hmupsi{i}{j})
				\right\} \\
				-\frac12 \sum_{j = 1}^p \E_q (1/\sigsqeps{j}) \T{\bD_L} \vect \{ \E_q (\Hpsi{i}{j}) \}
			\end{bmatrix},
\label{np_lik_zeta}
\end{equation}
where $ \Vpsi{j} \equiv [\begin{array}{ccc} \nupsi{1}{j} & \dots & \nupsi{L}{j} \end{array}]$,
$\hmupsi{i}{j} \equiv \VpsiT{j} \CT{i}{j} \C{i}{j} \numu{j}$ and $\Hpsi{i}{j} \equiv \VpsiT{j} \CT{i}{j} \C{i}{j} \Vpsi{j}$.

For each $j = 1, \dots, p$,
the message from $p (\bx \mid \bnu, \bzeta_1, \dots, \bzeta_n, \sigsqeps{1}, \dots, \sigsqeps{p})$ to $\sigsqeps{j}$
is an inverse-$\chi^2$ density function, with natural parameter vector
\begin{equation}
	\np{p (\bx \mid \bnu, \bzeta_1, \dots, \bzeta_n, \sigsqeps{1}, \dots, \sigsqeps{p})}{\sigsqeps{j}}
		\longleftarrow
			\begin{bmatrix}
				-\frac12 \displaystyle\sum_{i=1}^n n_i^{(j)} \\
				-\frac12 \displaystyle\sum_{i=1}^n \E_q \left\{ \T{\left(
					\bx_i^{(j)} - \C{i}{j} \bV^{(j)} \bzetatilde_i
				\right)} \left(
					\bx_i^{(j)} - \C{i}{j} \bV^{(j)} \bzetatilde_i
				\right) \right\}
			\end{bmatrix},
\label{np_lik_sigsqeps}
\end{equation}
where $\bV^{(j)} \equiv [\begin{array}{cccc} \numu{j} & \nupsi{1}{j} & \dots & \nupsi{L}{j} \end{array}]$.

Pseudocode for the multivariate functional principal component
Gaussian likelihood fragment is presented in Algorithm \ref{alg:mfpca_gauss_lik_frag}.
A derivation of all the relevant expectations and natural parameter vector updates is provided in Appendix~\ref{app:mfpca_gauss_lik_frag}. 

\begin{algorithm}
	\caption{
		Pseudocode for the multivariate functional principal component Gaussian likelihood fragment.
	}
	\label{alg:mfpca_gauss_lik_frag}
	\begin{algorithmic}[1]
		\Inputs $
			\{ \npq{\bnu^{(j)}} : j = 1, \dots, p \}, \quad
			\{ \npq{\bzeta_i} : i = 1, \dots, n \}, \quad
			\{ \npq{\sigsqeps{j}} : j = 1, \dots, p \}
		$
		\Updates
			\State Update posterior expectations.
				\Comment{see Appendix~\ref{app:mfpca_gauss_lik_frag}%
    }
			\For{$j = 1, \dots, p$}
				\State Update $\np{p (\bx \mid \bnu, \bzeta_1, \dots, \bzeta_n, \sigsqeps{1}, \dots, \sigsqeps{p})}{\bnu^{(j)}}$
					\Comment{equation \eqref{np_lik_nu}}
				\State Update $\np{p (\bx \mid \bnu, \bzeta_1, \dots, \bzeta_n, \sigsqeps{1}, \dots, \sigsqeps{p})}{\sigsqeps{j}}$
					\Comment{equation \eqref{np_lik_sigsqeps}}
			\EndFor
			\For{$i = 1, \dots, n$}
				\State Update $\np{p (\by \mid \bnu, \bzeta_1, \dots, \bzeta_n, \sigsqeps)}{\bzeta_i}$
					\Comment{equation \eqref{np_lik_zeta}}
			\EndFor
		\Outputs
			\begin{varwidth}[t]{\linewidth} $
				\{ \np{p (\bx \mid \bnu, \bzeta_1, \dots, \bzeta_n, \sigsqeps{1}, \dots, \sigsqeps{p})}{\bnu^{(j)}} : j = 1, \dots, p \},
			$\par$
				\{ \np{p (\bx \mid \bnu, \bzeta_1, \dots, \bzeta_n, \sigsqeps{1}, \dots, \sigsqeps{p})}{\bzeta_i} : i = 1, \dots, n \}
			$\par$
				\{ \np{p (\bx \mid \bnu, \bzeta_1, \dots, \bzeta_n, \sigsqeps{1}, \dots, \sigsqeps{p})}{\sigsqeps{j}} : j = 1, \dots, p \}
			$ \end{varwidth}
	\end{algorithmic}
\end{algorithm}

\subsection{Post-inference orthogonalisation}
\label{sec:post_vmp_orth}

Although the hierarchical model \eqref{bayes_mfpca} induces regularisation through component- and variable-specific prior variance formulations on the spline coefficients, the multivariate FPCA restrictions of orthonormal multivariate eigenfunctions and uncorrelated scores with non-increasing variances are not enforced. As a consequence, the variational inference-based decomposition must be altered to satisfy these conditions. Theorem
\ref{thm:orth_basis} guarantees that we can recover the orthogonal decomposition
through an appropriate sequence of orthogonalisations. %

In order to achieve an orthogonal mFPCA decomposition, we will generalise the post-processing steps  of \citet[Section~5]{nolan23}. The key posterior densities that we require for orthogonal Bayesian mFPCA  are 
$q (\bnu_j)$, $j = 1, \dots, p$, and
$q (\bzeta_i)$, $i = 1, \dots, n$, all of which are normal density functions. Firstly, the natural parameter vectors for
each $q (\bnu_j)$ and $q (\bzeta_i)$ can be determined via \eqref{etaq}. 
Next, the common parameters $\E_q (\bnu^{(j)})$
and $\Cov_q (\bnu^{(j)})$ for $q (\bnu^{(j)})$, and the common
parameters $\E_q (\bzeta_i)$ and $\Cov_q (\bzeta_i)$ for $q (\bzeta_i)$ can be computed from (C7), respectively (C10) of Appendix~\ref{app:mfpca_gauss_lik_frag}. %
In the remainder, we partition $\E_q (\bnu^{(j)})$ as $\E_q (\bnu^{(j)}) \equiv
\T{\{ \E_q (\numuT{j}), \E_q (\nupsiT{1}{j}), \dots, \E_q (\nupsiT{L}{j}) \}}$.

First construct the $n_g$-length vector $\bt_g \equiv \T{(t_1, \dots, t_{n_g})}$ of equidistant time points, where $t_1 = 0$ and $t_{n_g} = 1$, and establish the spline design matrix $\bC_g \equiv [\begin{array}{c c c c c} \bone_{n_g} & \bt_g & z_1 (\bt_g) & \dots & z_K (\bt_g) \end{array}]$, where $\bone_{n_g}$ is an $n_g$-length vector of ones. Then, the posterior estimate for each of the mean functions is $\muhat (\bt_g) \equiv \bC_g \E_q (\numu{j})$, for $j = 1, \dots, p$. Likewise, the variational Bayesian estimates of the latent functions are $\E_q \{ \psi_l^{(j)} (\bt_g) \} = \bC_g \E_q (\nupsi{l}{j})$, for $l = 1, \dots, L$ and $j = 1, \dots, p$.

Next define the vectors $\bpsi_l \equiv \T{(\T{\E_q \{ \psi_l^{(1)} (\bt_g) \}}, \dots, \E_q \{ \psi_l^{(p)} (\bt_g) \})}$, for $l = 1, \dots, L$. Then bind these vectors column wise to obtain the matrix $\bPsi \equiv [\begin{array}{ccc} \bpsi_1 & \cdots & \bpsi_L \end{array}]$ and establish the singular value decomposition $\bPsi = \bU_\psi \bD_\psi \T{\bV_\psi}$, with $\bU_\psi$ and $\bV_\psi$  unitary matrices and $\bD_\psi$ a diagonal matrix whose elements are the singular values of $\bPsi$. %

Now, establish the matrix $\bXi \equiv \T{[\begin{array}{ccc} \E_q (\bzeta_1) & \cdots & \E_q (\bzeta_N) \end{array}]}$. Then define $\bC_\zeta$ to be the $L \times L$ covariance matrix of the columns of $\bXi \bV_\psi \bD_\psi$ and establish its spectral decomposition $\bC_\zeta = \bQ \bLambda \T{\bQ}$, where $\bLambda$ is a diagonal matrix whose elements are the eigenvalues of $\bC_\zeta$ in descending order, and $\bQ$ is an orthogonal matrix containing the corresponding eigenvectors. %

Finally, define the matrices $\dot\bPsi \equiv \bU_\psi \bQ \bLambda^{1/2}$ %
and $\dot\bXi \equiv \bXi \bV_\psi \bD_\psi \bQ \bLambda^{-1/2}$. It can be seen that the columns of $\dot\bPsi$ are orthogonal and the columns of  $\dot\bXi$ are uncorrelated. 
Next, set the $l$th column of $\dot\bPsi$ as $\dot\psi_l (\bt_g)$ and the $i$th row of $\dot\bXi$ as $\dot\bzeta_i$. In addition, partition $\dot\psi_l (\bt_g)$ according to $\dot\psi_l (\bt_g) \equiv \T{\{ \dot\psi_l^{(1)} \T{(\bt_g)}, \dots, \dot\psi_l^{(p)} \T{(\bt_g)} \}}$, where each $\dot\psi_l^{(j)} (\bt_g)$ is an $n_g$-length vector.

While the columns of $\dot\bPsi$ are orthogonal vectors, we require orthonormal multivariate functions in $\Hsc$. We can approximate $\Vert \dot\psi_l \Vert_\Hsc$, the $\Hsc$-norm of $\dot\psi_l$, using numerical integration on \eqref{H_inn_prod}. This permits a posterior estimate on the $l$th multivariate eigenfunction over the vector $\bt_g$ as 
\[
    \psihat_l (\bt_g) = \frac{\dot\psi_l (\bt_g)}{\Vert \dot\psi_l \Vert_\Hsc}, \quad l = 1, \dots, L.
\]
This unit-norm constraint preserves identifiability with respect to scaling (up to a change of sign).

We may partition this vector as $\psihat_l (\bt_g) = \T{\{ \psihat_l^{(1)} \T{(\bt_g)}, \dots, \psihat_l^{(p)} \T{(\bt_g)} \}}$, where each $\psihat_l^{(j)} (\bt_g)$ is an \text{$n_g$-length} vector that is the posterior estimate of the $j$th element of the $l$th multivariate eigenfunction. In addition, the posterior estimate for each of the scores is given by $\hat\zeta_{il} = \Vert \dot\psi_l \Vert_\Hsc \dot\zeta_{il}$, and it now accounts for the contribution to the variance in the functional data, in descending order across $l = 1, \ldots L,$ thanks to the ordered diagonal elements of $\bLambda$ from the spectral decomposition of $\bC_\zeta$.%

The truncated  Karhunen–Loève expansion is left unchanged by this transformation, since
$$\dot\bPsi \dot\bXi^T = \bU_\psi \bQ \bLambda^{1/2}    \bLambda^{-1/2}\bQ^T\bD_\psi\bV_\psi^T\bXi^T = \bPsi \bXi^T.$$
In particular, setting $\xhat_i (\bt_g) \equiv \T{\{ \T{\xhat_i^{(1)} (\bt_g)}, \dots, \T{\xhat_i^{(p)} (\bt_g)} \}}$, the posterior trajectories from the variational algorithm satisfy
\begin{align*}
    \xhat_i (\bt_g) &= \muhat (\bt_g) + \bPsi \E_q (\bzeta_i) = \muhat (\bt_g) + \bU_\psi \bD_\psi \T{\bV_\psi} \E_q (\bzeta_i) \\ &= \muhat (\bt_g) + \bU_\psi \bQ \bLambda^{1/2} \bLambda^{-1/2} \T{\bQ} \bD_\psi \T{\bV_\psi} \E_q (\bzeta_i) \\ &= \muhat (\bt_g) + \dot\bPsi \dot\bzeta_i.
\end{align*}
Setting $\bA \equiv \diag (\Vert \dot\psi_1 \Vert_\Hsc, \dots, \Vert \dot\psi_L \Vert_\Hsc)$, we obtain $\xhat_i (\bt_g) = \muhat (\bt_g) + \dot\bPsi \bA^{-1} \bA \dot\bzeta_i$, which is equivalent to %
\begin{equation*}
    \xhat_i (\bt_g) = \muhat (\bt_g) + \sum_{l = 1}^L \zetahat_{il} \psihat_l (\bt_g), \quad i = 1, \dots, n.
\end{equation*}

\subsection{Adaptive selection of the number of spline functions and latent components}\label{sec:choice_K_L} %

Model (\ref{bayes_mfpca}) involves specifying (i) the dimension $K$ of the spline basis used to describe the mean and latent functions, and (ii) the number $L$ of latent components necessary to describe the variation in the functional data. In this section, we outline two possible approaches to select each of these quantities adaptively.

The choice of $K$ can be guided by empirical considerations aimed at balancing enough flexibility to fit the functional data and avoiding overfitting, especially when the number of observations is small. With this in mind, we propose adapting the rule of thumb proposed by \citet{ruppert2002selecting} to our multivariate irregular-grid setting. Specifically, we set variable-specific numbers of splines as $$K_j = \max\{\min(n_{obs,j}/4, 40), 7\},\qquad j = 1, \ldots, p,$$ where $n_{obs,j} = \mbox{median}\{n_{i}^{(j)}; i = 1,\ldots, n\}$. The lower bound ensures that the spline basis is sufficiently complex to capture essential data features, while the upper bound is %
based on noting that increasing the number of spline coefficients yields diminishing returns in terms of model accuracy and fit, but adds to the computational burden. %

A second approach is to learn $K$ from the data, using a model-choice approach, namely, treating $K$ as a model parameter and estimating its posterior distribution through approximations of the marginal likelihood for each model conditional on $K$. Here, we propose using the ELBO as a proxy of the marginal log-likelihood \citep[see, e.g., ][]{blei2017variational} and placing a discrete uniform prior on the number of spline coefficients:
$$p(K \mid \bx) \propto \exp\{ \log p(\bx \mid K)\} p(K) \approx  \exp\{ \text{ELBO}_K\} p(K),$$
where $p(K) =\text{Unif}(K_\text{min}, \ldots, K_\text{max})$, and $K_\text{min}$ and $K_\text{max}$ are user-specified hyperparameters. In our simulations of Section~\ref{sec:simul}, we use $K_\text{min} = 5$ and $K_\text{max} = 20$, which allows running the procedure in parallel on a 16-core machine; the posterior mass is %
well-within this range, however the support of the prior on $K$ can be extended as needed. 
The use of the ELBO provides a tractable means of model selection and posterior approximation, bypassing direct computation of the marginal likelihood. A similar approach has been employed by \citet{suarez17} for choosing the number of components $L$ in the context of univariate FPCA, but they approximate the marginal likelihood by %
running MCMC chains for each possible model. %

The rule-of-thumb approach has several advantages: it is simple, based on a well-established considerations, requires no extra computational resources %
and is tailored to the multivariate and irregular-grid settings, making it versatile for complex data structures. The model-based approach is more principled and learns the $K$ from the data. Both approaches are implemented in R package \texttt{bayesFPCA}, and the simulations presented in Section~\ref{sec:sim:choice} compare the two approaches. Note that %
we use O'Sullivan's penalised splines, which prevent overfitting and make inference relatively insensitive to reasonable choices of $K$.

A similar model-choice approach could be taken for setting the number of components $L$, whereby $L$ would also be assigned a uniform prior. Learning both $L$ and $K$ would then involve applying the variational algorithm to approximate the marginal likelihood $p(\bx \mid K, L)$ for a grid of $K$ and $L$, spanning the support of their respective prior distributions. Clearly, such ``double-grid search'' %
quickly becomes computationally cumbersome, even when exploiting parallelism. An alternative and elegant approach to choosing $L$ is via prior distributions, placed on scores or spline coefficients, that %
encode ordering constraints by increasing the degree of shrinkage with the component index $l$. %
Multiplicative gamma process shrinkage priors \citep{bhattacharya2011sparse, montagna2012bayesian},  cumulative shrinkage process priors \citep{legramanti2020bayesian, kowal2023semiparametric} and variants thereof  \citep{kowal2017bayesian, shamshoian2022bayesian} have been proposed based on this idea. 
In our case, thanks to the component-specific variances on the spline coefficients in (\ref{bayes_mfpca}), the variances of irrelevant latent functions are effectively shrunk, and therefore enforcing the required regularisation. However, instead of encouraging ordering of the components through the prior, we 
exploit the orthonormalisation procedure described in Section~\ref{sec:post_vmp_orth} and learn both this ordering and $L$ from the estimation of the proportion of variance explained by each FPC component, retaining the components %
that explain a non-negligible amount of the variation in the functional data. Specifically, exploiting the fact that $\Var (\zeta_{il}) = \lambda_l$, with $\lambda_1 \geq \cdots \geq \lambda_L$ in the multivariate Karhunen–Loève expansion, we form estimates $\hat{\lambda}_l$ by taking the empirical variance of the scores post-orthonormalisation and obtain the estimated proportion of variance explained (PVE) by the $l$th component as 
$\hat{\lambda}_l / \sum_{l' = 1}^{L_\text{max}} \hat{\lambda}_{l'}$, for $L_\text{max}$ sufficiently large. 
To programmatically estimate of the number of components in our numerical experiments, we perform a single run the algorithm with an upper bound $L_\text{max} = 10$ (also likely to be an overestimate in real settings), and retain the leading components until their cumulated PVE reaches $95\%$. This approach, or alike, %
is routinely employed in FPCA work such as \citet{ramsay05, greven11, happ18, li19} and \citet{li2023latent}, where $95\%$ is a common default threshold. %
Our simulations of Section~\ref{sec:sim:choice} compare this simple approach and the model-choice approach for $L$, and assess their performance in recovering the correct number of simulated components.

\section{Simulations}\label{sec:simul}

\subsection{Data generation, hyperparameter settings and performance metrics}\label{sec:data_gen}

We %
illustrate our approach in a series of simulation studies aimed at assessing its statistical and computational performance. We place particular emphasis on evaluating the benefits of pulling information from sparse and irregularly observed curves. Unless stated otherwise, for each numerical experiment, we generate synthetic multivariate response curves from  expansion (\ref{mfpca_decomp}), also adding a centered error term with variance unity. We use different numbers of observations $n_{i}^{(j)}$, for $i = 1, \ldots, n$, $j = 1, \ldots, p$, with time sampled uniformly over the interval $[0, 1]$ (subject- and variable-specific grids).  We evaluate performance for two different types of simulated mean function and eigenfunctions (orthonormal in $\Hsc \equiv L^2 ([0, 1])^p$), namely, periodic functions:
$$\mu^{(j)}(t) = (-1)^j 2\sin\left\{(2\pi+j)t\right\}, \qquad j = 1, \ldots, p\,,$$ 
$$\psi^{(j)}_{2l'-1}(t) = (-1)^j \sqrt{2/p}\cos\left(2l'\pi t\right), \quad \psi^{(j)}_{2l'}(t) = (-1)^j \sqrt{2/p}\sin\left(2l'\pi t\right), \qquad l' = 1, \ldots, L/2\,,$$
 for an even number of eigenfunctions $L$, or B-splines orthonormalised using the Gram–Schmidt method. Finally, we simulate the scores $\zeta_{il}$ independently from a centered Gaussian distribution with standard deviation $l^{-1/\alpha}$, $l = 1, \ldots L$,  for $\alpha \in \{1, 2, 8\}$ to cover different relative contributions %
 to the variation in the data across components. We also vary the total number of components $L \in \{1, \ldots, 8\}$.

We run our approach from its R package implementation (\texttt{bayesFPCA}); the package also includes functions to simulate data based on the above data generation procedure. 
We set the model hyperparameters to ensure uninformative specifications for the standard deviations, namely, $\sigma_{\beta} = A = 10^5$. 
In all experiments, we perform inference agnostically of the number of simulated components $L$, using the PVE-based procedure with $L_\text{max} = 10$, and learn $K$ from the data using the model-choice or rule-of-thumb procedure, as described in Section~\ref{sec:choice_K_L}. %
Moreover, Section~\ref{sec:sim:choice} is dedicated to assessing both the sensitivity of inference to these choices of $K$ and $L$, and the performance of our procedure for estimating $L$. 
Finally, we use a convergence tolerance of $\tau = 10^{-5}$ on the relative changes in the ELBO (variational objective function) %
as stopping criterion.

We evaluate estimation accuracy by calculating the root mean square error (RMSE) for the scores and the integrated squared error,
$\mbox{ISE}(f, \hat{f}) = \int_0^1 \vert f(x) - \hat{f}(x)\vert^2 \mbox{d}x$, for the mean function and eigenfunctions, where $f(\cdot)$ is the function used for data generation and $\hat{f}(\cdot)$ is its corresponding posterior estimate. The latent function estimates are based on the variational posterior means of the spline coefficients. 

\subsection{%
Accuracy of variational inference}\label{sec:acc}

We start by evaluating the accuracy of the MFVB and VMP algorithms through direct comparisons with MCMC for the same model. While variational procedures use a prescribed tolerance, MCMC sampling requires evaluating the chain's ability to explore the model space, which can be difficult for large problems. Since the two types of algorithms have different stopping rules and convergence diagnostics, comparing their accuracy and runtime can be challenging. %
To alleviate the risk of unfair comparisons, we conducted MCMC inference with the popular probabilistic programming language Stan \citep{carpenter2017stan}, using the default no-U-turn sampler (NUTS) with $2\,000$ iterations of which $1\,000$ were discarded as burn-in. We acknowledge that custom MCMC implementations for our model could be more efficient %
than Stan's general-purpose engine; however %
our focus is on providing a practical comparison using an off-the-shelf tool, while sidestepping the need to develop a tailored MCMC algorithm.

\begin{figure}[h!]
\centering
\includegraphics[scale=0.45]{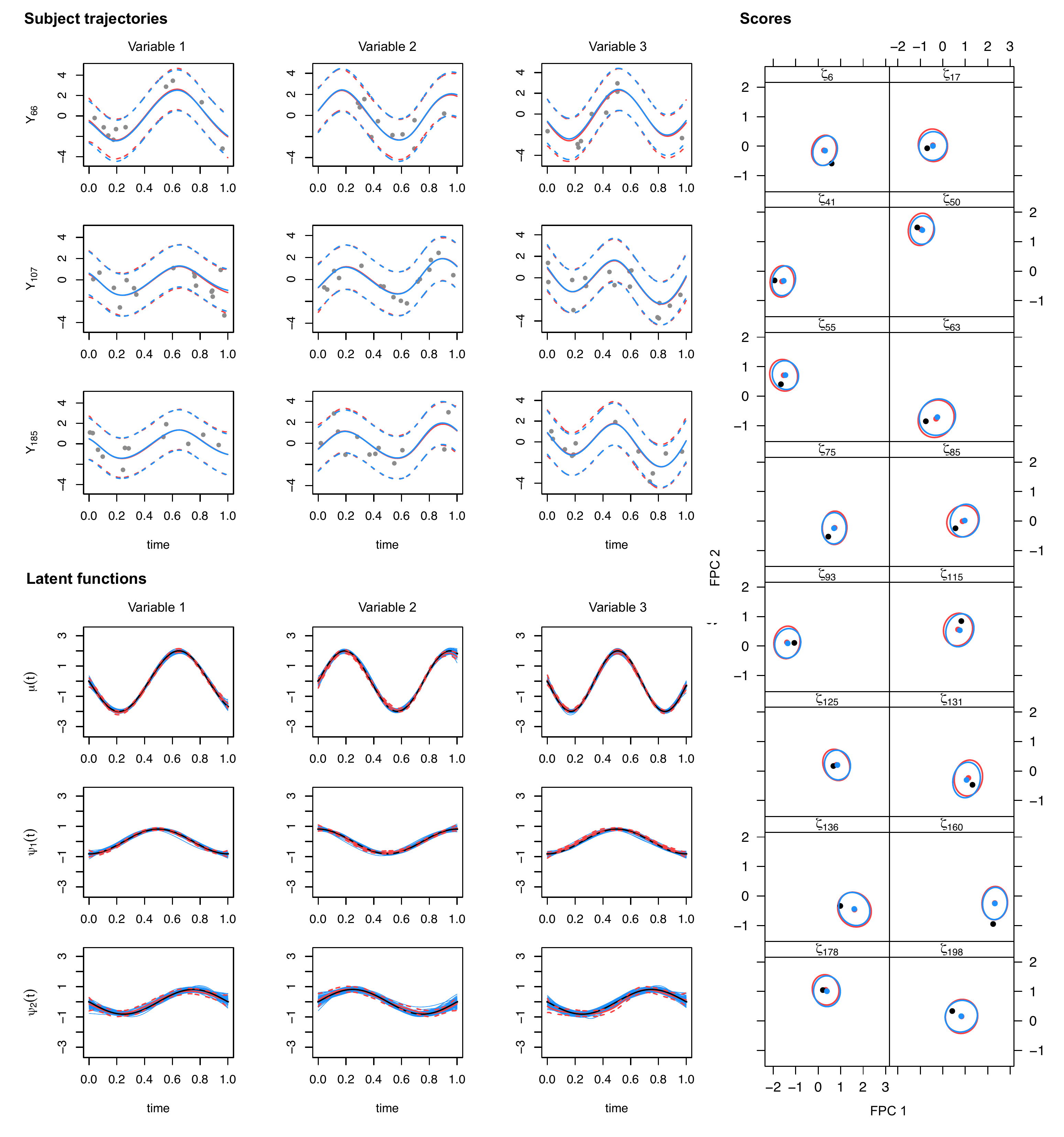}
\caption{MCMC and VMP estimates for a problem with $p = 3$ variables observed at an average of $15$ time points for $n = 200$ subjects; $L$ is learnt from the data by estimating the PVE with $L_\text{max} = 10$, and $K$ is set using  %
the rule of thumb 
described in Section~\ref{sec:choice_K_L}. 
Top left: estimated trajectories for a random subset of $3$ subjects, with posterior means (solid lines) and $95\%$ pointwise prediction bands (dashed). The lines corresponding to MCMC (red) and VMP (blue) inference overlap. Bottom left: mean and latent functions simulated (black) and estimated by MCMC inference (red) with $95\%$ credible bands (dashed) and by VMP inference (blue) for which estimates from $100$ replicates are overlaid. Right: scores simulated (black dots) and estimated by MCMC (posterior mean, red dots) and VMP (posterior mean, blue dots) inference, with $95\%$ credible contours, for a random subset of $16$ subjects.}\label{fig:vmp_vs_mcmc}
\end{figure}

Our approach to obtaining identifiable and orthonormalised estimates from the MCMC posterior summaries mirrors the technique employed for variational estimates. Rather than enforcing orthonormality during the sampling phase, we initially sample the parameters without constraints. Once convergence is reached, we apply the orthonomalisation steps described in Section~\ref{sec:post_vmp_orth} individually to each MCMC sample, after discarding the burn-in samples. Hence, after addressing potential sign changes in the basis functions, these transformed scores and eigenfunctions are directly comparable to the post-processed variational estimates.

We simulate problems with $p = 3$ variables, $n = 200$ subjects and a number of observations drawn uniformly from $\{10, \ldots, 20\}$ for each subject and each variable, and with $L = 2$ simulated latent components; this number of components is successfully retrieved by all methods, using PVE estimation with $L_\text{max} = 10$. %
Figure~\ref{fig:vmp_vs_mcmc} shows the reconstructed trajectories and scores for randomly selected subjects, as well as the estimated latent functions. %
indicating a very good agreement between the VMP and MCMC estimates. The same holds for comparisons between MFVB and MCMC estimates (Appendix~\ref{app:acc}). Variational inference is known to be prone to posterior variance underestimation, especially when poor mean-field factorisations are employed \cite[see, e.g.,][]{blei2017variational}; here however, the excellent agreement of the MCMC and variational  posterior intervals for the estimated trajectories and scores provides empirical evidence that is issue is not encountered under  the factorisation we employ. As expected the runtime is largely in favour of the variational procedure, with $20$ seconds for the VMP algorithm and $18$ minutes $28$ seconds for the MCMC algorithm on an Intel Xeon CPU, 2.60 GHz machine. 

A comprehensive comparison of the errors on the estimated scores and latent functions for a grid of problems with $n \in \{50, 100, 200, 300, 400, 500\}$ further shows %
strong agreement between the VMP, MFVB and MCMC, as well as a greater accuracy as $n$ increases (Figures~D2 \& D3 ). As our MFVB implementation tends to be faster than its VMP counterpart for a virtually indistinguishable statistical performance, our subsequent numerical experiments employ the MFVB algorithm; the computational advantages of variational inference over MCMC inference are discussed further in Section~\ref{sec:runtime}.%

\subsection{Selection of $K$ and $L$}\label{sec:sim:choice}

In Section~\ref{sec:choice_K_L}, we have presented approaches for selecting (i) the dimension $K$ of the spline basis used to represent the mean and latent functions, and (ii) the number of components $L$ explaining the variation in the functional data. In this section, we assess these approaches empirically, by inspecting the resulting errors on the scores and latent functions, as well as the ability to successfully retrieve the number of components simulated. 

We simulate a series of problems with numbers of eigenfunctions ranging from $1$ to $8$, corresponding to orthonormalised B-splines. In each case, we generate $100$ datasets with $p = 3$ variables, $n = 100$ subjects and $20$ observations per subject on average.

\begin{figure}[t!]
\centering
\includegraphics[scale=0.42]{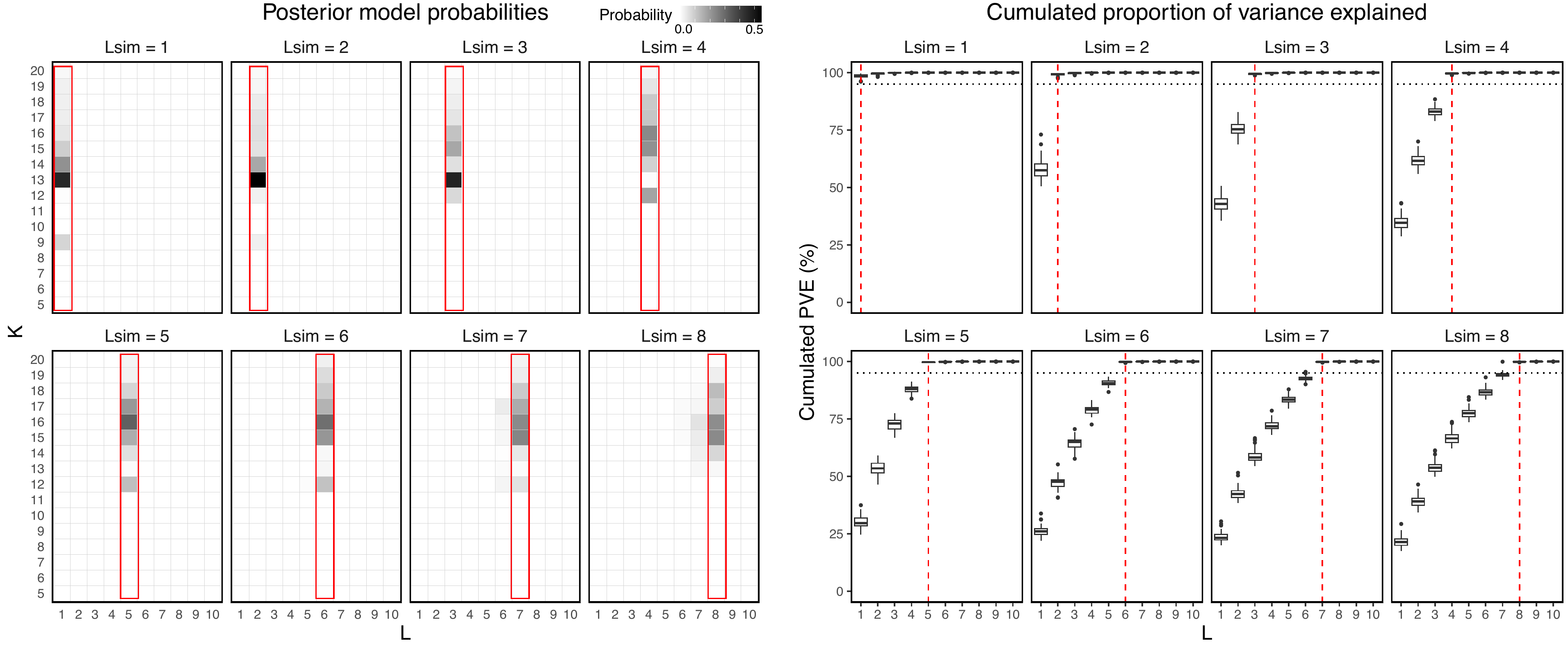}	
\caption{Adaptive selection of the number of latent functions, $L$, and spline basis functions, $K$, for problems with $L = 1, \ldots, 8$ simulated latent functions ($100$ data replicates per problem). Left: Posterior probabilities $p(K, L \mid \bx)$ estimated from the model-choice approach for learning $L$ and $K$, using discrete uniform prior distributions with support $\{1, \ldots, 10\}$ and $\{5, \ldots, 20\}$, respectively. The red contours indicate the number of  simulated components $L$ for each problem. Right: Estimated PVE obtained when inferring model parameters for $L_\text{max} = 10$ components (upper bound). The dotted horizontal lines indicate the $95\%$-PVE threshold and the dashed red vertical lines indicate the number of  simulated components $L$ for each problem.}\label{fig:choice_K_L}
\end{figure}

We start by evaluating the model-choice approach for choosing both $K$ and $L$. Recall that the marginal log-likelihood $p(\bx \mid K, L)$ is approximated by the ELBO$_{K, L}$, for $K = K_\text{min}, \ldots, K_\text{max}$ and $L = L_\text{min}, \ldots, L_\text{max}$ (here $K_\text{min} = 5, K_\text{max} = 20, L_\text{min} = 1, L_\text{max} = 10$), and model posterior probabilities are obtained assuming discrete uniform priors for both $K$ and $L$ with the above supports. Figure~\ref{fig:choice_K_L} (left) shows the posterior probabilities $p(K, L \mid \bx)$ averaged across the $100$ data replicates, for each problem with number of simulated components $L \in \{ 1, \ldots, 8\}$. This figure indicates that the marginal probabilities $p(L \mid \bx)$ concentrate on models with the correct $L$, while the marginal probabilities $p(K \mid \bx)$ are more spread out, covering values from  $K = 12$ to $18$. 

We next evaluate the PVE-based approach for choosing $L$, coupled with the model-choice approach for $K$, on the same simulated datasets: we infer the model parameters with $L_\text{max} = 10$ components and select $L$ as the minimum number of components whose estimated cumulated PVE exceeds $95\%$ (see Section~\ref{sec:choice_K_L}). %
Figure~\ref{fig:choice_K_L} (right) shows  that the correct $L$ is again selected in the vast majority of runs. Inspecting the variational estimates before orthonormalisation reveals that the appropriate regularisation is enforced by the model, since the spline coefficient variances corresponding to superfluous components are effectively shrunk to small values.  
The $95\%$-threshold rule is of course arbitrary; its relevance may be questioned when the contribution to the variance by the last components is small (see, e.g., panels with $L = 7$ and $L =8$ simulated components). In general, there is no reason why it would recover the correct number of components when the portion of the variance explained by the last components is $<5\%$. For this reason, in applications, we recommend choosing $L$ by visually inspecting scree plots of the cumulated PVE: as exemplified in Figure~\ref{fig:choice_K_L}, in our numerical experiments, %
the estimated PVE effectively plateaus after the correct $L$ is reached, with the subsequent components accounting for a negligible amount of the variation in the data. %

We also compare the errors (score RMSE and latent functions ISE) achieved under the PVE-based and model-choice approach for learning $L$, as well as under the rule of thumb adapted from \citet{ruppert2002selecting} and model-choice approach for learning $K$ (Section~\ref{sec:choice_K_L}). These results, reported in Appendix~\ref{app:sel_k_l}, indicate comparable errors, %
with no sign of overfitting. %
These results also suggest that the model choice and PVE-based approaches perform similarly well for recovering $L$. %
Therefore, in practice, we recommend using the PVE-based approach which doesn't require running the algorithm for every possible model. Note that in order to obtain reliable estimates of the PVE, $L_\text{max}$ should be an upper bound on $L$, which should ensure that the estimated eigenvalues corresponding to the last components are close to zero. 

Finally, we set to compare our approach with PVE-based selection to another functional framework, %
namely, the semiparametric functional factor modelling  (SFFM) approach of \citet{kowal2023semiparametric}, which derives the posterior distribution of $L$ from a cumulative shrinkage process prior formulation. We find that our approach does at least as well in estimating the number of components and producing small prediction errors (Appendix~\ref{app:kowal}).

\subsection{Comparison with frequentist multivariate FPCA}\label{sec:happ}

 \begin{table}[t!] %
 \begin{adjustwidth}{-0.8cm}{0cm}
\centering
\mysize
\begin{tabular}{cc c@{}c cc @{}ccc c@{}c c@{}c cc cc cc}
  \hline
Average  & \multicolumn{2}{c}{$\mu(t)$}&&  \multicolumn{2}{c}{$\Psi_1(t)$}&& \multicolumn{2}{c}{$\Psi_2(t)$}  &&  \multicolumn{2}{c}{$\zeta_1$}&& \multicolumn{2}{c}{$\zeta_2$}\\
\cline{2-3}\cline{5-6}\cline{8-9}\cline{11-12}\cline{14-15}
 $n_i^{(j)}$ & mFPCA & Happ && mFPCA & Happ && mFPCA & Happ && mFPCA & Happ && mFPCA & Happ    \\ 
  \hline
20 & \bf 0.81 (0.63) & 0.88 (0.61) && \bf 0.42 (0.32) & 0.80 (0.43) && \bf 1.37 (0.77) & 4.11 (2.97) && \bf 0.24 (0.04) & 0.28 (0.04) && \bf 0.22 (0.02) & 0.26 (0.04) \\ 
  40 & 0.74 (0.87) & \bf 0.62 (0.50) && \bf 0.27 (0.26) & 0.46 (0.32) &&\bf  0.73 (0.43) & 1.66 (0.71) && \bf 0.19 (0.06) & \bf 0.19 (0.03) && \bf 0.17 (0.03) & 0.19 (0.03) \\ 
  60 & 0.80 (0.98) & \bf 0.53 (0.48) && \bf 0.21 (0.23) & 0.33 (0.27) && \bf 0.55 (0.35) & 1.10 (0.48) && 0.18 (0.08) & \bf 0.16 (0.04) && \bf 0.15 (0.03) & 0.16 (0.03) \\ 
  80 & 0.75 (0.93) &\bf 0.39 (0.40) && \bf 0.17 (0.23) & 0.28 (0.22) && \bf 0.43 (0.31) & 0.85 (0.42) && 0.16 (0.07) & \bf 0.14 (0.04) && \bf 0.13 (0.03) & 0.14 (0.03) \\ 
  100 & \bf 0.78 (0.98) &   - && \bf 0.15 (0.17) &   - && \bf 0.35 (0.19) &   - && \bf 0.16 (0.08) &   - && \bf 0.12 (0.03) &   - \\ 
  120 & \bf 0.84 (1.18) &   - && \bf 0.14 (0.22) &   - && \bf 0.32 (0.25) &   - && \bf 0.16 (0.10) &   - && \bf 0.12 (0.05) &   - \\ 
  140 & \bf 0.78 (1.06) &   - && \bf 0.14 (0.21) &   - && \bf 0.30 (0.20) &   - && \bf 0.15 (0.10) &   - && \bf 0.11 (0.04) &   - \\ 
  160 & \bf 0.58 (0.95) &   - && \bf 0.14 (0.19) &   - && \bf 0.28 (0.21) &   - && \bf 0.13 (0.09) &   - && \bf 0.11 (0.04) &   - \\ 
  180 & \bf 0.54 (0.77) &   - && \bf 0.11 (0.18) &   - && \bf 0.23 (0.17) &   - && \bf 0.12 (0.08) &   - && \bf 0.11 (0.04) &   - \\ 
  200 & \bf 0.51 (0.73) &   - && \bf 0.13 (0.17) &   - && \bf 0.23 (0.20) &   - && \bf 0.12 (0.07) &   - && \bf 0.11 (0.04) &   - \\ 
  220 & \bf 0.62 (0.80) &   - && \bf 0.11 (0.17) &   - && \bf 0.21 (0.18) &   - && \bf 0.13 (0.09) &   - && \bf 0.10 (0.04) &   - \\ 
  240 & \bf 0.54 (0.80) &   - && \bf 0.089 (0.12) &    - && \bf 0.18 (0.13) &   - && \bf 0.12 (0.09) &   - && \bf 0.10 (0.04) &   - \\ 
  260 & \bf 0.54 (0.88) &   - &&\bf  0.095 (0.21) &    - && \bf 0.18 (0.22) &   - && \bf 0.12 (0.09) &   - && \bf 0.10 (0.05) &   - \\ 
   \hline
\end{tabular}
\end{adjustwidth}
\caption{Estimation errors obtained using our approach (with selection of $K$ and $L$ by model choice and PVE estimation, respectively) and Happ's approach. The integrated mean squared errors (ISE) $\times 100$ for the latent functions and root mean square errors (RMSE) for the scores are shown %
for a problem with $p = 3$ variables, $n = 100$ subjects, and average number of observations per variable and subject ranging from $20$ to $260$ (rows).  The median and interquartile range (IQR, parentheses) obtained from $200$ data replicates are shown, and the smallest median error of each row is highlighted in bold. For $\mu(t)$, $\Psi_1(t)$ and $\Psi_2(t)$, and for each of the data replicates, the per-variable ISE are computed and averaged across the $p$ variables. A ``-'' indicates scenarios where simulations for Happ's method fail to complete within $36$ hours (Intel Xeon CPU, 2.60 GHz).}\label{tab:happ_ise}
\end{table}

We next compare our Bayesian approach with the covariance-based frequentist multivariate approach proposed by \citet{happ18}. Here, we consider problems $n=100$ subjects, for whom $p = 3$ variables are measured longitudinally, with an average number of observations per subject and variable ranging from $20$ to $260$; we generate $200$ data replicates for each setting. For this simulation study, and all those presented in the remainder of the paper, we run our approach in parallel using model choice for setting $K$, and we estimate $L$ using the PVE-based procedure (Section~\ref{sec:choice_K_L}). Here, this procedure successfully recovers the number of simulated components for all replicates and for all settings, namely $L = 2$. In contrast, \citet{happ18}'s implementation does not allow learning $L$, so we run it assuming the correct number of components as known, giving their method an advantage. 

Table~\ref{tab:happ_ise} reports the ISE on the latent functions and the RMSE on the scores. Estimation accuracy is higher with our approach for the two eigenfunctions and comparable for the mean function, with a tendency for \citet{happ18}'s method to improve as the average number of observations increases. Similar observations hold for the estimation of the two sets of scores. %
This aligns with the fact that our model-based Bayesian approach better handles sparse observation settings, as it eliminates the need for estimating and smoothing covariances, unlike conventional decomposition frequentist procedures, such as used in \citet{happ18}'s approach. 
Moreover, while statistical performance becomes comparable for the two methods as the number of observations increases, the covariance-estimation requirement induces computational intractability for large covariance matrices, which prevents applications of \citet{happ18}'s method on problems where the average number of observations per variable and subject exceeds $80$.  We provide a more comprehensive discussion of this computational limitation in Section~\ref{sec:runtime}.

For the setting where %
\citet{happ18}'s method has the highest accuracy %
-- namely, with $80$ observations per subject and variable, on average -- the estimates obtained by the two approaches are visually very close (Appendix~\ref{app:freq}). %
An advantage of our approach is that posterior credible bands are readily obtained from the inferred variational posterior distributions. Such uncertainty estimates are unavailable using \citet{happ18}'s method. Pointwise bootstrap confidence bands for the eigenvalues and eigenfunctions can be obtained, but the computational cost associated with  the resampling becomes prohibitive for our problem sizes. %

\subsection{Borrowing strength in sparse settings}\label{sec:pulling}

We next examine the benefits of our hierarchical framework for handling estimation in complicated sparse-data settings, where borrowing information across variables and subjects is essential. We simulate a problem with $n=200$ subjects and $p = 6$ variables, of which the first is very scarcely observed (number of observations uniformly drawn from $\{5, \ldots, 10\}$), while the remaining five variables are observed more frequently (number of observations uniformly drawn from $\{50, \ldots, 75\}$). We then apply our mFPCA approach jointly on the six variables, and compare the posterior estimates of the scores with those of separate univariate FPCA runs. Specifically, we conduct a series of six variational inference runs, with $p = 1$ in (\ref{bayes_mfpca}), and rescale their posterior mean and standard deviation by a factor $\sqrt{p}$ to make them orthonormal in the multivariate Hilbert
space $\Hsc \equiv L^2 ([0, 1])^p$.

\begin{figure}[t!]
\centering
\includegraphics[scale=0.48]{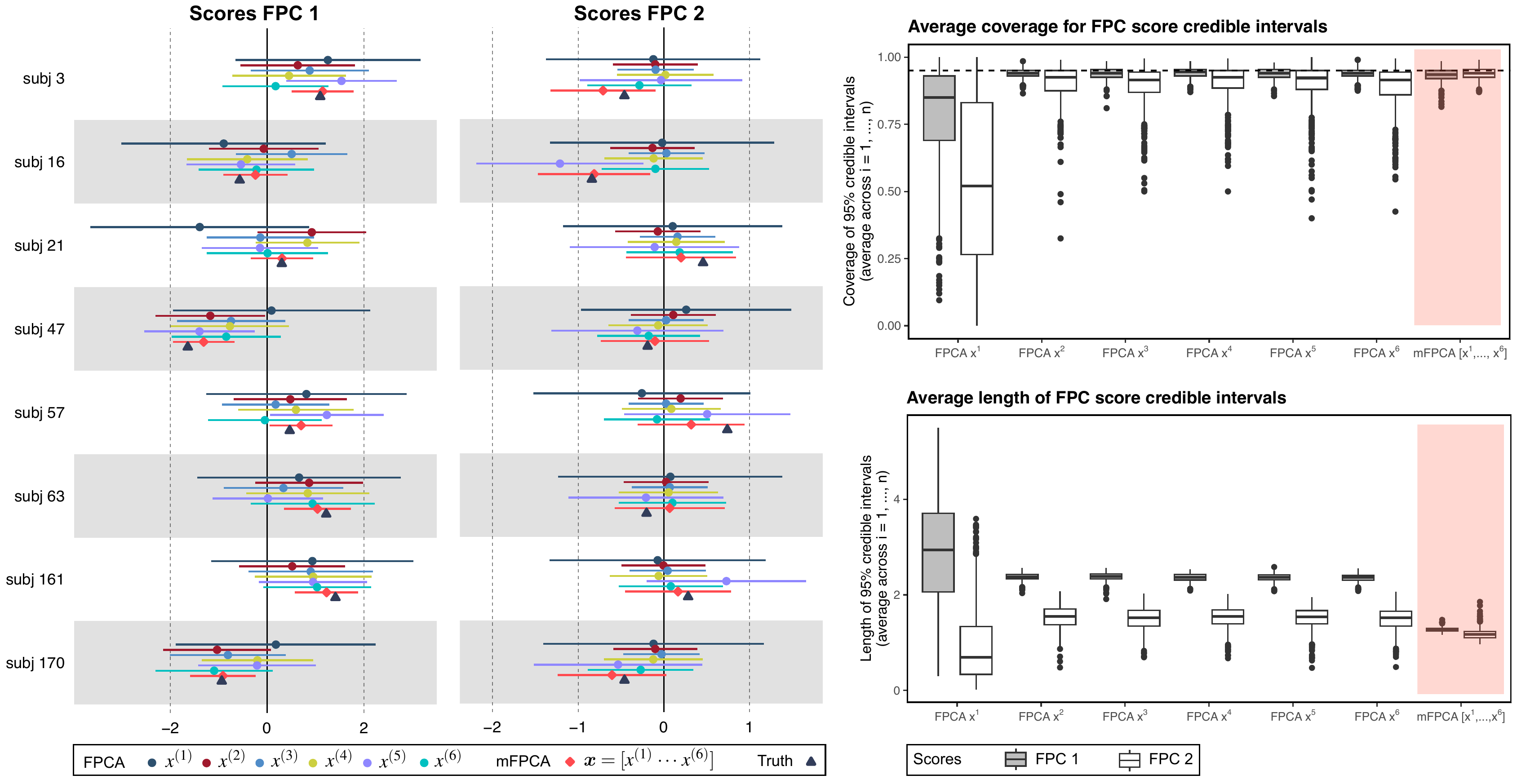}
\caption{Posterior means with $95\%$ credible intervals for the scores for the first two components, obtained from univariate and multivariate Bayesian FPCA for a problem with $n = 200$ subjects and $p = 6$ variables, of which the first, $x^{(1)}$, is infrequently observed ($500$ data replicates). Left: Estimates shown for a random subset of $8$ subjects (first data replicate). The mFPCA estimates are in light red, the true simulated values are indicated by black triangles and the remaining colors correspond to estimates obtained from univariate FPCA runs. Right: Empirical coverage (top) and interval length (bottom). The dashed horizontal line indicates the $95 \%$ threshold.}\label{fig:scores_pulling}
\end{figure}

In the previous section we have seen that the variational algorithm advantageously provides uncertainty quantification for the estimated scores. Here we further examine the effect of %
pulling information with joint modelling on the credible bands and posterior means of the scores, placing special emphasis on the first variable which is infrequently observed. %

Figure \ref{fig:scores_pulling} shows the estimated scores for the first and second principal components along with the $95\%$ credible intervals, %
for a random subset of $8$ subjects. %
The small number of observations collected for the first variable leads to very wide credible intervals for its corresponding scores (FPC~1 \& 2) estimated with univariate FPCA; these intervals cover zero. %
This suggests that $x^{(1)}$ is too infrequently observed in order for univariate FPCA to provide any %
useful results. The remaining five curves ($x^{(j)}, j = 2, \ldots, 6$) are more frequently observed, and therefore the credible intervals of their corresponding univariate scores are narrower, yet in a few instances the true score is not covered (for $j = 5$ subject 63 FPC~1, for $j=2, 3, 4, 6$ subjects 16 \& 57 FPC~2, and for $j = 2$ subject 21 FPC~2). The mFPCA intervals all contain the true simulated scores, and they are substantially narrower than their univariate counterparts for FPC~1.

Figure~\ref{fig:scores_pulling} also shows the empirical coverage and length of these credible intervals based on $500$ data replicates. The interval lengths (averaged across subjects) for both FPC~1 and~2 obtained by mFPCA are generally smaller than those obtained by univariate FPCA for the densely-observed variables $j = 2, \ldots, 6$. For the univariate FPCA of variable $j = 1$, intervals lengths are substantially larger for FPC~1, but are highly variable for FPC~2: for some runs, the FPC~2 scores are shrunk to zero, due to an insufficient number of observations on that variable and a lower proportion of the variance explained by the second component. The coverage obtained by mFPCA is comparable to the univariate runs for variables $j = 2, \ldots, 6$: it tends to be slightly worse for FPC~1 (mean $93.5\%$, sd $2.3\%$), but better for FPC~2 (mean $94.0\%$, sd $2.3\%$). The coverage achieved with univariate FPCA for the sparsely observed variable $j = 1$ is clearly insufficient in most runs (FPC~1: mean $78.5\%$, sd $19.2\%$; FPC~2: mean $53.3\%$, sd $31.2\%$). 

These results suggest that the %
pulling information across variables and subjects observed at irregular temporal grids is particularly beneficial in settings where some variables entail very few measurements. Another notable advantage of mFPCA is that it produces a single set of posterior estimates, under the form of scalar scores, for all variables (rather than six separate sets of scores here), thereby offering a very parsimonious summary of the temporal covariation in the multivariate data.

Finally, we conducted separate simulations to evaluate the empirical coverage and length of the prediction intervals for the response trajectories, along with the associated prediction errors. Benchmarks include the SFFM approach of \citet{kowal2023semiparametric} (which uses a semiparametric basis to model latent functions) and two fully parametric variants. The results, presented in Appendix~\ref{app:kowal}, show that our method achieves excellent coverage, with interval length comparable to those of the competing approaches. Hence, while the coverage for the score credible intervals, in the experiments of Figure~\ref{fig:scores_pulling}, could be insufficient for some univariate FPCA runs, the correct coverage results for the response trajectories here suggest that the independence assumptions assumed in our variational factorisation are reasonable, causing no noticeable underestimation of posterior variances.  Remarkably, our fully nonparametric approach also yields errors as low as those obtained with the fully parametric approaches for which the parametric forms of the simulated functions are taken as known. Moreover, across the six different scenarios tested with different numbers of latent functions, the prediction errors are similar or lower than those of the competing methods.

\subsection{Performance under model misspecification}\label{sec:sim:miss}

Borrowing information using joint approaches is sensible when the modelled variables are expected to covary over time. %
However assessing this can be difficult in practice. %
It is therefore important to evaluate the potential consequences on inference when this assumption is violated. In this section, we examine the robustness of our approach when the joint model is misspecified. %
Specifically, we generate $p$ variables from distinct univariate models, whose scores are simulated independently ($\rho = 0$), or with varying degrees of correlation ($\rho = \{0.2, 0.4, 0.6, 0.8\}$) up to complete correlation ($\rho = 1$). This last case corresponds to no misspecification, since the data share the same set of scores and therefore can be thought as generated from the joint model  (\ref{bayes_mfpca}). For each scenario, we compare the accuracy of our joint approach with its univariate counterpart, applied independently to each variable as in the previous section. %
We consider $100$ data replicates of a problem with $p = 6$ variables, $L = 3$ simulated latent components, $n = 50$ subjects, and with sparse observations drawn uniformly from $\{5, \ldots, 10\}$ for each curve.

\begin{figure}[h!]
\centering
\includegraphics[scale=0.53]{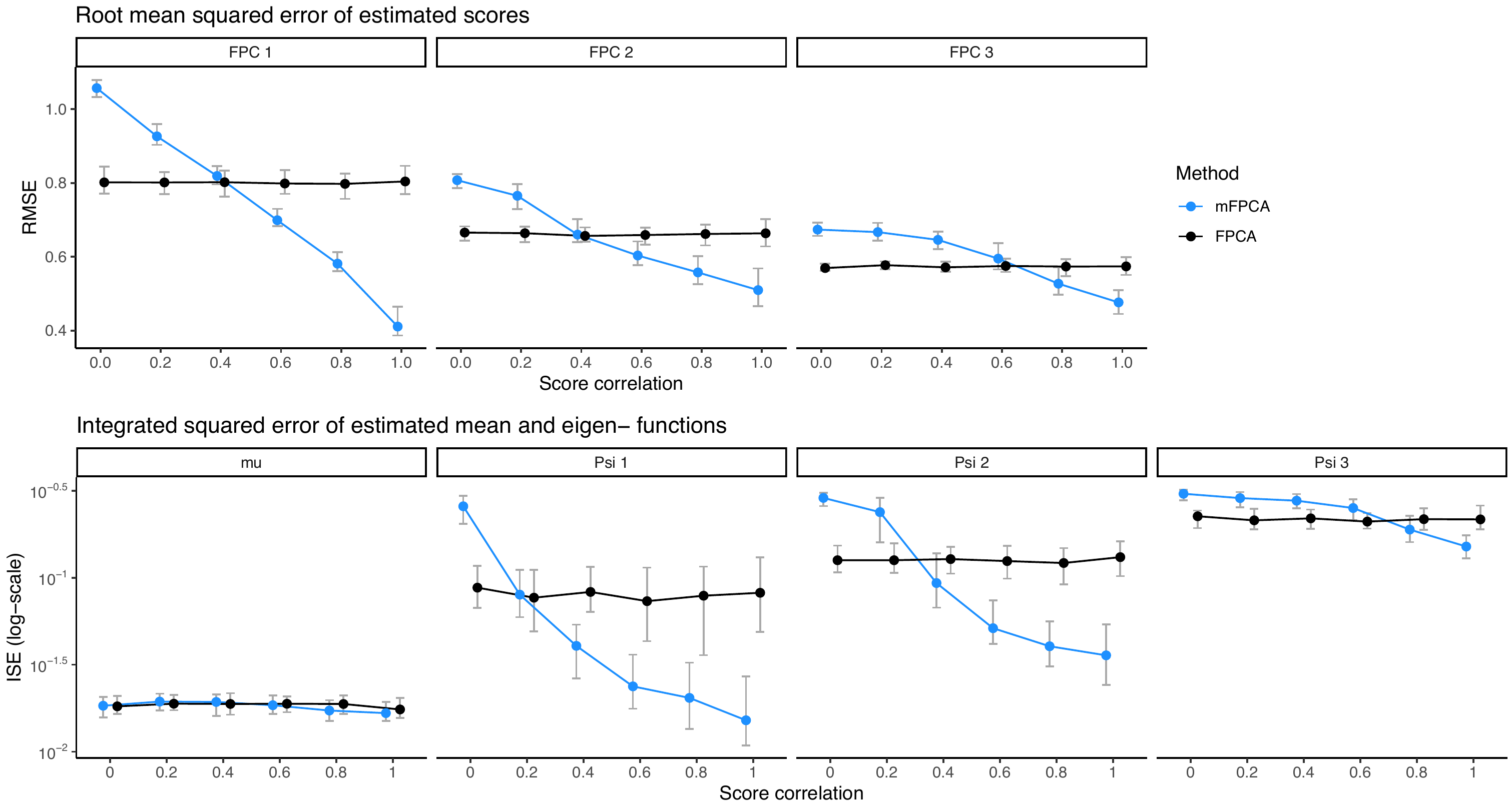}	
\caption{Comparison of the errors obtained using univariate and multivariate Bayesian FPCA for a problem with $p = 6$ variables, $L = 3$ simulated latent components, $n = 50$ subjects and numbers of observations drawn uniformly from $\{5,\ldots, 10\}$ for each variable and each subject. The data were simulated from univariate FPCA models with varying degrees of score correlation $\rho \in \{0, 0.2, 0.4, 0.6, 0.8, 1\}$ ($x$-axis). The median (dots) and first and third quartiles (grey) of the errors from $100$ data replicates are shown. Left: RMSE for the scores. For each data replicate, the univariate FPCA per-variable RMSE are computed and the average across all $p$ variables is shown. Right: ISE for the mean and eigen-functions. For each data replicate, and both univariate and multivariate FPCA, per-variable ISE are computed and the average across all $p$ variables is shown.}\label{fig:univ_vs_multiv}
\end{figure}

As anticipated, Figure~\ref{fig:univ_vs_multiv} indicates a deterioration on the error of the scores using mFPCA as the correlation of the scores %
weakens, to the point where the univariate model outperforms the joint model, for unrelated or weakly related variables where $\rho \in \{0, 0.2, 0.4\}$. The estimation of the latent functions tend to be more robust to the misspecification, because, unlike the scores, the mean function and eigenfunctions are variable specific, which confers great flexibility, %
even under different latent dynamics for the $p$ variables. 
The fact that the ISE for the first two eigenfunctions is smaller under the joint model than under the univariate model when the correlation is weak to moderate ($\rho~\in~\{0.4, 0.6\}$) can be attributed to the ``virtually larger sample sizes'' obtained by accounting jointly for the observations on all $p$ variables.  When the scores are more strongly correlated ($\rho \in \{0.8, 1\}$), there is a substantial reduction in the RMSE of the scores, which is particularly striking for the first component. This again demonstrates the benefits of borrowing strength across related variables. 

Altogether, these results suggest that the multivariate model is reasonably robust %
to misspecifications caused by a lack of common latent dynamics across the variables. Moreover, as shown in Section~\ref{sec:pulling}, when similar dynamics exist, joint modelling also tends to produce narrower credible intervals for the scores, which contain the true value.  %

\subsection{Computational performance}\label{sec:runtime}

Our variational algorithm has important computational advantages compared to other estimation %
approaches. These advantages concern both runtime and memory (RAM) usages, and can be attributed to the combination of two features: (1)
a direct, simultaneous inference of the Karhunen–Loève expansion, treating all parameters as unknown and estimating them jointly – this bypasses the need to model the covariance function, unlike with frequentist approaches which typically use a sequence of time- and memory-greedy smoothing and eigendecomposition steps to estimate large covariance functions; (2) 
a fast, deterministic inference approach, which scales to large numbers of subjects and time points, unlike more conventional Bayesian inference approaches based on MCMC inference.

\begin{figure}[t!]
\centering
\includegraphics[scale=0.44]{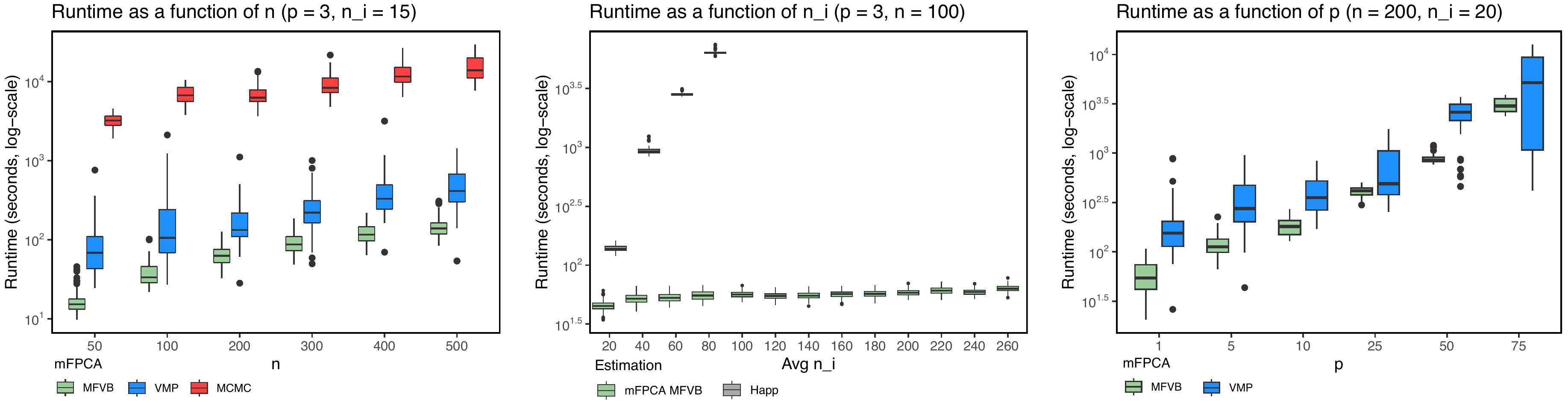}	
\caption{Runtime profiling (on the logarithmic scale) as a function of (i) the number of subjects $n$, (ii) the average number of observations per subject 
$n_i$, and (iii) the number of variables $p$, obtained on an Intel Xeon CPU, 2.60 GHz machine. Left:  study (i); comparison of the MFVB (green), VMP (blue) and MCMC (red) algorithms on the problems of Section~\ref{sec:acc} ($100$ replicates). Middle: study (ii); comparison of Happ's (grey) and our (blue) approaches on the problems of Section~\ref{sec:happ} ($200$ replicates). The results for Happ's method correspond to the four highest boxplots in grey in bins $20$ to $80$; the method does not complete within $36$ hours for problems with more than $80$ observations per variable and subject, hence no boxplots are displayed. Right:  study (iii); comparison of the MFVB (green) and VMP (blue) algorithms on problems with $n = 100$ subjects, average number of observations per subject $n_i = 20$ and varying numbers of variables $p$ ($100$ replicates). All mFPCA methods are run with PVE-based selection of $L$, using $L _{\text{max}} = 10$, and, for the choice of $K$,  either the rule of thumb (studies (i) and (iii)) or a 16-core implementation of the model-choice approach (study (ii)); see Section~\ref{sec:choice_K_L}.}\label{fig:runtime}
\end{figure}

Figure~\ref{fig:runtime} shows the runtime profiles as a function of (i) the number of subjects $n$, (ii) the average number of observations per subject $n_i$ and (iii) the number of variables $p$, presented on the logarithmic scale. Additionally, Appendix~\ref{app:runtime} presents the runtime as a function of the upper bound on the number of functional principal components used for inference, $L_\text{max}$. Profiling studies (i) and (ii) correspond to the simulations presented in Sections~\ref{sec:acc} and \ref{sec:happ}. 

Study (i) indicates that the variational algorithms are $1$ to $3$ %
orders of magnitude faster than a Stan's implementation of NUTS for the same model. As already noted, while Stan is a powerful tool, it may not be the most efficient choice for the specific class of models we consider. Here, however, our intention is to provide a practical comparison using an off-the-shelf tool, rather than to develop and benchmark a highly optimised MCMC algorithm. %
Our VMP implementation also tends to be slower than the MFVB implementation, due to a higher per-iteration cost. This might be explained by the fact that it involves the additional overhead of passing messages between nodes in the graphical model, while the MFVB updates are simpler; we also do not rule out the possibility that our code may benefit from optimisation.

Study (ii) indicates a large computational advantage of our MFVB algorithm compared the frequentist multivariate  approach of \citet{happ18}. %
In particular, the latter approach could not complete within $36$ hours for problems with an average number of observations per variable and subject exceeding $80$. Moreover, for \citet{happ18}'s approach, memory is also an important limiting factor, as the runs with $\leq 80$ observations required compute nodes with tens of GBs of RAM in order to store all matrix objects pertaining to the estimation of the covariance function.

Finally, study (iii) involves large problems, totalling up to $75 \times 200 \times 20 = 300\,000$ observations; the variational algorithms completed in less than 1.5 hour (MFVB: average 51 minutes; VMP: average 1 hour 28 minutes). 
The sole purpose of this profiling for large $p$ is to evaluate the scalability of our algorithms as, in practice, the shared-score assumption implied by the multivariate Karhunen–Loève expansion is unlikely to hold for a large number of variable $p$;  it implies that all these variables share the same underlying dynamics which can be 
unrealistic. %

Our variational inference implementation demonstrates excellent computational efficiency, not only when compared to Stan's generic implementation, but also against custom MCMC algorithms. Inspecting the runtimes reported in various Bayesian functional latent factor analysis articles, we found that the method of \citet{kowal2023semiparametric} %
ran in $\approx 5$ minutes or less on problems with a total number of observations of $\approx 2\, 000 - 3\,000$, \citet{montagna2012bayesian} required several hours for a $40\,000$-observation problem, and \citet{kowal2017bayesian} and \citet{goldsmith15} took up to 6 hours and 10 days, respectively, for problems with %
$400\,000-450\,000$ observations. When matched with problems of similar numbers of observations, our algorithm is $1$ to $2$ orders of magnitude faster than the above methods, except for \citet{kowal2017bayesian}, where it is only twice faster. The runtimes in \citet{kowal2017bayesian} are very good %
for an MCMC algorithm; it should also be noted that, in their experiments, the number of components $L$ is fixed to low values while we learn this number using $L_\text{max} = 10$. 
 Of course, these comparisons are only indicative as the runtimes reported in this literature concern implementations for models different to ours (most of which are models for a single functional variable).  In summary, the efficiency of our algorithm, combined with its accuracy and uncertainty quantification properties, positions it as highly competitive for large and complex datasets.

\section{The latent underpinnings of the immune response to SARS-CoV-2 infection}\label{sec:application}%

We return to the SARS-CoV-2 study presented in Section~\ref{sec:motivation}. As motivated there, %
since COVID-19 is a multi-system, heterogeneous disease, we %
will undertake to disentangle the patient variability of short- and long-term disease trajectories by applying mFPCA on  molecular markers across several biological systems. Such an analysis will hopefully help us interpret the shared latent dynamics driving disease severity and incomplete organismal recovery. %

Patient were categorised according to five clinical severity classes, based on their hospitalisation status and oxygen needs, namely, A, asymptomatic ($n=18$); B, mild symptomatic ($n=40$); C, hospitalised without supplemental oxygen ($n=50$); D, hospitalised with supplemental oxygen ($n=38$), and E, hospitalised with assisted ventilation ($n=69$). Time was measured from the first positive swab for patients of class A, and from symptom onset for patients of the remaining classes; to prevent biased inferences due to temporal shifts resulting from these different definitions, we focus our analysis on symptomatic patients, i.e., patients from classes B to E. We analyse data collected during the first $7$-weeks from symptom onset (acute and post-acute phase), examining five important molecular markers, namely,  C-reactive protein (CRP) levels, interleukin 10 (IL-10) cytokine levels, glycoprotein B (glyc-B) levels and levels of two metabolites from the kynurenine pathway,  quinolinic acid and tryptophan. We also exclude patients that had fewer than two measurements collected within the $7$-week time window for any given marker, leaving a total of $n = 82$ patients for analysis; the sparse observations recorded at different time points across subjects and variables calls for flexible hierarchical modelling with effective pooling of information.  Finally, we use one-time measurements from $45$ SARS-CoV-2 negative healthy controls (HC), and use the IQR of the HC measurements for the five parameters as a reference for normal parameter levels. %

\begin{figure}[t!]
\centering
\includegraphics[scale=0.45]{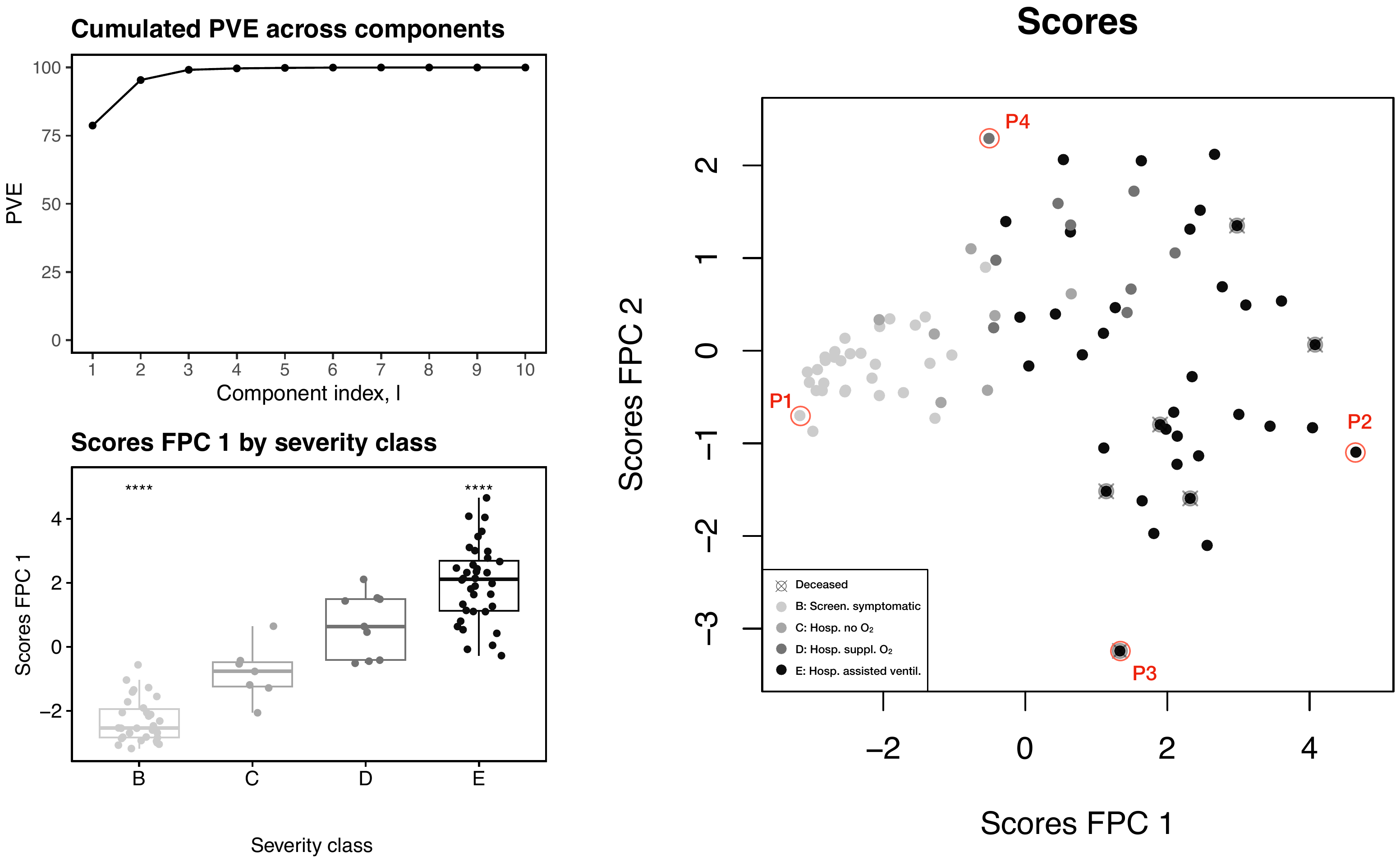}	
\caption{mFPCA analysis of the COVID-19 data. Top right: percentage of variance explained by the first $L_\text{max} = 10$ components. The first two components capture %
{$78.7\%$} and %
{$16.7\%$} of the total variance, respectively. Bottom right: FPC~1 scores (``severity scores'') as a function of the clinical severity classes B to E. Stars: one vs. all two-sided $t$-tests ($^{****}: p< 0.0001$, $^{***}: p< 0.001$, $^{**}: p < 0.01$ and $^{*}: p < 0.05$). Right: estimates of the FPC~1 scores ($x$-axis, ``severity scores'') and FPC~2 scores ($y$-axis, ``recovery scores''). The red labels P1, P2, P3 and P4 indicate the patients with the most extreme severity or recovery scores, and the grey levels indicate the clinical severity classes B (light grey), C (grey), D (dark grey) or E (black) to which each patient belongs; the gradient of grey points along the $x$-axis again reflects the association of the FPC~1 scores with the severity classes. The six points with a black cross indicate patients who died from COVID-19.}\label{fig:app_scores}
\end{figure}

\begin{figure}[t!]
\centering
\includegraphics[scale=0.81]{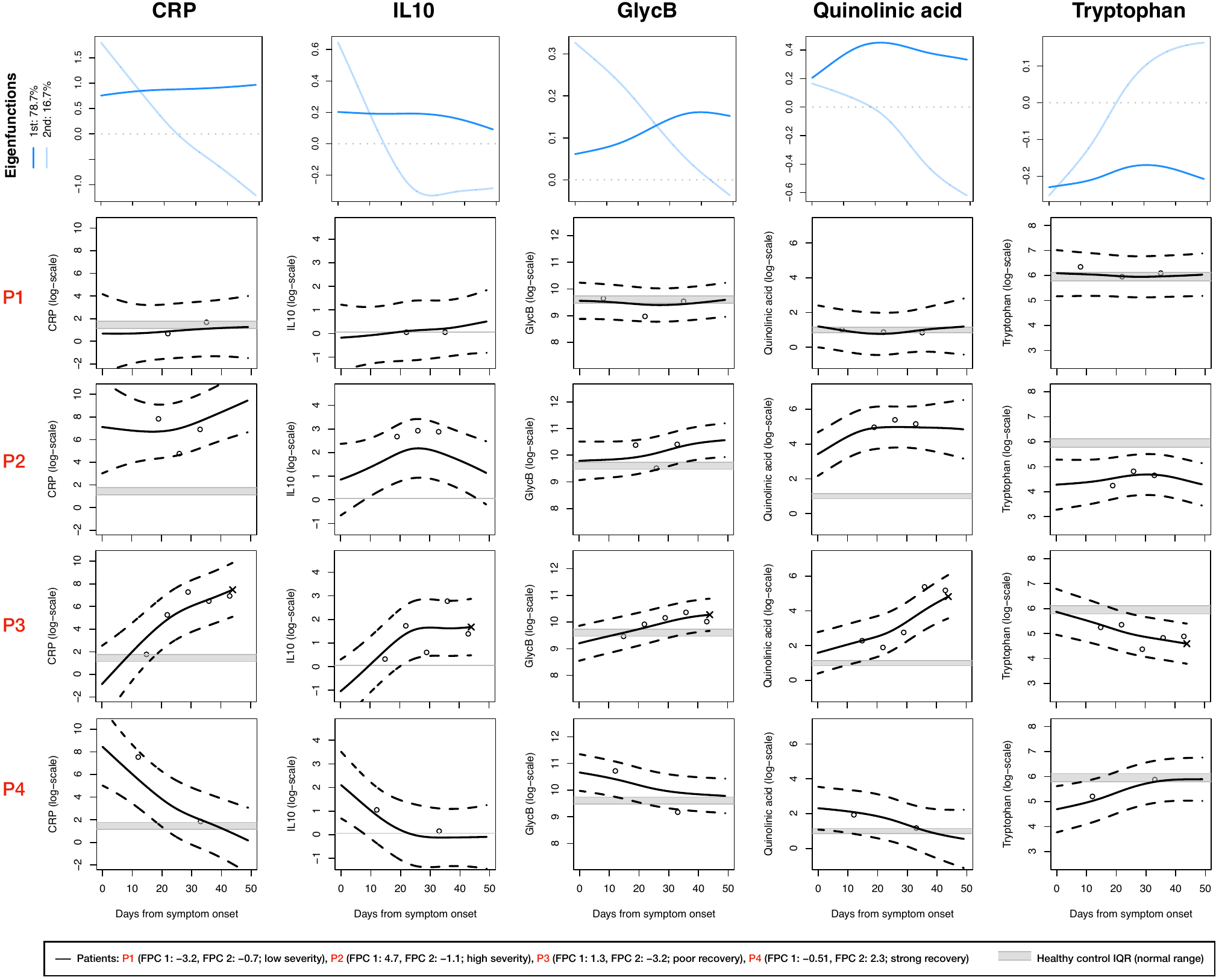}	
\caption{Eigenfunctions and patient trajectories from the mFPCA analysis of the COVID-19 data. Top row: first (blue) and second (light blue) eigenfunctions estimated for each marker and displayed over the first $7$ weeks post symptom onset. Bottom rows: estimated trajectories (posterior mean with $95\%$ prediction bands) for the four patients P1, P2, P3, P4 with most extreme scores as shown in Figure~\ref{fig:app_scores}. The scores of each patient for the first and second eigenfunctions are indicated in the legend. The horizontal grey band corresponds to the healthy control (HC) IQR, reflecting normal levels of the corresponding markers. }\label{fig:app_eigenfunctions}
\end{figure}

We perform inference using model choice for selecting the dimension of the spline basis $K$ and PVE estimation for learning the number of eigenfunctions $L$, with $L_\text{max} = 10$ (Section~\ref{sec:choice_K_L}). The posterior probability for $K$ is maximal for $K = 5$, and   
the first two components are sufficient to capture $>95\%$ of the variation ($78.7\%$ and %
$16.7\%$ respectively); we therefore focus on interpreting these two components.  %
Figure~\ref{fig:app_eigenfunctions} shows that %
these eigenfunctions, estimated for the five markers, have clear interpretations. Specifically, the first eigenfunction %
is above zero for CRP, IL-10, glyc-B and quinolinic acid, while it is below zero for tryptophan, over the entire $7$-week disease course. This means that patients with a positive score for the first eigenfunction tend to have positive deviations from the mean for the first four markers, and a negative deviation for tryptophan. This suggests that the scores corresponding to the first eigenfunction can serve as proxies for disease severity since it has been established that patients with severe COVID-19 infection tend to have increased levels of CRP, IL-10, glyc-B and quinolinic acid, and depleted tryptophan levels; in fact, these five markers have well-established roles in inflammation and immune regulation \citep{bergamaschi2021longitudinal, masuda2021integrative}. The second eigenfunction reflects parameter recovery over time, since it decreases for the first four markers and increases for tryptophan, meaning that patients with a positive score for the second eigenfunction tended to see their parameter disruption resolve over time, i.e., a decrease for the levels of the first four markers and an increase for tryptophan, toward normal levels.

Figure~\ref{fig:app_scores} displays the estimated patient scores corresponding to the two eigenfunctions. %
It suggests that the first set of scores reflects the clinical severity classes B to E, which corroborates their interpretation as proxies for disease severity, and this association is significant (anova $p < 0.001$). Moreover, the group of six patients that did not survive had significantly higher scores for the first eigenfunction and lower scores for the second eigenfunction on average, compared to random patient groups of same size ($p=0.0038$ and $p = 0.0085$, respectively) and four of these six patients had negative scores for the second eigenfunction (poor recovery). We thus hereafter refer to the scores corresponding to the first and second eigenfunctions as ``severity'' and ``recovery'' scores, respectively.

Figure~\ref{fig:app_eigenfunctions} shows the reconstructed parameter trajectories for four patients with extreme scores (referred to as ``P1'', ``P2'', ``P3'' and ``P4'' in Figure~\ref{fig:app_scores}). They also corroborate the interpretation on the scores, since the trajectories of patient P1, with the most favourable severity and recovery scores, largely overlap the normal HC range over the entire disease course. The trajectories of patient P2, with a high severity score, tend to stay well above (or below, for tryptophan) the normal range.  Finally, the trajectories of patient P3 (P4), with unfavourable (favourable) recovery scores, tend to deteriorate (resolve) over time, as expected.

\begin{table}[t]
\centering
\footnotesize
\begin{tabular}{rllll}
  \hline
mFPCA & \multicolumn{2}{c}{Severity scores} & \multicolumn{2}{c}{Recovery scores} \\ 
 \cline{1-1}
Dyspnoea & 0.35 & *** & 0.21 &  \\ 
  Cough & 0.31 & ** & 0.23 &  \\ 
  Chest Pain on exertion, palpitations or swollen ankles & 0.11 &  & 0.01 &  \\ 
  New leg swelling in one leg or shortness of breath with chest pain & 0.11 &  & 0.08 &  \\ 
  New skin rashes or sores & 0.24 & . & 0.26 & . \\ 
  Voice alteration & 0.22 & . & 0.19 &  \\ 
  Difficulties eating, drinking or swallowing & 0.19 &  & -0.02 &  \\ 
  Constant noisy breathing or throat whistling & 0.04 &  & 0.17 &  \\ 
  Anosmia or dysgeusia & 0.09 &  & 0.11 &  \\ 
  Difficulty to gain or maintain weight, loss of appetite & 0.03 &  & 0.06 &  \\ 
  New neurology in one or more limbs & 0.33 & ** & 0.07 &  \\ 
  New pain in one or more parts of the body & 0.5 & **** & 0.24 & . \\ 
  General muscle weakness, balance or range of movement of joints & 0.43 & *** & 0.16 &  \\ 
  Fatigue & 0.37 & *** & 0.28 & . \\ 
  Cognition: memory, concentration and thinking skills & 0.11 &  & 0.25 & . \\ 
    \hline
 Overall physical \& mental recovery: \\
 \cline{1-1}
  mFPCA (n = 40) & 0.41 & *** & 0.25 & * \\ 
  FPCA CRP (n = 51) & 0.31 & ** & 0.09 &  \\ 
  FPCA IL-10 (n = 40) & 0.19 &  & 0.23 &  \\ 
  FPCA glyc-B (n = 57) & 0.21 & * & 0.01 &  \\ 
  FPCA quinolinic acid (n = 57) & 0.27 &** & 0.08 &  \\ 
  FPCA tryptophan (n = 57) & 0.14 & & 0.11 &  \\ 
\end{tabular}
\caption{Association of the scores with long-COVID symptoms. Kendall's rank correlation and significance after Benjamini–Hochberg multiplicity adjustment (FDR: $^{****}$:~$< 0.1 \%$, $^{***}$:~$< 1\%$, $^{**}$:~$<5\%$, $^{*}:$~$< 10\%$, $^{.} :$~$< 15\%$) for the association of the severity and recovery mFPC scores with long-COVID symptoms ranked from $0$ (no symptom) to $5$ (extreme manifestation of the symptom). Kendall's rank correlation and significance with overall physical and mental recovery is also assessed for scores estimated by separate univariate FPCA analysis ($p$: $^{***}$:~$< 0.001$, $^{**}$:~$< 0.01$, $^{*}$:~$<0.05$); the sample size for the tests are indicated and correspond to the intersection of the number of patients used in each FPCA/mFPCA analysis and the number patients for which long-COVID questionnaires were collected.}\label{tab:symptoms}
\end{table}

To further illustrate the interpretation of the scores, we use survival data and questionnaires collected up to one year after infection to ask whether the latent dynamics underlying the disease courses over the first $7$ weeks post symptom onset are associated with fatalities and specific long-COVID symptoms. The six patients who did not survive were all under assisted ventilation (class E) and died at $19$,  $20$,  $44$, $109$, $153$ and $158$ days post symptom onset, respectively.  We explore the association of the scores with survival using Bayesian %
Cox proportional hazards modelling, which we implement using %
the R package \texttt{brms}. Running $4$ chains with $2\,000$ iterations of which the first $1\,000$ iterations are discarded as burn-in, we %
obtain a posterior mean estimate of $0.46$ for the association of the severity scores with survival, with $95\%$ credible interval $(0.02, 1.00)$,  suggesting a probable positive association. The $95\%$ credible interval of the recovery scores survival association estimate covers zero, i.e., $(-1.59, 0.05)$, suggesting that peak severity during the acute and post-acute phases may have a greater prognostic value than the actual disease dynamics. Table~\ref{tab:symptoms} next shows associations of the severity and recovery scores with specific long-COVID symptoms that were evaluated for a subset of $67$ patients on a scale from $0$ (no symptom) to $5$ (extreme manifestation of the symptom). In particular, using Kendall's rank correlation tests with Benjamini–Hochberg multiplicity correction, severity scores are associated with six long-term symptoms, namely, dyspnoea, cough and four neurological symptoms (FDR $<5\%$), while there is  weaker evidence of associations of recovery scores with fatigue, pain and cognition defects (FDR $<15\%$). Moreover, both sets of severity and recovery scores are associated with an ``Overall physical and mental recovery'' category recorded in all patient questionnaires. %

Finally, we explore the extent to which the analysis benefits from borrowing information across the five markers using our mFPCA approach, by inspecting whether added biological insights are obtained compared to separate univariate FPCA analyses. To this end, we perform independent applications of the univariate version of our Bayesian approach ($p = 1$) for each of the five markers. Except for the IL-10 analysis, the number of patients analysed is larger than for the multivariate analysis as it is not limited by the threshold $\geq 2$ on the number of observations across all five markers (i.e., we have CRP: $n=96$, IL-10: $n=82$, glyc-B: $n=106$, quinolinic acid: $n=106$ and tryptophan: $n=106$).   Interestingly, in each univariate analysis, the scores corresponding to the first two eigenfunctions have the same interpretation as in the multivariate analysis, that is, they are proxies of severity and recovery, respectively. However, as shown in Table~\ref{tab:symptoms}, inspecting their association with long-term recovery indicates that, while %
all sets of univariate severity scores, except for IL-10 and tryptophan, are %
associated with the ``Overall physical and mental recovery'' category, none of the univariate recovery scores are associated with it. This may suggest that the joint analysis better captures the latent dynamics underlying recovery patterns, since associations with the mFPCA severity and recovery scores are both highly significant, as discussed above. Finally, as emphasised in the different simulation studies, the fact that mFPCA provides single set of scalar scores, FPC~1 and FPC~2, common to all five markers, allows effectively summarising their temporal covariation, unlike with univariate FPCA which yields separate sets of scores for the different markers, making it challenging to achieve a %
synthetic picture.

\section{Discussion}\label{sec:conclusion}

We have presented a hierarchical modelling framework for multivariate functional principal component analysis (mFPCA) using variational message passing (VMP) and mean-field variational Bayes (MFVB) inference, which addresses important challenges posed in complex real-world observational settings, by flexibly pooling information from limited data and infrequently observed curves. %
This is, to the best of our knowledge, the first Bayesian approach to mFPCA. %

We model the temporal covariation of multivariate curves via the model hierarchy, namely, via shared scores which allow borrowing strength across related longitudinal measurements. In addition to enhancing statistical power, this model-based approach 
circumvents the estimation of large covariance and cross-covariance matrices, and thus enables mFPCA in (i) sparse and irregular sampling settings and (ii) sizable real-data settings which are typically beyond the scope of %
existing frequentist mFPCA approaches.  Thanks to the model's conjugacy properties, all mean-field variational inference updates are obtained in closed-form. Additionally, our   %
variational message passing implementation, based on the same posterior factorisation, permits an elegant modularisation of the algorithm algebra under the form of fragments. Thanks to this principle, we could seamlessly repurpose fragments obtained in our previous work \citep{nolan23} and combine them with a newly-derived multivariate functional principal component Gaussian likelihood fragment. 
This new fragment, on its own, constitutes a novel contribution, with utility extending beyond the present work context: its algebraic derivation and computer code are now readily available to statisticians willing to employ VMP inference for any models involving multivariate Gaussian likelihood components. 

Our detailed simulation studies indicate that our variational implementation (whether based on MFVB and VMP) is both accurate and computationally efficient: %
it produces comparable or lower estimation errors on the scores and latent functions %
compared to MCMC inference on the same model and to the frequentist approach of \citet{happ18}, while being orders of magnitude faster. 
Our experiments also suggest that the variational lower bound (ELBO) can conveniently serve as proxy for the marginal log-likelihood in model-choice procedures to learn the numbers of spline coefficients (and eigenfunctions). Crucially, we have also seen how our hierarchical model enables estimation from data with very sparsely observed curves ($5$-$10$ observations), exploiting information about other related curves with sampling grids that differ across variables and subjects.

In addition to its accuracy, flexibility and  computational convenience, our Bayesian mFPCA framework %
possesses important advantages %
in terms of interpretability and
uncertainty quantification. 
The ``shared-score'' parametrisation offers a parsimonious representation of the major modes of joint variation of the curves, and %
our variational procedure approximates the full posterior distribution of the scores, making it straightforward to construct credible boundaries. %
These scores therefore constitute a principled, subject-level scalar summary of the multivariate curves that can be used for further analysis tasks, such as regression or clustering. We have illustrated such a use case 
in the COVID-19 study where we %
inspected the association of patient-specific ``severity'' (FPC~1) and ``recovery'' (FPC~2) scores estimated from related molecular markers, %
with survival outcomes and long-term symptoms. Specifically, our mFPCA framework highlighted complex coordinated dynamics across the inflammatory, immune and metabolic systems, and suggested that the patients' molecular status during the acute and post-acute phase is interlinked with incomplete clinical recovery up to one year post disease onset. This should prompt further research on organismal recovery from COVID-19, to pinpoint the specific inflammatory and immune mechanisms mobilised early in the disease course, understand their possible %
 long-term consequences, and help formulate therapeutic recommendations applicable soon after infection to prevent adverse outcomes. 

Thanks to its versatility and scalability, our framework %
is readily applicable to any study that involves multiple parameters measured longitudinally, possibly with scarce and irregular observations across subjects and functional curves. We anticipate that frameworks like ours will gain relevance in the near future for acquiring early personalised insights into disease risk, development and monitoring, especially when applied to routinely collected blood test data (complete blood count or CBC) %
available from electronic health records.

There are several avenues for methodological development. One of them concerns extensions to high-dimensional curves, a setting which is expected to gain prominence; for instance, in genomics, there is growing interest in collecting longitudinal observations on a large number of genes (up to $20\, 000$ within the human genome) %
to understand how dynamic gene expression programs drive disease formation. In this context, coupling sparse prior formulations with shared scores or shared latent function assumptions could prove relevant.  Another interesting question, motivated by the application to the COVID-19 data, is the formulation of a Bayesian FPCA-based joint survival model, to account for the survival information in the estimation of the FPC scores, while propagating the uncertainty associated with their estimation into the survival analysis. One natural specification %
 would be to link %
 the longitudinal mFPCA model and the survival model %
 through the %
 FPC scores although this would require %
 integrating over scores in the survival model. %

\section*{Software}

\noindent The R package \texttt{bayesFPCA} is available at \url{https://github.com/hruffieux/bayesFPCA}. The source code accompanying this article is available at \url{https://github.com/hruffieux/VB-mFPCA-paper-code}.

\section*{Data}

\noindent Data from the NIHR CITIID COVID-19 Cohort is available at (\url{https://www.covid19cellatlas.org/patient/citiid/}).

\section*{Funding}

\noindent This research was supported by the Wellcome Collaborative Award 219506/Z/19/Z (T.N.), the UK Medical Research Council programme MRC MC UU 00002/10 (S.R.), the Alan Turing Institute, London, UK (S.R.) and the Lopez–Loreta Foundation (H.R.).

\section*{Acknowledgements}

\noindent 
We thank Daniel Temko for insightful discussions on the model formulation, and Christoph Hess, Glenn Bantug, 
Julien Wist, Aimee Hanson, 
Paul Lyons, 
Kenneth Smith and Jeremy Nicholson for their valuable input about the SARS-CoV-2 data. We thank NIHR BioResource volunteers for their participation, and gratefully acknowledge NIHR BioResource centres, NHS Trusts and staff for their contribution. We thank the National Institute for Health and Care Research, NHS Blood and Transplant, and Health Data Research UK as part of the Digital Innovation Hub Programme. The views expressed are those of the author(s) and not necessarily those of the NHS, the NIHR or the Department of Health and Social Care. \\

\noindent \emph{Conflict of Interest:} None declared.

\section*{\LARGE Appendix}
\appendix
\renewcommand\thefigure{\thesection.\arabic{figure}}

\section*{Matrix algebraic background}
\label{sec:matrix}

We define the $\vect$ and $\vech$ operators,
which are well-established \citep[e.g.]{gentle07}.
For a $d_1 \times d_2$ matrix, the $\vect$ operator concatenates the columns of the matrix from left to right.
For a $d_1 \times d_1$ matrix, the $\vech$ operator concatenates the columns of the matrix after removing
the above diagonal elements. For example, suppose that $\bA = [\begin{array}{c c} \T{(2, -3)} & \T{(-1, 1)} \end{array}]$.
Then $\vect (\bA) = \T{(2, -3, -1, 1)}$ and $\vech (\bA) = \T{(2, -3, 1)}$.
For a $d^2 \times 1$ vector
$\ba$, $\vect^{-1} (\ba)$ is the $d \times d$ matrix such that $\vect \{\vect^{-1} (\ba)\} = \ba$. Additionally, the matrix
$\bD_d$ is the duplication matrix of order $d$, and it is such that $\bD_d \vech (\bA) = \vect (\bA)$ for a
$d \times d$ symmetric matrix $\bA$. Furthermore, $\bD_d^{+} \equiv (\T{\bD_d} \bD_d)^{-1} \T{\bD_d}$ is the
Moore-Penrose inverse of $\bD_d$, where $\bD_d^+ \vect (\bA) = \vech(\bA)$.

For a set of $d$ matrices $\{ \bM_i \}_{i = 1, \dots, d}$, we define:

\[
	\stack_{i = 1, \dots, d} (\bM_i)
		\equiv
			\begin{bmatrix}
				\bM_1 \\
				\vdots \\
				\bM_d
			\end{bmatrix}
	\quad \text{and} \quad
	\blockdiag_{i = 1, \dots, d} (\bM_i)
		\equiv
			\begin{bmatrix}
				\bM_1 & \bO & \cdots & \bO \\
				\bO & \bM_2 & \cdots & \bO \\
				\vdots & \vdots & \ddots & \vdots \\
				\bO & \bO & \cdots & \bM_d
			\end{bmatrix},
\]

\noindent with the first of these definitions requiring that each $\bM_i$ has the same number of columns.


\section{Proof of Proposition \ref{propn:nu}}
\label{app:proof_propn_nu}

First, set $\btheta \equiv \{ \bnu, \bzeta_1, \dots, \bzeta_n, \bsigma_\epsilon^2, \ba_\epsilon, \bsigma_\mu^2,
\ba_\mu, \bsigma_{\psi_1}^2, \dots, \bsigma_{\psi_L}^2, \ba_{\psi_1}, \dots, \ba_{\psi_L} \}$, the entire parameter set.
Then, the optimal posterior density function for $\bnu$ in \eqref{mf_restrn} that minimises the Kullback–Leibler divergence of
its left-hand side from its right-hand side is computed according to

\begin{equation}
	q (\bnu) = C_1 \exp \left\{ \E_{q (\btheta \setminus \bnu)} \log p (\bx, \btheta) \right\},
\label{q_nu_comp}
\end{equation}

\noindent where $C_1$ is a constant of proportionality. Next, note that

\begin{equation}
	\log p (\bx, \btheta) =
		\sum_{j = 1}^p \left\{
			\sum_{i = 1}^n \log p (\bx_i^{(j)} \mid \bnu^{(j)}, \bzeta_i, \sigsqeps{j})
			+ \log p (\bnu^{(j)} \mid \sigsqeps{(j)})
		\right\} + f (\btheta \setminus \bnu),
\label{log_full_post_nu}
\end{equation}

\noindent where $\bx^{(j)} \equiv \T{(\bx_1^{(j) \intercal}, \dots, \bx_n^{(j) \intercal})}$, for $j = 1, \dots, p$, and
$f (\btheta \setminus \bnu)$ represents an arbitrary function that does not depend on $\bnu$. Substituting
\eqref{log_full_post_nu} into \eqref{q_nu_comp}, we have

\[
	q (\bnu)
		= C_2 \prod_{j = 1}^p \exp \left[
			\E_{q (\btheta \setminus \bnu^{(j)})} \left\{
				\sum_{i = 1}^n \log p (\bx_i^{(j)} \mid \bnu^{(j)}, \bzeta_i, \sigsqeps{j})
				+ \log p (\bnu^{(j)} \mid \sigsqeps{(j)})
			\right\}
		\right]
		= \prod_{j = 1}^p q (\bnu^{(j)}),
\]

\noindent where $C_2$ is an arbitrary constant of proportionality.


\section{Exponential family form}
\label{app:exp_fam_form}

We first describe the exponential family form for the normal distribution.
For a $d \times 1$ multivariate normal
random vector $\bx \sim \normal (\bmu, \bSigma)$, the probability density function of $\bx$ can be shown to satisfy

\begin{equation}
	p(\bx) = \exp \left\{ \T{\bT_{\vect} (\bx)} \bdeta_{\vect} - A_{\vect} (\bdeta_{\vect}) - \frac{d}{2} \log (2 \pi) \right\},
\label{vec_repn}
\end{equation}

\noindent where $\bT_{\vect} (\bx) \equiv \T{\{ \T{\bx}, \T{\vect ( \bx \T{\bx} )} \}}$ is the vector of sufficient statistics
and $\bdeta_{\vect} \equiv \T{( \T{\bdeta_{\vect, 1}}, \T{\bdeta_{\vect, 2}} )} \equiv \T{[ \T{( \bSigma^{-1} \bmu )},
-\frac12 \T{\{ \vect (\bSigma^{-1}) \}} ]}$ is the natural parameter vector.
The function $A_{\vect} (\bdeta_{\vect}) =
-\frac14 \T{\bdeta}_{\vect, 1} \{ \vect^{-1} (\bdeta_{\vect, 2}) \}^{-1} \bdeta_{\vect, 1}
- \frac12 \log | -2 \vect^{-1} (\bdeta_{\vect, 2}) |$ is the log-partition function.
The inverse mapping of the natural parameter vector is \citep[equation~S.4]{wand17}

\begin{equation}
	\bmu = -\frac12 \left\{ \vect^{-1} (\bdeta_{\vect, 2}) \right\}^{-1} \bdeta_{\vect, 1} \quad
	\text{and} \quad
	\bSigma = -\frac12 \left\{ \vect^{-1} (\bdeta_{\vect, 2}) \right\}^{-1}.
\label{gauss_vec_comm_params}
\end{equation}

\noindent We will refer to the representation of the multivariate normal probability density function in
\eqref{vec_repn} as the \emph{vec-based representation}.

Alternatively, a more storage-economical
representation of the multivariate normal probability density function is the \emph{vech-based representation}:

\[
	p (\bx) = \exp \left\{ \T{\bT_{\vech} (\bx)} \bdeta_{\vech} - A_{\vech} (\bdeta_{\vech}) - \frac{d}{2} \log (2 \pi) \right\},
\]

\noindent where the vector of sufficient statistics, the natural parameter vector and the log-partition function are,
$\bT_{\vech} (\bx) \equiv \T{\{ \T{\bx}, \T{\vect ( \bx \T{\bx} )} \}}$,
$\bdeta_{\vech} \equiv \T{( \T{\bdeta_{\vech, 1}}, \T{\bdeta_{\vech, 2}} )} \equiv \T{[ \T{( \bSigma^{-1} \bmu )}, \allowbreak
-\frac12 \T{ \T{\bD_d} \{ \vect (\bSigma^{-1}) \}} ]}$ and
$A_{\vech} (\bdeta_{\vech}) = -\frac14 \T{\bdeta}_{\vech, 1} \{ \vect^{-1} (\bD_d^{+ \intercal} \bdeta_{\vech, 2}) \}^{-1}
\bdeta_{\vech, 1} - \frac12 \log | -2 \vect^{-1} (\bD_d^{+ \intercal} \bdeta_{\vech, 2}) |$, respectively.
The inverse mapping of the natural parameter vector under the vech-based representation is

\begin{equation}
	\bmu = -\frac12 \left\{ \vect^{-1} (\bD_d^{+ \intercal}\bdeta_{\vech, 2}) \right\}^{-1} \bdeta_{\vech, 1} \quad
	\text{and} \quad
	\bSigma = -\frac12 \left\{ \vect^{-1} (\bD_d^{+ \intercal}\bdeta_{\vech, 2}) \right\}^{-1}.
\label{gauss_vech_comm_params}
\end{equation}

The other major distribution within the exponential family that is pivotal for this article is the inverse-$\chi^2$ distribution.
A random variable $x$ has an inverse-$\chi^2$ distribution with shape parameter $\xi > 0$ and scale parameter
$\lambda > 0$ if the probability density function of $x$ is

\[
	p(x) = 
		\frac{(\lambda/2)^{\xi/2}}{\Gamma (\xi/2)}
		x^{-(\xi + 2)/2} \exp \left( -\frac{\lambda}{2 x} \right) \ind (x > 0),
\]

\noindent where the vector of sufficient statistics, the natural parameter vector and the log-partition function are
$\bT (x) \equiv \T{( \log (x), 1/x )}$, $\bdeta = \T{(\eta_1, \eta_2)} = \T{\{-\frac12 (\xi + 2), -\frac{\lambda}{2} \}}$
and $A(\bdeta) \equiv \log \{ \Gamma (\xi/2) \} - \frac{\xi}{2} \log (\lambda/2)$, respectively. Note that
$\Gamma (z) \equiv \int_0^\infty u^{z - 1} e^u du$ is the gamma function,
$\ind (\cdot)$ is the indicator function, $\zeta > 0$ is the scale parameter and $\lambda > 0$ is the shape parameter.
The inverse mapping of the natural parameter vector is
$\xi = -2 \eta_1 - 2$ and $\lambda = -2 \eta_2$.


\section{Derivation of the multivariate functional principal component Gaussian likelihood fragment}
\label{app:mfpca_gauss_lik_frag}

From \eqref{bayes_mfpca}, 
we have, for $i = 1, \dots, n$ and $j = 1, \dots, p$, 

\begin{equation}
	\log p (\bx_i^{(j)} | \bnu^{(j)}, \bzeta_i, \sigsqeps{j}) =
		-\frac{n_i^{(j)}}{2} \log (\sigsqeps{j})
		- \frac{1}{2\sigsqeps{j}} \left|\left|
			\bx_i^{(j)} - \C{i}{j} \left( \numu{j} + \sum_{l=1}^L \zeta_{il} \nupsi{l}{j} \right)
		\right|\right|^2
		+ \const
\label{log_bayes_mfpca_mod}
\end{equation}

First, we establish the natural parameter vector for each of the optimal posterior density functions. These natural
parameter vectors are essential for determining expectations with respect to the optimal posterior distribution.
From equation (10) of \citet{wand17}, we deduce that the natural parameter vector for $q (\bnu^{(j)})$, $j = 1, \dots,
p$ is

\[
	\npq{\bnu^{(j)}} =
		\np{p (\bx \mid \bnu, \bzeta_1, \dots, \bzeta_n, \sigsqeps{1}, \dots, \sigsqeps{p})}{\bnu^{(j)}}
		+ \np{\bnu^{(j)}}{p (\bx \mid \bnu, \bzeta_1, \dots, \bzeta_n, \sigsqeps{1}, \dots, \sigsqeps{p})},
\]

\noindent the natural parameter vector for $q (\bzeta_i)$, $i = 1, \dots, n$, is

\[
	\npq{\bzeta_i} =
		\np{p (\bx \mid \bnu, \bzeta_1, \dots, \bzeta_n, \sigsqeps{1}, \dots, \sigsqeps{p})}{\bzeta_i}
		+ \np{\bzeta_i}{p (\bx \mid \bnu, \bzeta_1, \dots, \bzeta_n, \sigsqeps{1}, \dots, \sigsqeps{p})},
\]

\noindent and the natural parameter vector for $q(\sigsqeps{j})$, $j = 1, \dots, p$ is

\[
	\npq{\sigsqeps{j}} =
		\np{p (\bx \mid \bnu, \bzeta_1, \dots, \bzeta_n, \sigsqeps{1}, \dots, \sigsqeps{p})}{\sigsqeps{j}}
		+ \np{\sigsqeps{j}}{p (\bx \mid \bnu, \bzeta_1, \dots, \bzeta_n, \sigsqeps{1}, \dots, \sigsqeps{p})}.
\]

\noindent Next, we consider the updates for standard expectations that occur for each of
the random variables and random vectors in
\eqref{log_bayes_mfpca_mod}. For each $\bnu^{(j)}$, we need to determine the mean vector $\E_q (\bnu^{(j)})$
and the covariance matrix $\Cov_q (\bnu^{(j)})$. The expectations are taken with respect to the normalisation
of

\[
	\msg{p (\bx \mid \bnu, \bzeta_1, \dots, \bzeta_n, \sigsqeps{1}, \dots, \sigsqeps{p})}{\bnu^{(j)}} (\bnu^{(j)}) \
	\msg{\bnu^{(j)}}{p (\bx \mid \bnu, \bzeta_1, \dots, \bzeta_n, \sigsqeps{1}, \dots, \sigsqeps{p})} (\bnu^{(j)}),
\]

\noindent which is a multivariate normal density function with natural parameter vector $\npq{\bnu^{(j)}}$.
From \eqref{gauss_vec_comm_params}, we have

\begin{align}
\begin{split}
	\E_q (\bnu^{(j)})
		&\longleftarrow
			-\frac12 \left[
				\vect^{-1} \left\{
					\left( \npq{\bnu^{(j)}} \right)_2
				\right\}
			\right]^{-1} \left( \npq{\bnu^{(j)}} \right)_1 \\
	\text{and} \quad
	\Cov_q (\bnu^{(j)})
		&\longleftarrow
			-\frac12 \left[
				\vect^{-1} \left\{
					\left( \npq{\bnu^{(j)}} \right)_2
				\right\}
			\right]^{-1}.
\end{split}
\label{mom_lik_nu}
\end{align}

\noindent Furthermore, the mean vector has the form

\begin{equation}
	\E_q (\bnu^{(j)}) \equiv \T{
		\left\{ \T{\E_q (\numu{j})}, \T{\E_q (\nupsi{1}{j})}, \dots, \T{\E_q (\nupsi{L}{j})} \right\}
	},
\label{exp_lik_nu}
\end{equation}

\noindent and the covariance matrix has the form

\begin{equation}
	\Cov_q (\bnu) \equiv \begin{bmatrix}
		\Cov_q (\numu{j}) & \Cov_q (\numu{j}, \nupsi{1}{j}) & \dots & \Cov_q (\numu{j}, \nupsi{L}{j}) \\
		\Cov_q (\nupsi{1}{j}, \numu{j}) & \Cov_q (\nupsi{1}{j}) & \dots & \Cov_q (\nupsi{1}{j}, \nupsi{L}{j}) \\
		\vdots & \vdots & \ddots & \vdots \\
		\Cov_q (\nupsi{L}{j}, \numu{j}) & \Cov_q (\nupsi{L}{j}, \nupsi{1}{j}) & \dots & \Cov_q (\nupsi{L}{j}) \\
	\end{bmatrix}.
\label{cov_lik_nu}
\end{equation}

\noindent Similarly, for each $i = 1, \dots, n$, we need to determine the optimal mean vector and covariance matrix
for $\bzeta_i$, which are $\E_q (\bzeta_i)$ and $\Cov_q (\bzeta_i)$, respectively. The expectations are taken with
respect to the normalisation of

\[
	\msg{p (\bx \mid \bnu, \bzeta_1, \dots, \bzeta_n, \sigsqeps{1}, \dots, \sigsqeps{p})}{\bzeta_i} (\bzeta_i) \
	\msg{\bzeta_i}{p (\bx \mid \bnu, \bzeta_1, \dots, \bzeta_n, \sigsqeps{1}, \dots, \sigsqeps{p})} (\bzeta_i),
\]

\noindent which is a multivariate normal density function with natural parameter vector $\npq{\bzeta_i}$.
According to \eqref{gauss_vech_comm_params},

\begin{align}
\begin{split}
	\E_q (\bzeta_i)
		&\longleftarrow
			-\frac12 \left[
				\vect^{-1} \left\{
					\bD_L^{+ \intercal}
					\left( \npq{\bzeta_i} \right)_2
				\right\}
			\right]^{-1} \left( \npq{\bzeta_i} \right)_1 \\
	\text{and} \quad
	\Cov_q (\bzeta_i)
		&\longleftarrow
			-\frac12 \left[
				\vect^{-1} \left\{
					\bD_L^{+ \intercal}
					\left( \npq{\bzeta_i} \right)_2
				\right\}
			\right]^{-1}, \quad \text{for $i = 1, \dots, n$.}
\end{split}
\label{exp_lik_zeta}
\end{align}

\noindent Finally, for each $j = 1, \dots, p$, we need to determine $\E_q (1/\sigsqeps{j})$, with the expectation taken with
respect to the normalisation of

\[
	\msg{p (\bx \mid \bnu, \bzeta_1, \dots, \bzeta_n, \sigsqeps{1}, \dots, \sigsqeps{p})}{\sigsqeps{j}} (\sigsqeps{j}) \
	\msg{\sigsqeps{j}}{p (\bx \mid \bnu, \bzeta_1, \dots, \bzeta_n, \sigsqeps{1}, \dots, \sigsqeps{p})} (\sigsqeps{j}).
\]

\noindent This is an inverse-$\chi^2$ density function, with natural parameter vector $\npq{\sigsqeps{j}}$.
According to Result 6 of \citet{maestrini20},

\[
	\E_q (1/\sigsqeps{j})
		\longleftarrow
			\frac{
				\left( \npq{\sigsqeps{j}} \right)_1 + 1
			}{
				\left( \npq{\sigsqeps{j}} \right)_2
			}.
\]

\noindent Now, we turn our attention to the derivation of the message passed from $p (\bx \mid \bnu, \bzeta_1, \dots, \bzeta_n,
\sigsqeps{1}, \dots, \sigsqeps{p})$ to $\bnu^{(j)}$, $j = 1, \dots, p$. Notice that

\begin{equation}
	\C{i}{j} \left( \numu{j} - \sum_{l=1}^L \zeta_{il} \nupsi{l}{j} \right) = (\T{\bzetatilde_i} \otimes \C{i}{j}) \bnu^{(j)}.
\label{theta_simplification}
\end{equation}

\noindent Therefore, as a function of $\bnu^{(j)}$, \eqref{log_bayes_mfpca_mod} can be re-written as

\begin{align*}
	\log p (\bx_i^{(j)} \mid \bnu^{(j)}, \bzeta_i, \sigsqeps{j})
		& = -\frac{1}{2\sigsqeps{j}} \left|\left|
			\bx_i^{(j)} - (\T{\bzetatilde_i} \otimes \C{i}{j}) \bnu^{(j)}
		\right|\right|^2 + \tni{\bnu^{(j)}} \\
		& = \T{\begin{bmatrix}
			\bnu^{(j)} \\
			\vect (\bnu^{(j)} \bnu^{(j) \intercal})
		\end{bmatrix}} \begin{bmatrix}
			\frac{1}{\sigsqeps{j}} \T{(\T{\bzetatilde_i} \otimes \C{i}{j})} \bx_i^{(j)} \\
			-\frac{1}{2\sigsqeps{j}} \vect \left\{
				(\bzetatilde_i \T{\bzetatilde_i}) \otimes (\CT{i}{j} \C{i}{j})
			\right\}
		\end{bmatrix} \\ &\qquad + \tni{\bnu^{(j)}}.
\end{align*}

\noindent According to equation (8) of \citet{wand17}, the message from the factor $p (\bx \mid \bnu, \bzeta_1, \dots,
\bzeta_n, \sigsqeps{1}, \dots, \sigsqeps{p})$ to $\bnu^{(j)}$ is

\[
	\msg{p (\bx \mid \bnu, \bzeta_1, \dots, \bzeta_n, \sigsqeps{1}, \dots, \sigsqeps{p})}{\bnu^{(j)}} (\bnu^{(j)}) \propto
		\exp \left\{
			\T{\begin{bmatrix}
				\bnu^{(j)} \\
				\vect (\bnu^{(j)} \bnu^{(j) \intercal})
			\end{bmatrix}}
			\np{p (\bx \mid \bnu, \bzeta_1, \dots, \bzeta_n, \sigsqeps{1}, \dots, \sigsqeps{p})}{\bnu^{(j)}}
		\right\},
\]

\noindent which is proportional to a
multivariate normal density function. The update for the message's natural parameter vector,
in \eqref{np_lik_nu}, is dependent upon
the mean vector and covariance matrix of $\bzetatilde_i$, which are

\begin{equation}
	\E_q (\zetatilde_i) = \T{\{ 1, \T{\E_q (\bzeta_i)} \}} \quad
	\text{and} \quad
	\Cov_q (\zetatilde_i) = \blockdiag \left\{ 0, \Cov_q (\bzeta_i) \right\}, \quad
	\text{for $i = 1, \dots, n$,}
\label{exp_lik_zeta_tilde}
\end{equation}

\noindent where $\E_q (\bzeta_i)$ and $\Cov_q (\bzeta_i)$ are defined in \eqref{exp_lik_zeta}. Note that
a standard statistical result allows us to write

\begin{equation}
	\E_q (\bzetatilde_i \T{\bzetatilde_i}) =
		\Cov_q (\bzetatilde_i) + \E_q (\bzetatilde_i) \T{\E_q (\bzetatilde_i)}, \quad \text{for $i = 1, \dots, n$.}
\label{E_q_outer_zeta}
\end{equation}

\noindent Next, notice that

\begin{equation}
	\sum_{l=1}^L \zeta_{il} \nupsi{l}{j} = \Vpsi{j} \bzeta_i
\label{zeta_simplification}
\end{equation}

\noindent Then, for each $i = 1, \dots, n$, the log-density function
in \eqref{log_bayes_mfpca_mod} can be represented as a function of $\bzeta_i$ by

\begin{align*}
	\log p (\bx_i^{(j)} \mid \bnu^{(j)}, \bzeta_i, \sigsqeps{j})
		& = -\frac{1}{2\sigsqeps{j}} \left|\left|
			\bx_i^{(j)} - \C{i}{j} \numu{j} - \C{i}{j} \Vpsi{j} \bzeta_i
		\right|\right|^2 + \tni{\bzeta_i} \\
		& = \T{
			\begin{bmatrix}
				\bzeta_i \\
				\vech (\bzeta_i \T{\bzeta_i})
			\end{bmatrix}
		} \begin{bmatrix}
			\frac{1}{\sigsqeps{j}} (\VpsiT{j} \CT{i}{j} \bx_i^{(j)} - \hmupsi{i}{j}) \\
			-\frac{1}{2 \sigsqeps{j}} \T{\bD_L} \vect (\Hpsi{i}{j})
		\end{bmatrix} \\ &\qquad + \tni{\bzeta_i},
\end{align*}

\noindent According to equation (8) of \citet{wand17}, the message from the factor
$p (\bx \mid \bnu, \bzeta_1, \dots, \bzeta_n, \sigsqeps{1}, \dots, \sigsqeps{p})$ to $\bzeta_i$ is

\[
	\msg{p (\bx \mid \bnu, \bzeta_1, \dots, \bzeta_n, \sigsqeps{1}, \dots, \sigsqeps{p})}{\bzeta_i} (\bzeta_i) \propto
		\exp \left\{
			\T{\begin{bmatrix}
				\bzeta_i \\
				\vech (\bzeta_i \T{\bzeta_i})
			\end{bmatrix}}
			\np{p (\bx \mid \bnu, \bzeta_1, \dots, \bzeta_n, \sigsqeps{1}, \dots, \sigsqeps{p})}{\bzeta_i}
		\right\},
\]

\noindent which
is proportional to a multivariate normal density function. The message's natural parameter vector update, in 
\eqref{np_lik_zeta}, is dependant on the following expectations that are yet to be determined:

\[
	\E_q (\Vpsi{j}), \quad \E_q (\Hpsi{i}{j}), \quad \text{and} \quad \E_q (\hmupsi{i}{j}), \quad i = 1, \dots, n \quad
	\text{and} \quad j = 1, \dots, p.
\]

\noindent Now, we have,

\begin{equation}
	\E_q (\Vpsi{j}) = \begin{bmatrix}
		\E_q (\nupsi{1}{j}) & \dots & \E_q (\nupsi{L}{j})
	\end{bmatrix},
\label{E_q_Vpsi}
\end{equation}

\noindent where, for $l = 1, \dots, L$, $\E_q (\nupsi{l}{j})$ is defined by \eqref{mom_lik_nu} and \eqref{exp_lik_nu}.
Next, $\E_q (\hmupsi{i}{j})$ is an $L \times 1$ vector, with $l$th component being

\begin{equation}
	\E_q (\hmupsi{i}{j})_l =
		\tr \{ \Cov_q (\numu{j}, \nupsi{l}{j}) \CT{i}{j} \C{i}{j} \}
		+ \T{\E_q (\nupsi{l}{j})} \CT{i}{j} \C{i}{j} \E_q (\numu{j}), \quad
	l = 1, \dots, L,
\label{exp_lik_hmupsi}
\end{equation}

\noindent which depends on sub-vectors of $\E_q (\bnu^{(j)})$ and sub-blocks of $\Cov_q (\bnu^{(j)})$ that are defined
in \eqref{exp_lik_nu} and \eqref{cov_lik_nu}, respectively. Finally, $\E_q (\Hpsi{i}{j})$ is an $L \times L$ matrix,
with $(l, l')$ component being

\begin{equation}
	\E_q (\Hpsi{i}{j})_{l, l'} =
		\tr \{ \Cov_q (\nupsi{l'}{j}, \nupsi{l}{j}) \CT{i}{j} \C{i}{j} \}
		+ \T{\E_q (\nupsi{l}{j})} \CT{i}{j} \C{i}{j} \E_q (\nupsi{l'}{j}), \quad
	l, l' = 1, \dots, L.
\label{exp_lik_Hpsi}
\end{equation}

\noindent The final message to consider is the message from $p (\bx \mid \bnu, \bzeta_1, \dots, \bzeta_n, \sigsqeps{1}, \dots,
\sigsqeps{p})$ to
$\sigsqeps{j}$, for $j = 1, \dots, p$. As a function of $\sigsqeps{j}$, \eqref{log_bayes_mfpca_mod} takes the form

\begin{align*}
	\log p (\bx_i^{(j)} \mid \bnu^{(j)}, \bzeta_i, \sigsqeps{j})
		& = -\frac{n_i^{(j)}}{2} \log (\sigsqeps{j}) - \frac{1}{2 \sigsqeps{j}} \left|\left|
			\bx_i^{(j)} - \C{i}{j} \bV^{(j)} \bzetatilde_i
		\right|\right|^2 \\ & \qquad + \tni{\sigsqeps{j}} \\
		& = \T{
			\begin{bmatrix}
				\log (\sigsqeps{j}) \\
				\frac{1}{\sigsqeps{j}}
			\end{bmatrix}
		} \begin{bmatrix}
			-\frac{n_i^{(j)}}{2} \\
			-\frac12 \left|\left| \bx_i^{(j)} - \C{i}{j} \bV^{(j)} \bzetatilde_i \right|\right|^2
		\end{bmatrix} + \tni{\sigsqeps{j}},
\end{align*}

\noindent According to equation (8) of \citet{wand17}, the message from $p (\bx \mid \bnu, \bzeta_1, \dots, \bzeta_n,
\sigsqeps{1}, \dots, \sigsqeps{p})$ to $\sigsqeps{j}$ is

\[
	\msg{p (\bx \mid \bnu, \bzeta_1, \dots, \bzeta_n, \sigsqeps{1}, \dots, \sigsqeps{p})}{\sigsqeps{j}} (\sigsqeps{j}) \propto
		\exp \left\{
			\T{\begin{bmatrix}
				\log (\sigsqeps{j}) \\
				1/\sigsqeps{j}
			\end{bmatrix}}
			\np{p (\bx \mid \bnu, \bzeta_1, \dots, \bzeta_n, \sigsqeps{1}, \dots, \sigsqeps{p})}{\sigsqeps{j}}
		\right\},
\]

\noindent which is proportional
to an inverse-$\chi^2$ density function. The message's natural parameter vector, in \eqref{np_lik_sigsqeps}, depends
on the mean of the square norm $\Vert \bx_i^{()j} - \C{i}{j} \bV^{(j)} \bzetatilde_i \Vert^2$, for $i = 1, \dots, n$.
This expectation takes the form

\begin{align*}
	\E_q \left(
		\left|\left| \bx_i^{j} - \C{i}{j} \bV^{(j)} \bzetatilde_i \right|\right|^2
	\right) =
		& \bx_i^{(j) \intercal} \bx_i^{(j)} - 2 \T{\E_q (\bzetatilde_i)} \T{\E_q (\bV^{(j)})} \CT{i}{j} \bx_i^{(j)} \\
		& + \tr \left[
			\left\{ \Cov_q (\bzetatilde_i) + \E_q (\bzetatilde_i) \T{\E_q (\bzetatilde_i)} \right\} \E_q (\bH_i^{(j)})
		\right],
\end{align*}

\noindent where we introduce the matrices

\begin{equation}
	\bH_i^{(j)} \equiv \begin{bmatrix}
		\hmu{i}{j} & \hmupsiT{i}{j} \\
		\hmupsi{i}{j} & \Hpsi{i}{j}
	\end{bmatrix}, \quad
	\text{for $i = 1, \dots, n$ and $j = 1, \dots, p$},
\label{H_mat}
\end{equation}

\noindent and vectors

\begin{equation}
	\hmu{i}{j} \equiv \numuT{j} \CT{i}{j} \C{i}{j} \numu{j}, \quad
	\text{for $i = 1, \dots, n$ and $j = 1, \dots, p$}.
\label{hmu_vec}
\end{equation}

\noindent For each $i = 1, \dots, n$, the mean vector $\E_q (\bzetatilde_i)$ and $\Cov_q (\bzetatilde_i)$ are defined in
\eqref{exp_lik_zeta_tilde}. However, $\E_q (\bV^{(j)})$ and $\E_q (\bH_i^{(j)})$, $i = 1, \dots, n$ and $j = 1, \dots, p$,
are yet to be determined. We then have,

\[
	\E_q (\bV^{(j)}) = \begin{bmatrix}
		\E_q (\numu{j}) & \E_q (\nupsi{1}{j}) & \dots & \E_q (\nupsi{L}{j})
	\end{bmatrix},
\]

\noindent where the component mean vectors are defined by \eqref{exp_lik_nu}.
For each $i = 1, \dots, n$ and $j = 1, \dots, p$, the expectation of $\bH_i^{(j)}$, defined in \eqref{H_mat},
with respect to the optimal posterior distribution is

\[
	\E_q (\bH_i^{(j)}) \equiv \begin{bmatrix}
		\E_q (\hmu{i}{j}) & \T{\E_q (\hmupsi{i}{j})} \\
		\E_q (\hmupsi{i}{j}) & \E_q (\Hpsi{i}{j})
	\end{bmatrix},
\]

\noindent where $\hmu{i}{j}$ is defined in \eqref{hmu_vec} with expected value

\[
	\E_q (\hmu{i}{j}) \equiv
		\tr \{ \Cov_q (\numu{j}) \CT{i}{j} \C{i}{j} \}
		+ \T{\E_q (\numu{j})} \CT{i}{j}\C{i}{j} \E_q (\numu{j}).
\]

\noindent Furthermore, $\E_q (\hmupsi{i}{j})$ and $\E_q (\Hpsi{i}{j})$ are defined in \eqref{exp_lik_hmupsi} and
\eqref{exp_lik_Hpsi}, respectively.

\noindent The multivariate FPCA Gaussian likelihood fragment, summarised in Algorithm \ref{alg:mfpca_gauss_lik_frag}, 
is a
proceduralisation of these results.

\section{Additional simulation results}\label{app:sim}

This section contains additional results from the simulation studies.

\subsection{Addendum on the accuracy of the variational algorithm}\label{app:acc}

\begin{figure}[H]
\centering
\includegraphics[scale=0.46]{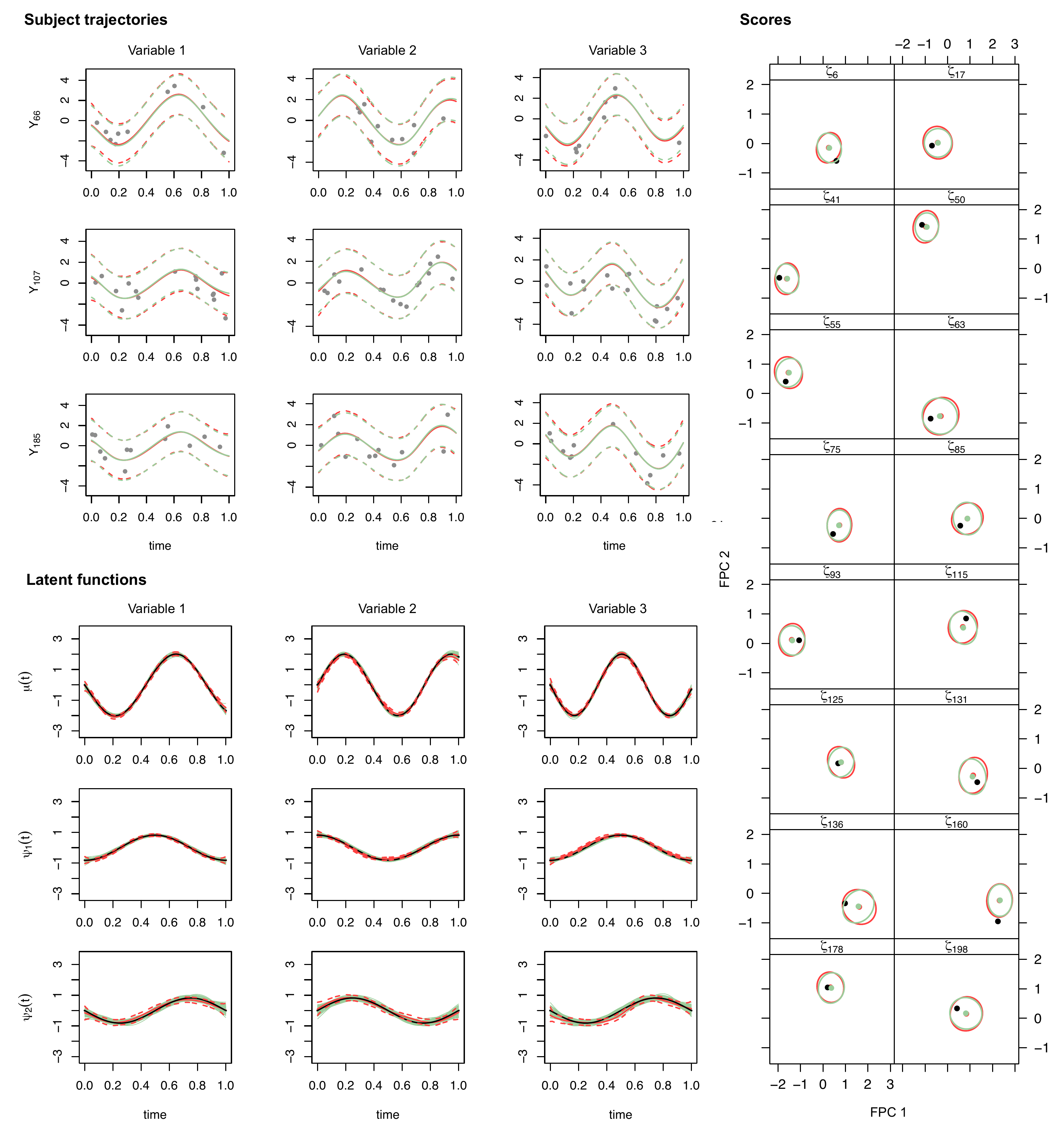}	
\caption{MCMC and MFVB estimates for a problem with $p = 3$ variables observed at an average of $15$ time points for $n = 200$ subjects; $L$ is learnt  by estimating the PVE with $L_\text{max} = 10$, and $K$ is set using   
the rule of thumb described in Section~\ref{sec:choice_K_L} of the main text. Top left: estimated trajectories for a random subset of $3$ subjects, with the posterior mean (solid lines) and $95\%$ pointwise prediction bands (dashed). The lines corresponding to MCMC (red) and MFVB (green) inference overlap. Bottom left: mean and eigenfunctions simulated (black) and estimated by MCMC inference (red) with $95\%$ credible bands (dashed) and by MFVB inference (green) for which estimates from $100$ replicates are overlaid. Right: scores simulated (black dots) and estimated by MCMC (posterior mean, red dots) inference and MFVB (posterior mean, green dots), with $95\%$ credible contours, for a random subset of $16$ subjects.}\label{fig_sm:acc}
\end{figure}

Figure \ref{fig_sm:acc} shows a comparison between the MCMC and MFVB subject trajectories, scores and latent functions, for the problem described in Section~\ref{sec:acc}. 
It indicates strong agreement, similarly as in the comparison between VMP and MCMC estimates.

\begin{figure}[h!]
\centering
\includegraphics[scale=0.52]{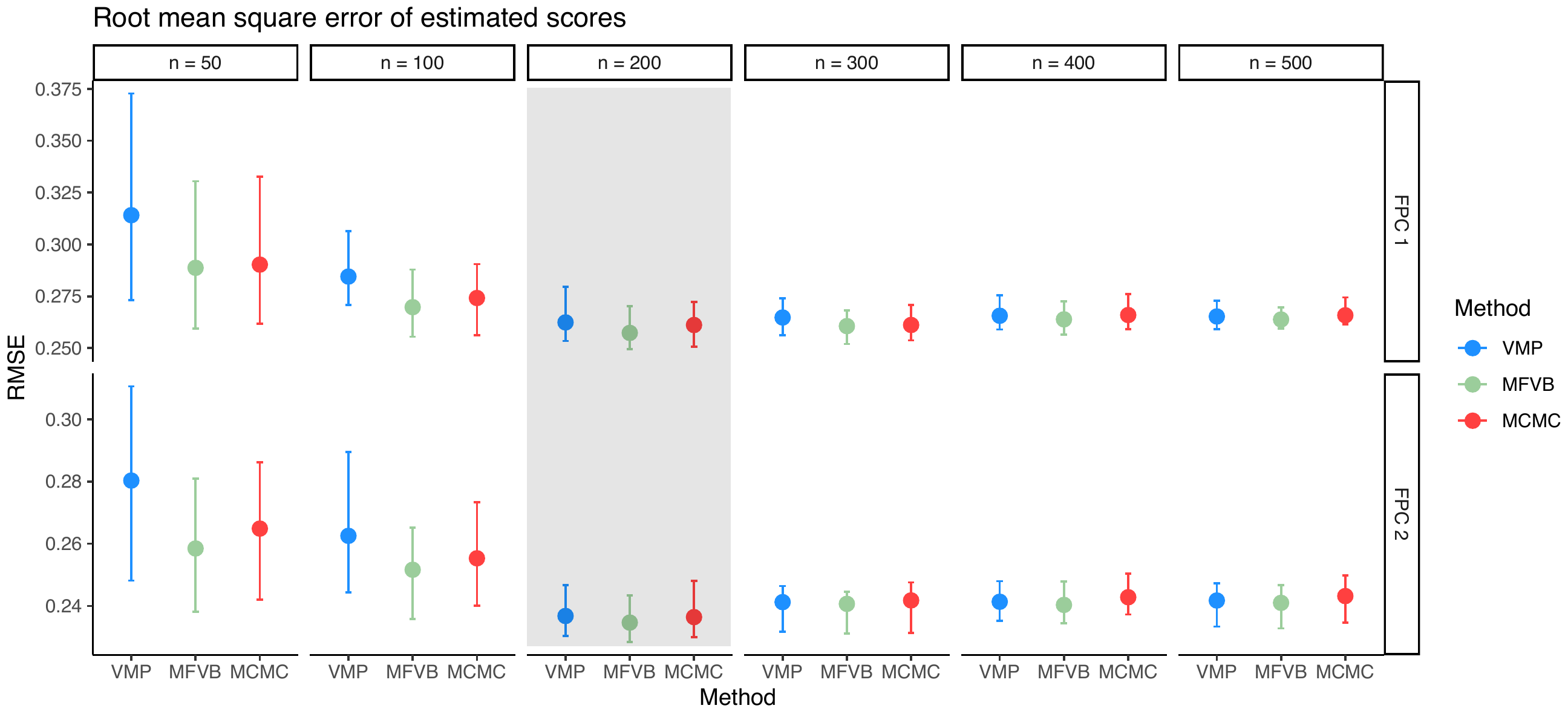}	
\caption{Root mean square error (RMSE) of scores (FPC 1 first row, FPC 2 second row) for a problem with $p = 3$ variables observed at an average of $15$ time points for a grid of  $n \in \{50, 100, 200, 300, 400, 500\}$ subjects, using VMP (blue), MFVB (green) and MCMC (red) inference. Each scenario is based on $100$ data replicates and the scenario shown on Figure~\ref{fig:vmp_vs_mcmc} 
is shaded in grey.}\label{fig:vmp_vs_mcmc_rmse}
\end{figure}

\begin{figure}[h!]
\centering
\includegraphics[scale=0.52]{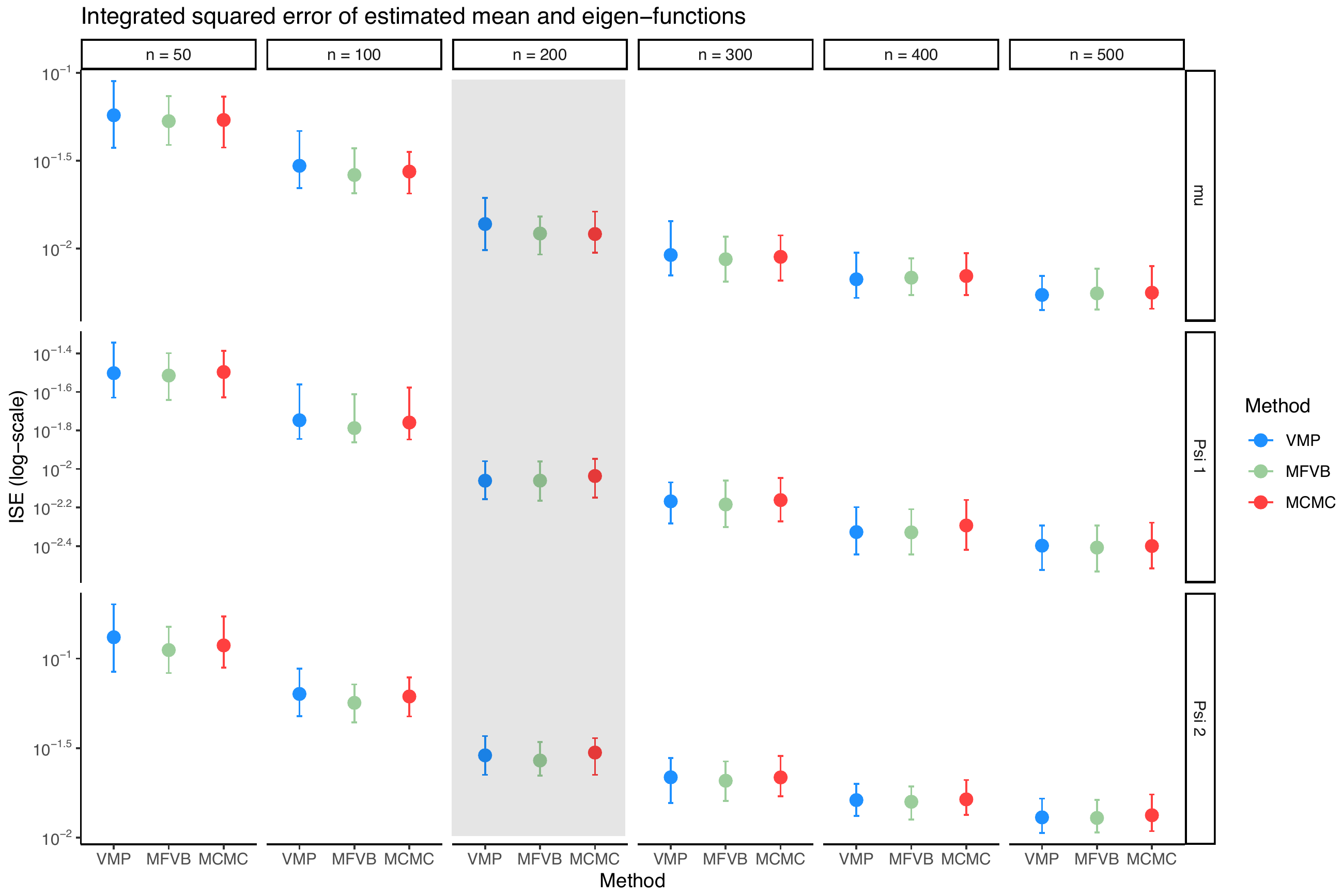}		
\caption{Integrated squared error (ISE) of the mean function (first row) and two eigenfunctions (second and third rows) for a problem with $p = 3$ variables observed at an average of $15$ time points for a grid of  $n~\in~\{50, 100, 200, 300, 400, 500\}$ subjects, using VMP (blue), MFVB (green) and MCMC (red) inference. Each scenario is based on $100$ data replicates and the scenario shown on Figure~\ref{fig:vmp_vs_mcmc} 
is shaded in grey. For each replicate, per-variable ISE are computed and averaged across all $p$ variables.}\label{fig:vmp_vs_mcmc_ise}
\end{figure}

\noindent Figures~\ref{fig:vmp_vs_mcmc_rmse} and \ref{fig:vmp_vs_mcmc_ise} show a comparison of the errors on the estimated scores and latent functions for a grid of problems with $n \in \{50, 100, 200, 300, 400, 500\}$, simulating $100$ replicates for each.  The VMP, MFVB and MCMC errors are very similar, and all decrease as $n$ increases. Additionally, as we expect, when $n$ is small, the variability of the errors is greater (which is better seen for the scores, whose RMSE is presented on the original scale, rather than for the latent functions whose ISE is shown on the logarithmic scale). 

\subsection{Addendum on the selection of $K$ and $L$}\label{app:sel_k_l}

\begin{figure}[h!]
\centering
\includegraphics[scale=0.48]{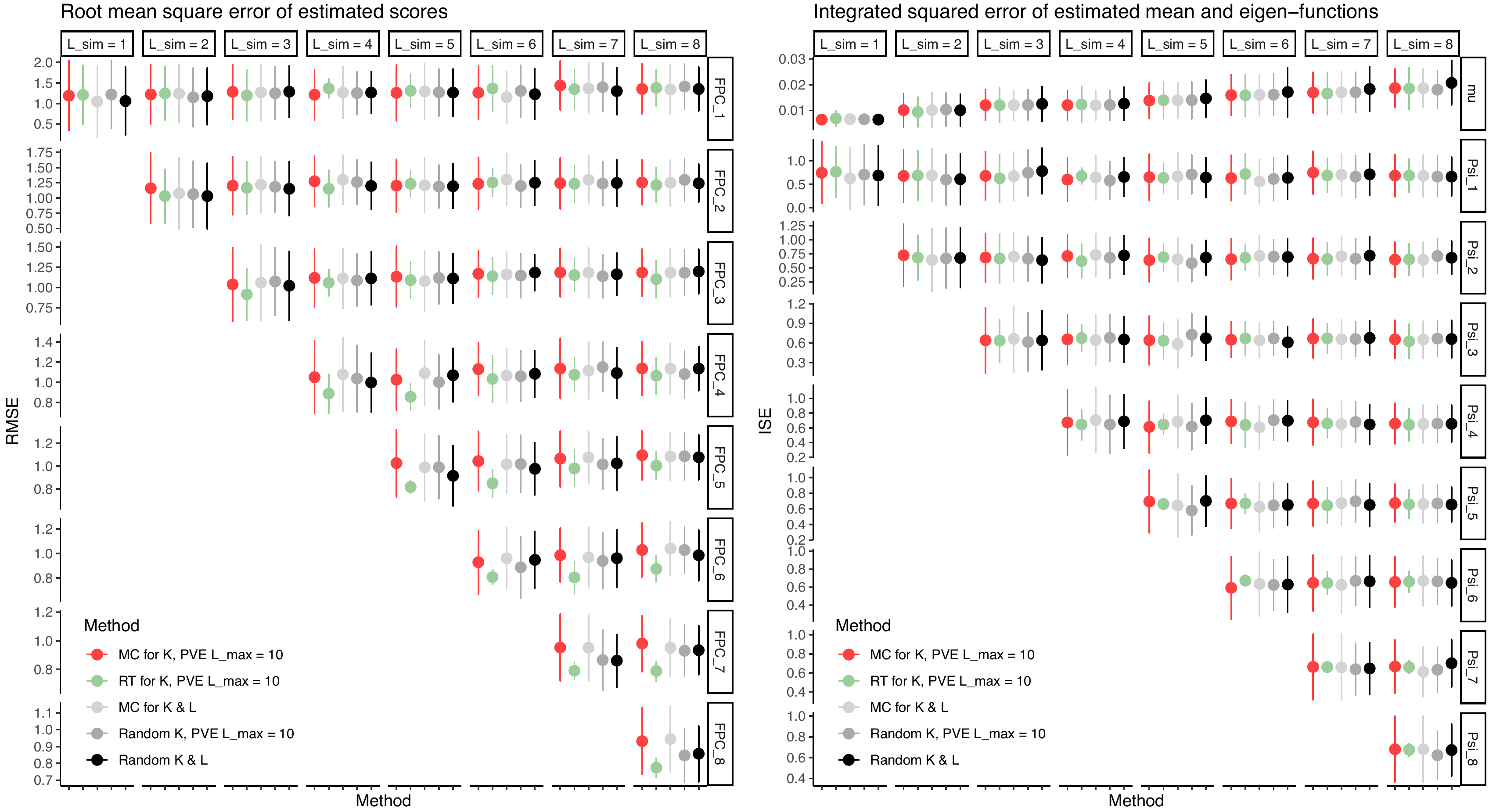}	
\caption{Estimation errors for a problem with $p = 3$ variables, $n = 100$ subjects, $20$ observations per subject on average, and with numbers of simulated components ranging from $L = 1$ to $L = 8$ (columns). The procedures for settings $K$ and $L$ are: model choice for $K$ and PVE-based selection for $L$ (red); rule of thumb adapted from \citet{ruppert2002selecting} for $K$ and PVE-based selection for $L$ (green); model choice for both $K$ and $L$ (light grey); random choice for $K \in \{5, \ldots, 20\}$ and PVE-based selection for $L$ (dark grey);  random choice for $K \in \{5, \ldots, 20\}$ and $L_\text{max} \in \{L, \ldots, 10\}$. Left: RMSE on the scores. Right: ISE on the mean function (first row) and eigenfunctions (ISE).}\label{fig_sm:errors_choice}
\end{figure}

Figure~\ref{fig_sm:errors_choice} shows the errors on the scores and latent functions achieved by the method when the model is estimated with different choices of numbers of latent functions $L$ and spline basis functions for representing these latent functions $K$. It suggests that the different approaches to learning $K$ and $L$, described in Section~\ref{sec:choice_K_L} of the main text, result in similar errors, with no sign of overfitting, even compared with random choices $K \in \{5, \ldots, 20\}$ and $L_\text{max} \in \{L, \ldots, 10\}$. The data correspond to $100$ replicates  based on the simulation setup used in Section~\ref{sec:sim:choice}.

\subsection{Comparisons with parametric and semiparametric approaches}\label{app:kowal}

In this section, we present a comparison of our modelling framework with the semiparametric functional factor model (SFFM) approach, proposed by \citet{kowal2023semiparametric}. Specifically, we assess:
\begin{itemize}
\item the recovery of the correct number of components $L$;
\item the empirical coverage and length of the curve's prediction intervals;
\item the associated prediction errors,
\end{itemize}
for the SFFM approach which models the latent function with a semiparametric basis (``SFFM'' and a finite mixture model variant ``SFFM (fmm)''), two models based on fully parametric bases (``PFFM'' and ``PFFM+gp''); we direct the reader to the authors' paper for more detail about these models.

\begin{figure}[t!]
\centering
\includegraphics[scale=0.255]{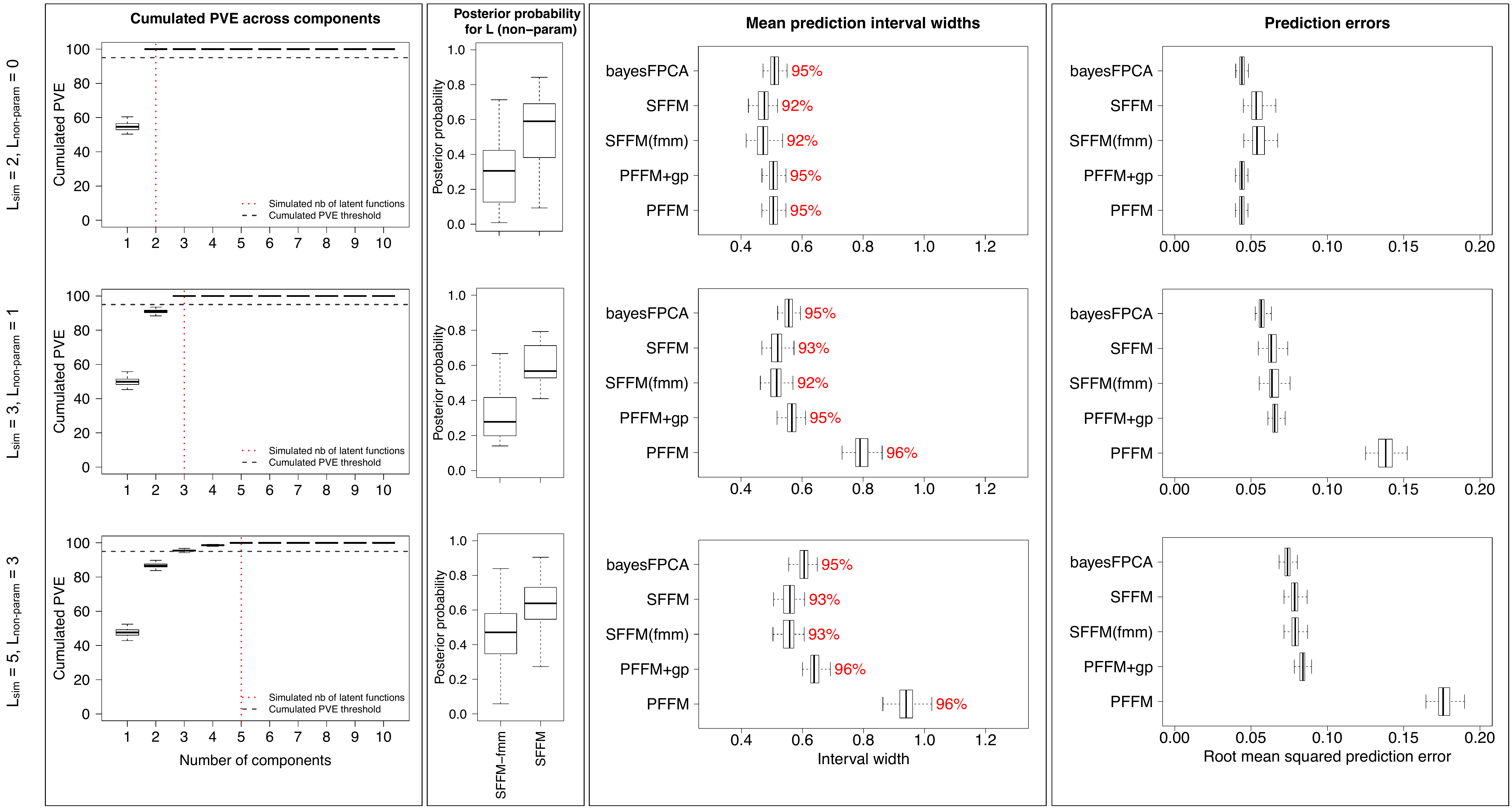}	
\caption{Comparison of Bayesian FPCA (bayesFPCA) with the semiparametric method of \citet{kowal2023semiparametric} (``SFFM''  and ``SFFM (fmm)'') and fully parametric variants (`PFFM'' and ``PFFM+gp''), for different numbers of simulated components $L$ of which $L_\text{nonparam}$ are nonparametric under the ``linear template'' (first row:  $L = 2$, $L_\text{nonparam} = 0$; second row $L = 3$, $L_\text{nonparam} = 1$; third row $L = 5$, $L_\text{nonparam} = 3$). First column: Estimation of the cumulated PVE by bayesFPCA; Second column: Posterior probability for the simulated $L_\text{nonparam}$, as estimated by SFFM; third column: Prediction interval length and empirical coverage (using training/test set split) by the different methods; fourth column: Prediction errors  (using training/test set split) by the different methods. }\label{fig_sm:kowal_1}
\end{figure}

\begin{figure}[h!]
\centering
\includegraphics[scale=0.255]{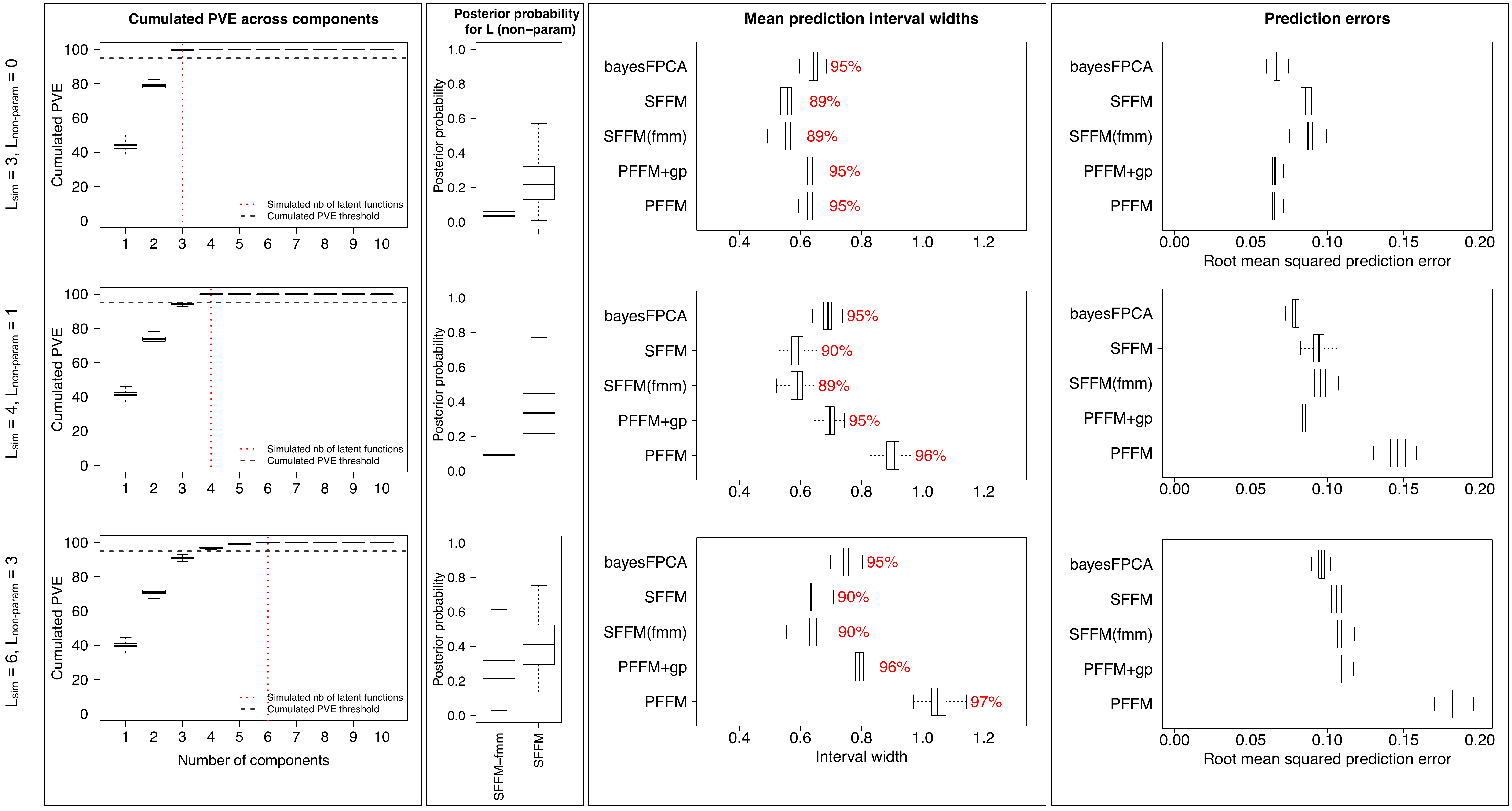}	
\caption{Comparison of Bayesian FPCA (bayesFPCA) with the semiparametric method of \citet{kowal2023semiparametric} (``SFFM''  and ``SFFM (fmm)'') and fully parametric variants (`PFFM'' and ``PFFM+gp''), for different numbers of simulated components $L$ of which $L_\text{nonparam}$ are nonparametric under the ``linear template'' (first row:  $L = 3$, $L_\text{nonparam} = 0$; second row $L = 4$, $L_\text{nonparam} = 1$; third row $L = 6$, $L_\text{nonparam} = 3$). The column description is the same as in Figure~\ref{fig_sm:kowal_1}.}\label{fig_sm:kowal_2}
\end{figure}

We simulate data under the two data-generation scenarios used in \citet{kowal2023semiparametric}. Specifically, we generate semiparametric latent functions using, for the parametric factors (whose basis is treated as known, in the SFFM and PFFM variants): (i) a ``linear template'', spanned by $1$ and $t$ (Figure~\ref{fig_sm:kowal_1}), and (ii) a ``Nelson \& Siegel template'', consisting of the basis defined by \citet{nelson1987parsimonious}:
$$g_1(t ; \gamma)=1,\quad g_2(t; \gamma) = \{1 - \exp(-t \gamma)\}/(t\gamma),\quad g_3(t ; \gamma)=g_2(t ; \gamma) - \exp(-t\gamma),\quad \gamma >0,$$
commonly employed in yield curve models (Figure~\ref{fig_sm:kowal_2}). Moreover, to the parametric factors, a number $L_\text{nonparam}~\in~\{0, 1, 3\}$ of unknown nonparametric factors are added (thus corresponding to a total of $L \in \{2, 3, 5\}$ for the linear template and $L \in \{3, 4, 6\}$ for the Nelson \& Siegel template). Therefore, a total of $2\times 3$ different simulated scenarios is considered. The number of subjects in each scenario is $n = 200$, with $n_i = 15$ observations per subject on average. Moreover SFFM and PFFM focus on univariate factor modelling, we therefore compare them to our implementation with $p = 1$ (``bayesFPCA'').

Figures~\ref{fig_sm:kowal_1} and~\ref{fig_sm:kowal_2}  show that the $95\%$ cumulated-PVE threshold recovers the correct $L$, when the number of components is $<5$, but provides an underestimate for the two scenarios $L=5$ and $L = 6$; however, as recommended in Section~\ref{sec:sim:choice}, the inspection of the index $l$ after which the cumulated PVE levels off provides a more effective way of identifying this number. The SFFM's approach obtains the posterior distribution of $L$ from a cumulative shrinkage process prior formulation. The correct $L$ is recovered with median probability greater $0.5$ for the linear template problems but is lower for the Nelson \& Siegel problems.

The prediction interval coverage achieved by bayesFPCA is excellent across all scenarios. The corresponding interval widths are comparable to those obtained by the PFFM methods, and slightly larger than those obtained by the SFFM method; however the SFFM coverage tends to be insufficient for the Nelson \& Siegel problems. 

Finally, the prediction errors are consistently similar or lower than those of the competing methods in all scenarios tested. In particular, and quite strikingly, in the two simulation scenarios involving pure parametric factors only ($L_{\text{nonparam}} = 0$), our fully nonparametric approach bayesFPCA achieves prediction errors lower  than SFFM and comparable to PFFM, although the basis of the latent parametric functions was known to the latter two approaches. This suggests that, although our modelling approach does not encode the parametric structure of the latent functions, it is still able able to reconstruct them effectively, with no consequence on estimation accuracy.

\subsection{Addendum on the comparison with frequentist multivariate FPCA}\label{app:freq}

\begin{figure}[h!]
\centering
\includegraphics[scale=0.53]{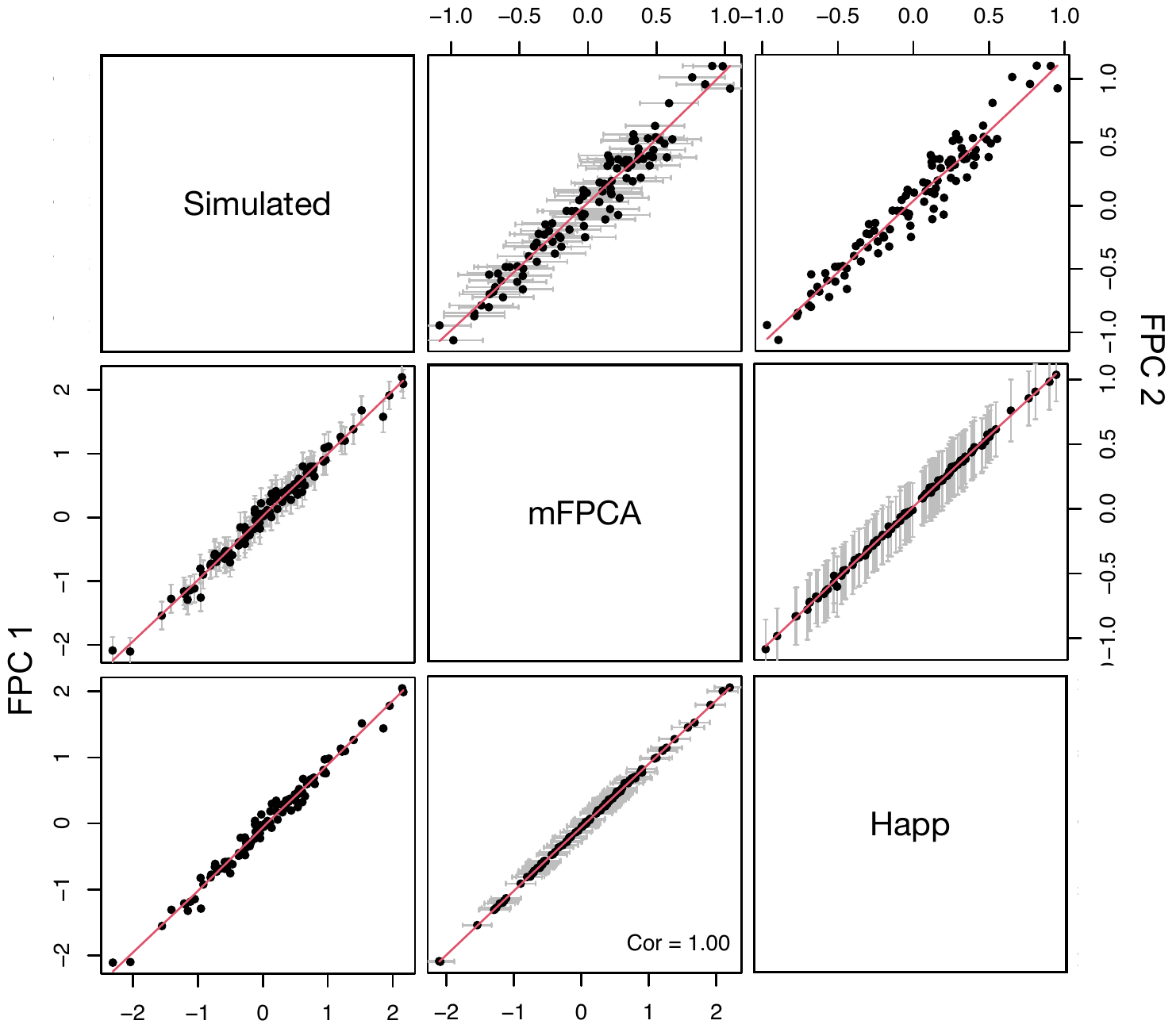}	

\caption{Comparison of scores simulated and estimated with Happ's  and our approach for one data replicate of a problem with $80$ observations per subject and variable on average. 
The scores corresponding to the first (resp. second) eigenfunction are shown on the lower (resp. upper) diagonal part. $95\%$ credible intervals are shown for estimates obtained with our approach (grey); such uncertainty quantification is unavailable for Happ's estimates.}\label{fig:happ}
\end{figure}

\begin{figure}[h!]
\centering
\includegraphics[scale=0.56]{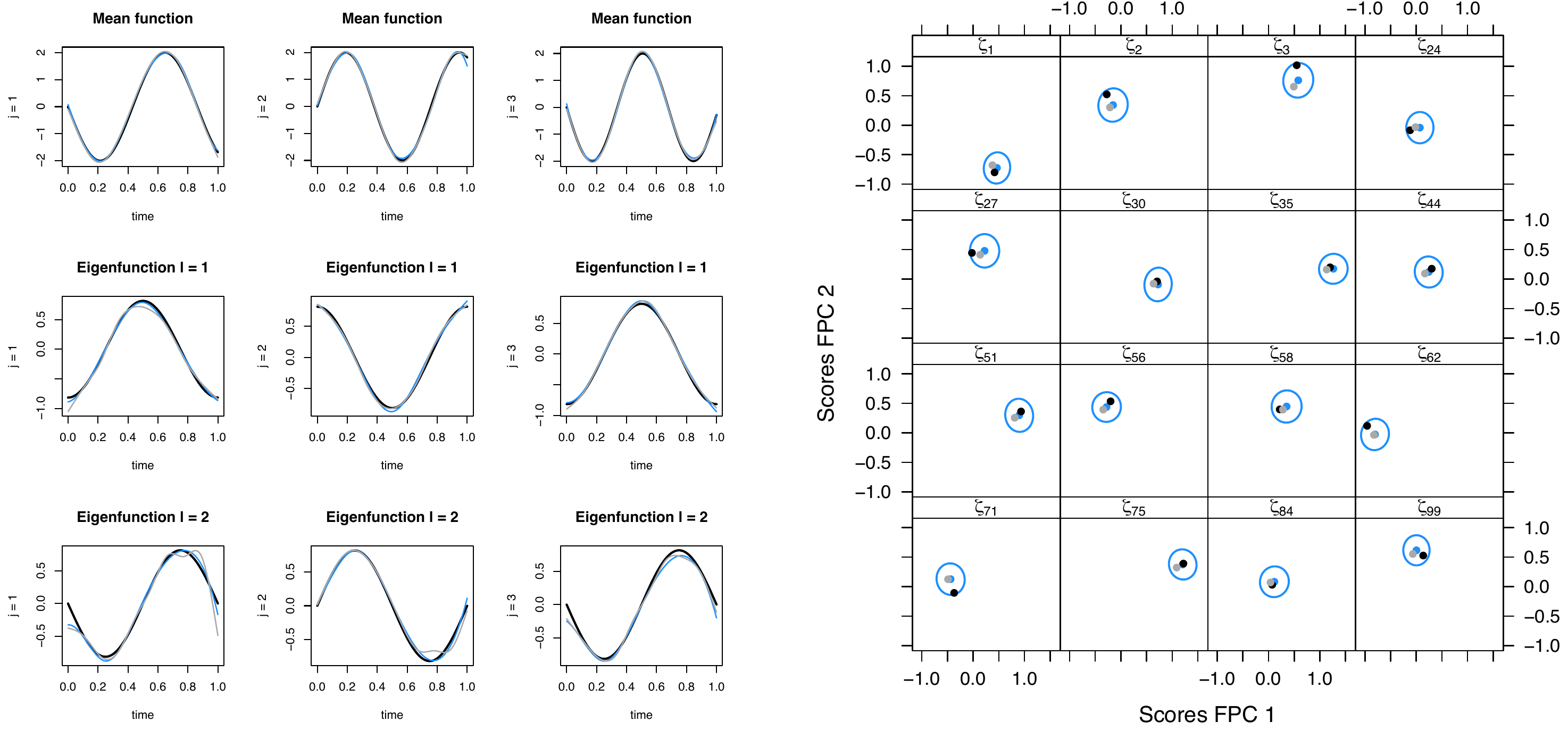}	
\caption{Comparison of FPCA estimates obtained with Happ's and our approaches, for a problem with $p = 3$ variables, $n = 100$ subjects, and an average of $80$ observations per variable and subject. Left: mean and eigen- functions simulated (black) and estimated by Happ's (grey) and our (blue) approach. Right: scores simulated (black dots) and estimated by Happ's approach (grey dots) and our approach (posterior mean, blue dots, and $95\%$ credible contours), for a random subset of $16$ subjects.}\label{fig_sm:happ}
\end{figure}

Figure~\ref{fig:happ} compares the estimated scores with the true scores for one replicate of the setting where \citet{happ18}'s method has the highest accuracy according to the simulation results of Section~\ref{sec:happ}, 
namely, with $80$ observations per subject and variable, on average. The two methods perform similarly well; note that our approach permits quantifying the uncertainty around the scores thanks to posterior credible ellipses readily available from the estimated VMP posterior distributions.

\subsection{Addendum on the runtime profiling}\label{app:runtime}

\begin{figure}[h!]
\centering
\includegraphics[scale=0.56]{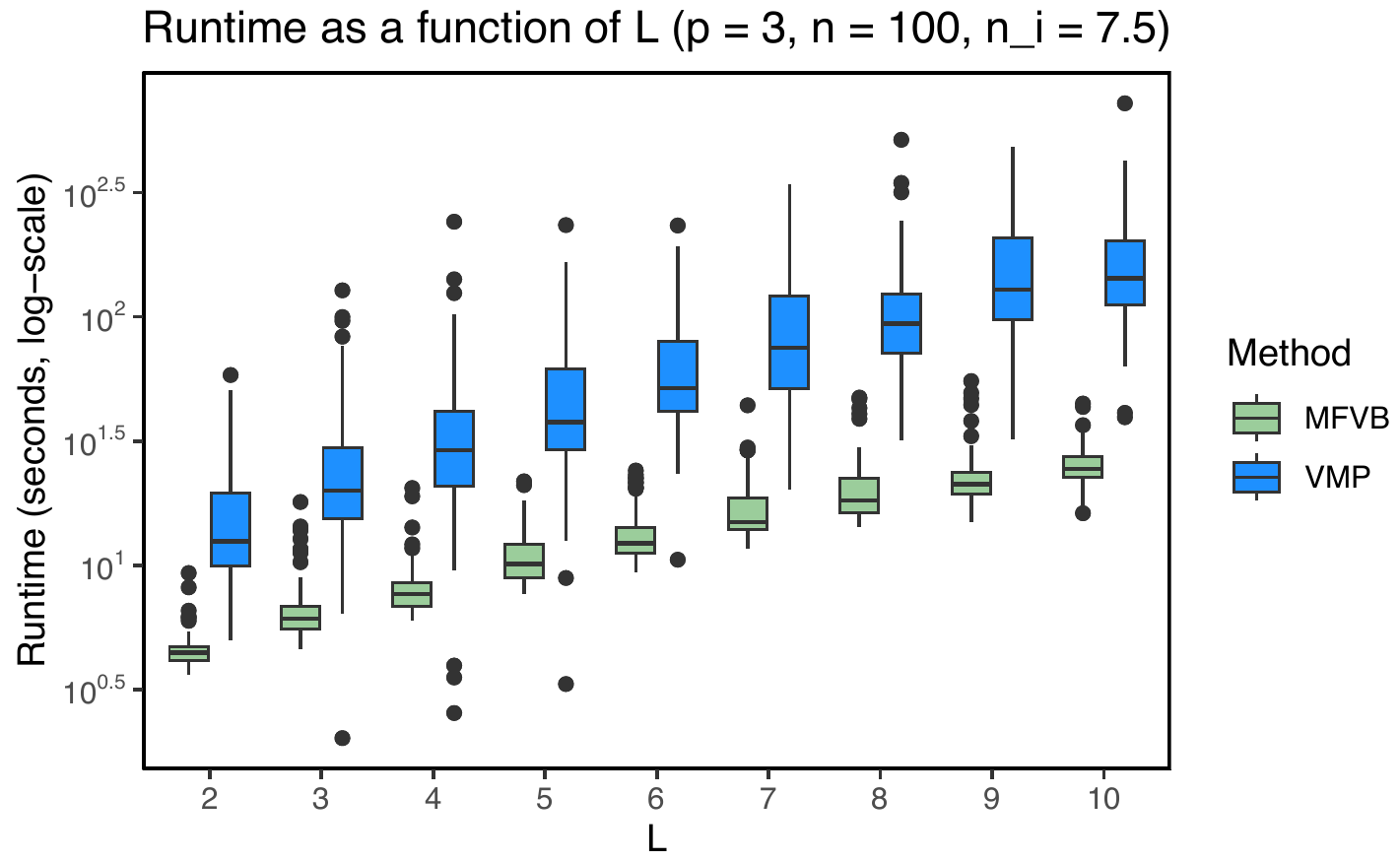}	

\caption{Runtime profiling (on the logarithmic scale) as a function of the number of estimated components, $L_\text{max}$, obtained on an Intel Xeon CPU, 2.60 GHz machine. Problem with $p = 3$ variables, $n = 100$ subjects, and $n_i = 7.5$ observations per subject on average ($100$ replicates).}\label{fig_sm:runtime}
\end{figure}

Figure~\ref{fig_sm:runtime} shows the runtime of the MFVB and VMP algorithms as a function of the upper bound on the number of components, $L_\text{max}$ (see Section~\ref{sec:runtime} of the main text for details on the runtime profiling setting).

\bibliography{bibliography}
\bibliographystyle{plainnat}

\end{document}